\documentclass[prb,twocolumn,amssymb,floatfix,amsmath,letterpaper]{revtex4}

\usepackage{amsfonts}
\usepackage{bm}
\usepackage{color}
\usepackage{graphicx}

\newcommand{\yy}{\hat{y}}
\newcommand{\xx}{\hat{x}}
\newcommand{\zz}{\hat{z}}
\newcommand{\aaa}{\hat{a}}
\newcommand{\bbb}{\hat{b}}
\newcommand{\bb}{\vec{b}}
\newcommand{\s}{{\bm{s}}}
\newcommand{\pp}{{\bm{p}}}
\newcommand{\XX}{X}
\newcommand{\YY}{Y}
\newcommand{\ZZ}{Z}
\newcommand{\ket}[1]{\left|#1\right\rangle}

\newcommand{\mat}[4]{\left(\begin{smallmatrix}#1 & #2\\#3 & #4\end{smallmatrix}\right)}
\newcommand{\matl}[4]{\left(\begin{matrix}#1 & #2\\#3 & #4\end{matrix}\right)}
\newcommand{\Zz}{\ensuremath{\mathbb{Z}_2}}
\newcommand{\Zn}{\ensuremath{\mathbb{Z}_n}}
\newcommand{\rr}{\vec{r}}
\newcommand{\bAx}{X}
\newcommand{\bAy}{Y}
\newcommand{\cyc}{\gamma}
\newcommand{\cy}{\cyc^1}
\newcommand{\cyy}{\cyc^2}
\newcommand{\sug}{\Sigma_g}
\newcommand{\su}{\Sigma}
\newcommand{\bbx}{X_1}
\newcommand{\bby}{X_2}
\newcommand{\omx}{\omega^{(1)}}
\newcommand{\omy}{\omega^{(2)}}
\newcommand{\om}{\omega}

\begin{document}

\title{Changing topology by topological defects in three-dimensional topologically ordered phases}

\author{Andrej Mesaros$^1$, Yong Baek Kim$^2$ and Ying Ran$^1$}

\affiliation{$^1$Department of Physics, Boston College, Chestnut Hill, MA 02467, USA}
\affiliation{$^2$Department of Physics, University of Toronto, Toronto, Ontario M5S 1A7, Canada}

\begin{abstract}
A hallmark feature of topologically ordered states of matter is the dependence of ground state degeneracy (GSD) on the topology of the manifold determined by the global shape of the system. Although the topology of a physical system is practically hard to manipulate, recently it was shown that in certain topologically ordered phases, topological defects can introduce extra topological GSD. Here the topological defects can be viewed as effectively changing the topology of the physical system. Previous studies have been focusing on two spatial dimensions with point-like topological defects. In three dimensions, line-like topological defects can appear. They are closed loops in the bulk that can be linked and knotted, effectively leading to complex three dimensional manifolds in certain topologically ordered states. This paper studies the properties of such line-defects in a particular context: the lattice dislocations. We give an analytical construction, together with support from exact numerical calculations, for the dependence of the GSD on dislocations of certain doubled versions of the exactly solvable Kitaev's toric code models in both two and three dimensions. We find that the GSD of the 3d model depends only on the total number of dislocation loops, no matter how they are linked or knotted. The results are extended to $\Zn$ generalizations of the model. Additionally, we consider the phases in which the crystalline orders are destroyed through proliferation of double dislocations. The resulting phases are shown to host topological orders described by non-Abelian gauge theories.
\end{abstract}

\date{\today}

\maketitle

\tableofcontents

\section{Introduction}

The introduction of topological order as the underlying principle of the fractional quantum Hall effect (FQH)~\cite{Wen:1995p6287} was central to the development of this new paradigm in modern condensed matter physics, which parallels the ubiquitous symmetry breaking mechanism. Topologically ordered phases, by definition, are quantum states of matter having long-range quantum entanglement, which can be characterized via the topological entanglment entropy\cite{Kitaev:2006p7455,Levin:2006p7456}. Topologically ordered phases can exist in both two and three spatial dimensions.

In topologically ordered phases, non-trivial excitations such as fractional charges may arise and realize anyonic statistics in two spatial dimensions~\cite{WEN:1991p6377}. Tunneling of virtual pairs of such fractional charges across the system can lead to a degenerate ground state, which implies non-trivial ground state degeneracy (GSD) that depends on the topology of the underlying physical system~\cite{Oshikawa:2006p5869},  which is another hallmark of topological order\cite{WEN:1991p6377}. A system with excitations is typically also described by degenerate states, and spatial manipulations of excitations correspond to unitary transformations (Abelian or non-Abelian) of the degenerate states, potentially realizing quantum computational operations~\cite{Kitaev:1997p6232,Kitaev:2003p6185,Nayak:2008p6311}. Recent years saw experimental implementations of anyonic excitations and other ingredients of quantum computation in photonic, superconducting, ultracold atoms, and most promisingly in non-Abelian FQH systems [for a review, see Refs.\onlinecite{Ladd:2010p6403,Stern:2010p6400}].

In particular, the specific dependence of the ground state degeneracy on the real space topology of the physical model is a crucial signature of topological order, but is hard to probe directly\cite{Wen:1990p5870}. Recently, a novel proposal to circumvent this constraint was made\cite{Barkeshli:2012p7361} for a FQH system on a (2d) lattice. That proposal is based on the mapping of a flat band with Chern number $N$ onto $N$ independent quantum Hall systems (``layers''), while these layer degrees of freedom are connected by the two-dimensional lattice translations. Lattice dislocations (translational topological defects) are then shown to act as ribbons connecting the layers, effectively setting the quantum Hall system on a higher genus surface, although the physical system shape is unchanged. Further, the dislocations themselves carry a non-trivial quantum dimension even in an Abelian FQH system\cite{Barkeshli:2012p7361}. In earlier studies\cite{Kitaev:2006p6266,Bombin:2010p6412}, it was also shown how lattice dislocations in the 2d Toric Code model can change the anyonic excitations of the model. This series of works demonstrate qualitatively new features of topological defects (point-defects) in topologically ordered phases in two spatial dimensions; namely these defects can carry extra GSD.

However, in three spatial dimensions, topological defects can be either point-like or line-like. In three spatial dimension, a new ingredient in the problem is that the line-like defects can be linked or knotted. For example, crystalline dislocations in a three-dimensional lattice are line-like defects which form closed loops. Therefore a dislocation itself can have a complicated topology if it is knotted or linked with other dislocation loops.

Natural questions emerge: Are there qualitatively new features associated with line-like topological defects (e.g., dislocations) in 3d topologically ordered phases? Can the line-like defects carry extra GSD? In addition, because the GSD is a topological property of the system, GSD can only be a topological invariant of the links and/or knots of the dislocation loops in 3d. Is the GSD a trivial or non-trivial topological invariant of the links/knots?

It is interesting to speculate what would happen if the GSD was indeed a non-trivial topological invariant of the links/knots. In this case, the non-trivial topological invariant is distinct from the conventional topological invariants of links/knots (such as the Jones polynomial\cite{Jones:1985p7711,Witten:1989p7712}): The conventional topological invariants of links/knots are mappings from links/knots to a complex number. But here, the possible non-trivial topological invariant of the links/knots is a mapping from links/knots to a Hilbert space associated with the GSD. We note that this new type of topological invariants of links/knots has attracted some interest in mathematics\cite{Khovanov:1999p7713} and physics\cite{Witten:2011p7714}.

At this point, it is worth mentioning that qualitatively new features of topological defects in topological phases without long-range quantum entanglement (and thus \emph{not} topologically ordered) were shown quite some time ago. Examples include Majorana fermions bound to vortices of p+$i$p topological superconductors\cite{Ivanov:2001p5868}, and helical modes bound to dislocations in 3d topological insulators\cite{Ran:2009p1115}. Even general understanding for such phenomena is available in the context of non-interacting fermion systems\cite{Teo:2010p4577}. However, our understanding of topological defects in topologically ordered phases is probably far from complete.

Motivated by these questions, in this paper we attempt to obtain a general understanding, to a certain level, of the extra GSD induced by dislocation loops in certain topologically ordered phases in 3d. We mainly focus on a 3d topologically ordered phase described by $Z_2\times Z_2$ gauge group, in which the lattice translation symmetry interchanges the two $Z_2$ gauge sectors. Generalization to other topologically ordered phases with $Z_n\times Z_n$ gauge groups is also presented. To study the effects of dislocation defects in these 3d phases, we use exactly solvable models as a powerful tool.

In the past, important insights about topological order came from classes of exactly solvable models in two\cite{Kitaev:2003p6185,Kitaev:2006p6266,Levin:2005p6399,Wen:2003p7601,Wang:2010p6408,Yao:2011p6419} and three\cite{Hamma:2005p6133,Ryu:2009p6255,Mandal:2009p6256,Levin:2005p6399} dimensions. The paradigmatic exactly solvable model is the two-dimensional ``toric code'' \cite{Kitaev:2003p6185} (2dTC) defined through interactions of spin-$\frac{1}{2}$ sites on a square lattice. This model was used to originally introduce the concept of topological quantum computation using anyons\cite{Kitaev:2003p6185}. Signatures of its Abelian anyonic excitations have been observed and further manipulation schemes are developed\cite{Pachos:2009p6404,Lu:2009p6407,Xue:2011p6405,Zhang:2008p6406,Brennen:2009p6238}. Parallel to such efforts, the toric code has been generalized to three dimensions\cite{Hamma:2005p6133} (3dTC). Both 2dTC and 3dTC are described at low energies by a $\Zz$ gauge theory, describing therefore $\Zz$ topological order. However, the 3dTC is predicted to stay topologically ordered even at finite temperatures due to its dimensionality\cite{Castelnovo:2008p6134}. The 3dTC is still a relatively simple model, and therefore we can study in detail its properties on the three-dimensional lattice.

We calculate the effect of arbitrary dislocation loops in an exactly solvable three-dimensional doubled Kitaev toric code\cite{Hamma:2005p6133}. The model system consists of two copies (``flavors'') of the 3dTC, displaced by half of lattice constant along one direction (say $\yy$), which allows dislocation lines with the minimal Burgers vector $\bb=\yy/2$ to connect the two flavors (sublattices) and therefore induce non-trivial topologies of the total (three-dimensional) manifold on which the model exists. We find that our constructive analytical results, which are corroborated by exact numerical calculations, show that this Abelian model is \textit{not} sensitive to the knotting and linking of dislocation lines, but can only distinguish the total number of dislocation loops in the crystal lattice. The topological ground state degeneracy in a lattice with periodic boundary conditions scales simply with the number of dislocation loops ($k\geqslant 1$) as $|GSD|=2^6\cdot 2^{k-1}$ ($|GSD|=2^6$ when dislocations are absent). We generalize this result to the doubled $\Zn$ toric code in three dimensional periodic lattice in the presence of dislocations, where the ground state degeneracy is $|GSD|=n^6\cdot n^{k-1}$, again independent of linking and knotting of the dislocation loops. This implies a zero-temperature entropy contribution which is extensive in the number of dislocation loops, and we will show it seems to persist in a realistic system with open boundaries (i.e. without periodicity).

We find somewhat similar results in the two-dimensional doubled Kitaev toric code, where a dislocation loop takes the form of a pair of pointlike edge dislocations (with Burgers vectors $\bb$ and $-\bb$). The GSD of the lattice with periodic boundary conditions (equals $2^4$ in absence of dislocations) is also not influenced by a single dislocation pair, but scales with number of pairs $k\geqslant 1$ as $|GSD|=2^4\cdot 4^{k-1}$. Further, $|GSD|=n^4\cdot n^{2k-2}$ for doubled $\Zn$ toric code in periodic system in two dimensions.

A further fundamental question arises: What is the state of matter obtained by making the dislocations dynamical excitations? In 2d, we consider a state where ``double dislocations'' ($|\bb|=1$, therefore not mixing flavors) proliferate and melt the lattice. Then, a tunneling event of a $\bb=\yy/2$ dislocation pair introduces a $\Zz$ twist across which the flavors are exchanged. We show that the resulting theory is described by a non-Abelian group $G_n\equiv(\Zn\times\Zn)\rtimes\Zz$ by analyzing the ground state degeneracy on the torus. We further derive the degeneracy on arbitrary genus oriented surfaces, which reveals the quasiparticle content of the theory. In the case $n=2$, i.e. the melting of simplest doubled toric code, both the degeneracy and the quasiparticle content explicitly match the properties of topological states with (non-Abelian) dihedral group $D_4$, as is already indicated by the isomorphism $D_4\simeq(\Zz\times\Zz)\rtimes\Zz$. In 3d, we focus on the melted state on a three-torus, i.e. a three-dimensional melted lattice with periodic boundary conditions. Also, we only consider the original toric code, i.e. $n=2$, and find the ground state degeneracy of the melted state, which exactly matches the degeneracy of the $D_4$ topologically ordered state on the three-torus. Therefore, there is very strong evidence that the condensation of double dislocations, which carry no flux nor charge in the Abelian $\Zn\times\Zn$ theory, gives rise to a non-Abelian gauge theory in both two and three dimensions. The defect condensation promotes a global symmetry (the $\Zz$ flavor) to a gauge symmetry, and can be seen as a physical way to connect phases that exhibit different topological order and different global symmetry, i.e. connect different symmetry enriched topological phases.

This paper is organized as follows. The majority of the discussion is focused on the doubled Kitaev toric codes, and we leave the detailed analysis of the generalized doubled $\Zn$ toric code for Appendix~\ref{sec:appendixZn}. We start with the complete analysis of the doubled 2dTC model in Section~\ref{sec:doubl-two-dimens}. The model description including dislocations, and proof of the ground state degeneracy are presented in detail, and also serve as a warm-up for the 3d case. The exact numerical results for the 2d model are presented jointly with the 3d case in Section~\ref{sec:numerical-results}. In Section~\ref{sec:doubl-three-dimens} we introduce the three-dimensional lattice model and construction of dislocation loops. Some details of the construction are relegated to Appendix~\ref{sec:appendixsurgery}. We present the behavior of stabilizer constraints, the string and membrane operators, and the exact numerical results for GSD in Sections~\ref{sec:numerical-results},~\ref{sec:numer-analys-stab}. Section~\ref{sec:cont-loops-membr} introduces the rules for string operators we found by analyzing the lattice. This facilitates the presentation of our main analytical results for arbitrarily linked dislocations with, and without, knotting in Sections~\ref{sec:gsd-stabilizers-k} and \ref{sec:dislocation-knotting}, respectively. The detailed analysis from the dual perspective of membranes is presented in Appendix~\ref{sec:membr-oper-persp}. Section \ref{sec:emergent-non-abelian} presents the analysis of the theory of melted lattice, with emerging non-Abelian gauge theory. We close with a discussion of $\Zn$ models (which is studied in detail in Appendix~\ref{sec:appendixZn}), the influence of system boundaries, as well as some implications of our study.

\section{The two-dimensional doubled toric code}
\label{sec:doubl-two-dimens}

The 2dTC is a very well known\cite{Kitaev:2003p6185} model that has $Z_2$ topological order. We describe our doubled variant in detail below.
%%%%%%%%%%%%%%%%%%%%%%%%%%%%%%%%%%%%%%%
\begin{figure*}
%  \begin{widetext}  
  \centering
\includegraphics[width=1\textwidth]{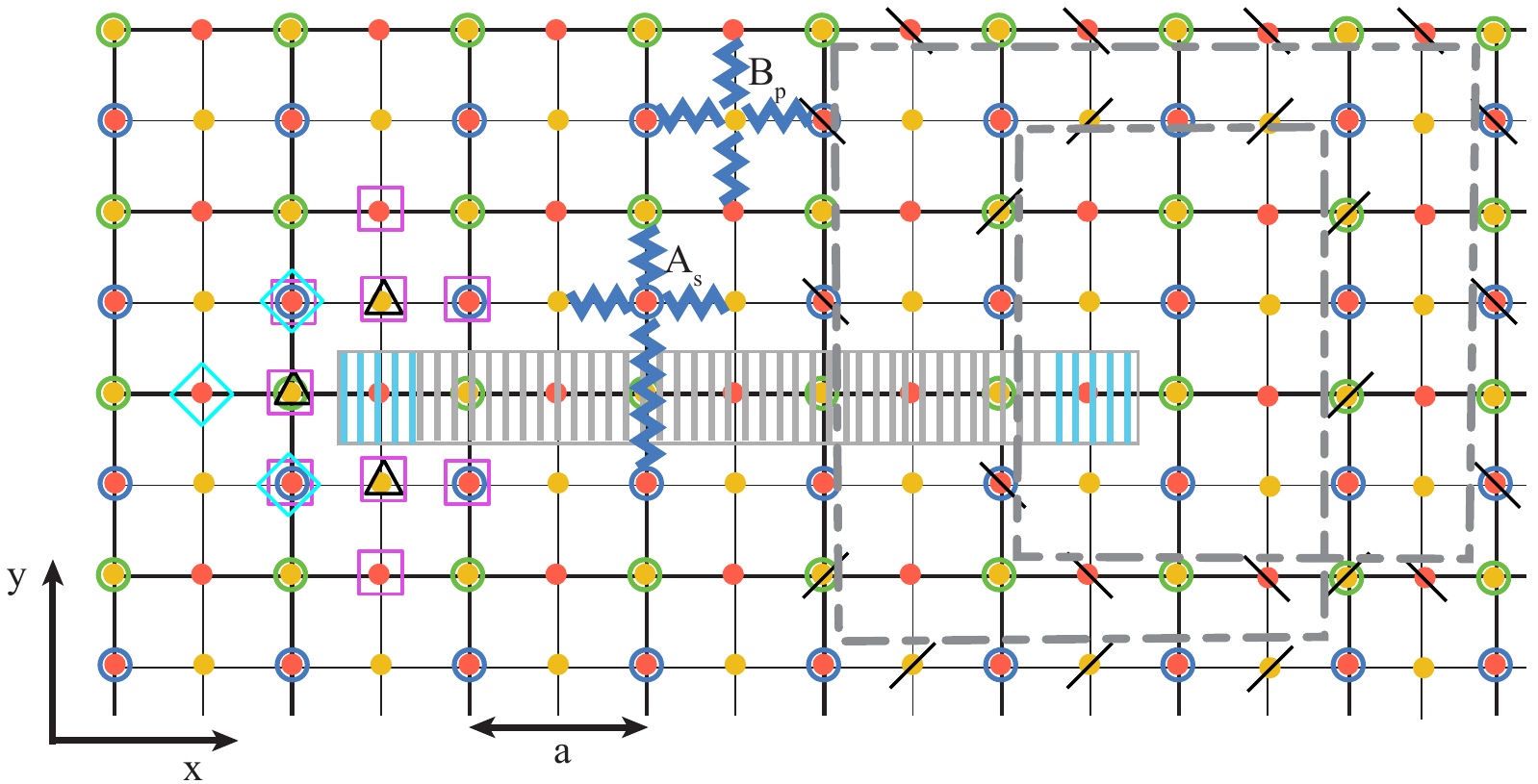}
\caption{The 2d doubled toric code with a dislocation pair. Red/yellow dots are spins of flavor $\alpha/\beta$, two copies displaced by $a\yy/2$. Green/blue circles mark centers of $\alpha/\beta$ star operators. Removing all sites within the shaded area (the Volterra, ``$\alpha\beta$'', line) leaves two dislocations (positioned at blue shaded sites). $A_s$ is a trivially  locally repaired star operator (acting with $\sigma^x$ on neighboring spins labeled by wavy lines). Near the dislocation site, non-trivial local surgery reinstates commuting stabilizers: Star marked by black triangle now acts on three light blue diamond spins only; Two plaquettes marked by black triangles are replaced by the merged 9-spin stabilizer (acting on violet square spins). If the dislocation was sitting at a star operator site (example not shown), the two new stabilizers would look the same, except that the assignment star/plaquette would be exchanged. Dashed line follows a lattice string loop $\mathbb{P}$, giving operator $\Sigma_{2d}$ acting through $\sigma^z$ on crossed-out spins. $\mathbb{P}$ is constructed locally step-by-step, always hopping between two star lattice sites and acting on the spin site they share. Notice the flavor change (flavor is also marked by slant of cross-out lines) when crossing the $\alpha\beta$ line, leading to the obligatory even winding around the dislocation. Multiplying $\Sigma_{2d}$ with the plaquette $B_p$ locally changes the shape of the loop $\mathbb{P}$, but not the (possible) independence of string loop $\Sigma_{2d}$ from the set of plaquette stabilizers. The $\Sigma_{2d}$ is actually trivial, because the loop is contractible. One must explicitly use the repaired plaquettes near the dislocation site to show this.}
\label{fig:2d0}
 % \end{widetext}
\end{figure*}
%%%%%%%%%%%%%%%%%%%%%%%%%%%%%%%%%%%%%%% 

The square lattice of one 2dTC copy has lattice constant $a\equiv 1$ and spin$-\frac{1}{2}$'s on lattice links. The second copy of the 2dTC is displaced by half the lattice constant along $\yy$ direction, this direction being vertical in Fig.~\ref{fig:2d0}. The two copies are called ``flavors'', having flavor index $f=\alpha,\beta$. When we discuss one copy of the model, i.e. one fixed flavor, it is understood that we consider the copy of the square lattice assigned to that copy, having $a=1$. Star operators (circles of the same color in Fig.~\ref{fig:2d0}) are positioned at lattice sites and act only on the four neighboring spins (Fig.~\ref{fig:2d0}):
\begin{equation}
  \label{eq:10d} A^f_{\s}=\sigma^x_{\s+\xx/2}\sigma^x_{\s+\yy/2}\sigma^x_{\s-\xx/2}\sigma^x_{\s-\yy/2},
\end{equation}
where $\sigma^i_R$ is the $i$-th Pauli matrix acting on the spin at position $R$, while $s$ is a lattice site position in the $f$ flavored lattice. Each square plaquette of the lattice centered on $\pp\equiv \s+(\xx/2+\yy/2)/2$ for some $\s$, carries a plaquette operator $B$ acting on 4 neighboring spins:
\begin{equation}
  \label{eq:2d} B^f_{\pp}=\sigma^z_{\pp+\xx/2}\sigma^z_{\pp+\yy/2}\sigma^z_{\pp-\xx/2}\sigma^z_{\pp-\yy/2}.
\end{equation}
The centers of plaquettes belonging to flavor $\alpha (\beta)$ are easily recognized as positioned on spins of flavor $\beta (\alpha)$, when there is no star operator (circle) on that spin already, see Fig.~\ref{fig:2d0}. The displacement of two copies by $\yy/2$ (vertically) creates alternating columns of star and plaquette operators in the total 2d lattice.

The Hamiltonian
\begin{equation}
H=-\sum_{f\in\{\alpha,\beta\}}\left(\sum_\s A^f_\s+\sum_\pp B^f_\pp\right)
\end{equation}
is formed by the star and plaquette operators (``stabilizers''), defined on a lattice with periodic boundary conditions. Since all stabilizers commute, the ground state degeneracy follows from counting the number of independent stabilizers.\cite{Gottesman:1996p7600,Nielsen:2000p7597} In a lattice of $N$ squares (per flavor), there are $2N$ spins, $N$ plaquettes, and $N$ stars, so that the number of \textit{locally} unconstrained spin$-\frac{1}{2}$ degrees of freedom is
\begin{equation}
  \label{eq:3d}
  N_{spin}-N_{stab}=0
\end{equation}
The non-trivial GSD follows from (spatially) \textit{global} constraints obeyed by the stabilizers. For each flavor, there are 2 global constraints: the product of all stars is the identity operator, as well as the product of all plaquettes,
\begin{align}
  \label{eq:13d}
  \prod_s A^f_s&=\openone\\
  \prod_p B^f_p&=\openone ,
\end{align}
because Pauli matrices act twice on each spin. The two unconstrained spin$-\frac{1}{2}$ give
\begin{equation}
  \label{eq:12d}
|GSD_{ideal2d}|=2^2\cdot 2^2.
\end{equation}

Importantly for our purpose, the GS manifold can be explicitly labeled using additional operators. There are two types (per flavor) of string operators\cite{Kitaev:2003p6185}. Consider a path (``string'') $\mathbb{P}_f$ along the $f$-flavored lattice edges, connecting two star positions: all spins $i$ on the links within the path are $i\in\mathbb{P}_f$. Similarly, a dual path (string) $\bar{\mathbb{P}}_f$ connects plaquette centers in the dual lattice of flavor $f$. The string operators are then:
\begin{align}
  \label{eq:4d}
  \Sigma_{2d}^f(\mathbb{P}_f)&=\prod_{i\in\mathbb{P}_f}\sigma^{z}_i\\
  \Xi_{2d}^f(\bar{\mathbb{P}}_f)&=\prod_{i\in\bar{\mathbb{P}}_f}\sigma^{x}_i.
\end{align}
If a string forms a closed loop, the string operator commutes with all stabilizers, and therefore with $H$, because a string without endpoints must share an even number of spins with any stabilizer. Further, a string shape $\mathbb{P}/\bar{\mathbb{P}}$ on the lattice can be deformed by multiplying the string operator $\Sigma_{2d}/\Xi_{2d}$ by $B/A$ stabilizers (see Fig.~\ref{fig:2d0}). Such an operation preserves the dependence or independence of the string operator from the stabilizers. Then, $\Sigma_{2d}/\Xi_{2d}$ is the trivial operator if $\mathbb{P}/\bar{\mathbb{P}}$ is contractible on the lattice in that way. The only non-contractible closed strings (loops) span the periodic system. These string loops are therefore independent from the set of stabilizers, and can be used to label the degenerate GS. A $\Sigma^f_{2d}$ loop spanning the $\XX$ ($\YY$) direction has an odd number of intersections with a $\Xi^f_{2d}$ string spanning $\YY$ ($\XX$), which causes their anticommutation. Due to their smooth deformations (using stabilizers), independent string loops are exhausted by topologically inequivalent ones. Any loop with multiple windings along the system can be deformed into a product of the elementary $\XX$  and $\YY$ spanning ones. Therefore, a single flavor has two pairs of anticommuting $\Sigma_{2d},\Xi_{2d}$ (the $\XX,\YY$ and $\YY,\XX$ spanning pairs), giving the expected total GSD from Eq.~(\ref{eq:12d}).

We can now move on to a lattice with dislocations. Generally speaking, dislocations in 2d are pointlike, only of edge type, and are operationally obtained by removing a finite line (the Volterra line) of lattice sites and repairing the lattice between the line's endpoints (see Fig.~\ref{fig:2d0}). The endpoints mark two dislocations, translational topological defects\cite{Kleinert:1989p3863}. We only consider dislocations with Burgers vector $\bb=\yy/2$, and therefore only remove \textit{single lines of sites} lying along the $\xx$ direction in Fig.~\ref{fig:2d0}. The lattice is locally repaired seamlessly across the Volterra line away from its endpoints, and so the line itself is unphysical\cite{Kleinert:1989p3863}, and only its endpoints matter. The Burgers vector topology says that any loop encircling a single dislocation will experience the ``jump'' due to the removed sites, even though it is impossible to say where the removed sites were.

In the case of the 2dTC, our choice of $\bb$ leads to the mixing of the two copies of the model (the two flavors $f$) explicitly through the local repair of stabilizers near the removed line (example of $A_\s$ in Fig.~\ref{fig:2d0}). However, strictly speaking, the global assignment of flavor becomes impossible due to the topological obstruction. We can however always assign flavors locally (starting from a desired point), and use it to build up arbitrary operators of fixed flavor step-by-step from the selected point, using the repaired local lattice links. The switch of flavors induced by a dislocation will always appear when we follow a string back to its starting point. Equivalently, and conveniently, we instead retain the global flavor assignment of the original lattice \textit{together} with an explicit choice of the removed site (Volterra) line.

\begin{figure}
%  \begin{widetext}  
  \centering
\includegraphics[width=0.45\textwidth]{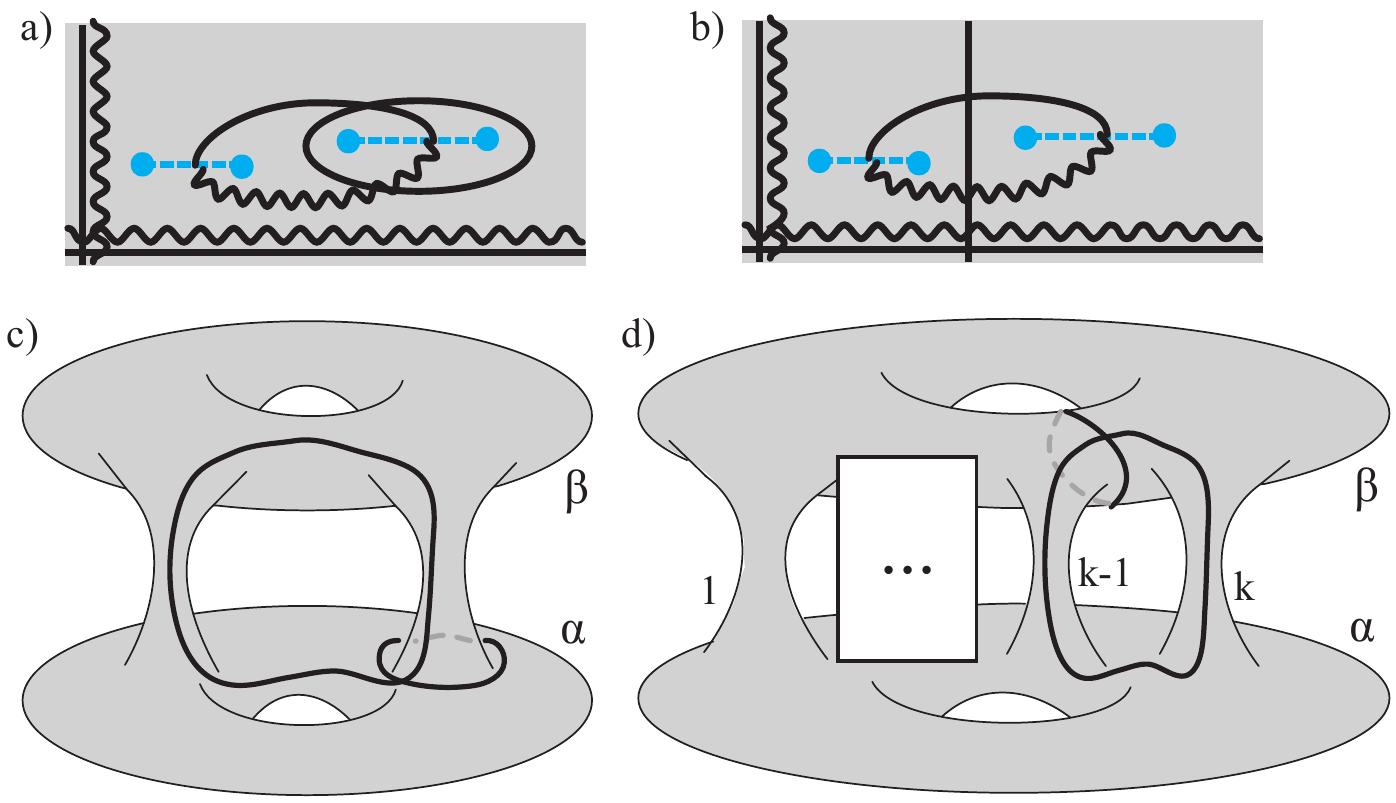}
\caption{The GSD resolving non-contractible $\Sigma_{2d}$ strings in the 2dTC (the $\Xi_{2d}$ strings are topologically the same). The Volterra $\alpha\beta$ line is dashed blue, $\alpha$/$\beta$ flavored strings are straight/wavy black lines, dislocations are blue dots. a) Two dislocation pairs and the relevant strings, periodic boundary conditions are used. b) Another, equivalent, set of GSD resolving strings. c) The underlying manifold of the model. Two tori are the two flavor copies of the system, and the dislocation induced $\sigma^z$ strings from (a) are shown. d) The $2+(k-1)$ genus manifold on which the $k$ dislocation pair model lives, with the 2 new $\Sigma_{2d}$ strings (of the type as in (b)) added per each new pair after the first.}
\label{fig:2d}
 % \end{widetext}
\end{figure}

We can now repeat the counting of degrees of freedom from Eqs.(\ref{eq:3d}),~(\ref{eq:13d}). Around the dislocation positions the lattice is heavily distorted, and so the stabilizers there have to be carefully and explicitly reconstructed such that they commute with all others, as shown in Fig.~\ref{fig:2d0}. Along the Volterra line (shaded region of Fig.\ref{fig:2d0}), each removed spin is also labeling a removed stabilizer. At a dislocation point however one stabilizer is lost (two original $\alpha,\beta$ stabilizers next to it merge into one). However, also two global stabilizer constraints are lost. Namely, the new stabilizers near the defects are still local but act on spins of both flavor, because of which now the flavor separated global constraints of Eq.~(\ref{eq:13d}) are merged 
\begin{align}
  \label{eq:6d}
  \prod_{f,s\in S}A^f_s\prod_{s'\in D_s}\tilde{A}_{s'}&=\openone\\
    \prod_{f,p\in P}B^f_p\prod_{p'\in D_p}\tilde{B}_{p'}&=\openone,
  \end{align}
  where the sets $S$, $P$, $D_s$ and $D_p$ exhaust all stabilizers in the dislocated lattice: original stabilizers and ones repaired along the Volterra line are in the sets $S$ and $P$ (having well defined local flavor), while $D_s,D_p$ are the sets of heavily modified stabilizers near the dislocation cores. In a lattice with single dislocation pair therefore the changes in counting unconstrained degrees of freedom cancel, leaving the GSD intact. For $k$ dislocation pairs, compared to ideal lattice, we get additional $2k$ degrees of freedom, but still only lose $2$ global constraints. We therefore get additional $2k-2$ unconstrained degrees of freedom, predicting
  \begin{equation}
    \label{eq:11d}
    |GSD_{2d}|=|GSD_{ideal2d}|\cdot 2^{2k-2}=2^4\cdot 4^{k-1},
  \end{equation}
for $k\geq1$.
Exact numerical lattice calculation as introduced in Sections~\ref{sec:numerical-results},~\ref{sec:numer-analys-stab} shows that
\begin{equation}
  \label{eq:8d}
  |GSD_{2dnum}|=2^4\cdot 4^{k-1},\;\;k\geqslant 1,
\end{equation}
at least for $k\leqslant 3$, and it also corroborates the independent string operators we will present in Fig.~\ref{fig:2d}. The following discussion is based on a nice geometric interpretation in Ref.\onlinecite{Barkeshli:2012p7361}.

The neat feature of the 2dTC is that we can visualize the topology of the model manifold entirely. The 2dTC starts as two separate copies of the two-torus $T^2$, one per flavor. The $\pi_1(T^2)=\mathbb{Z}^2$ gives the $\XX$ and $\YY$ spanning non-contractible loops per flavored torus.

Next we introduce a single dislocation pair. Because of the local flavor change upon crossing the Volterra line of removed sites (as explained above), we call this line the ``$\alpha\beta$ line". As an example, Fig.~\ref{fig:2d0} shows an explicit string loop which winds around a dislocation. The string operator is constructed by definition, in a step-by-step manner: Starting from a star location and hopping to a neighboring star with which it shares a link, the spin site on this link is acted on by $\sigma^z$ and becomes a step in the path $\mathbb{P}$ of the string Eq.~(\ref{eq:4d}). The string is forced to wind (at least) twice around the defect, and in general an even number of times, because it must change flavor an even number of times to be able to close into a loop.

Generally, knowing how strings change flavor upon crossing the $\alpha\beta$ line between dislocations, we see that the $\alpha\beta$ line represents a connection between the two flavored tori. This connection is a smooth tube, because we can show from the lattice model (e.g. using Fig.~\ref{fig:2d0}) that a loop encircling the $\alpha\beta$ line (therefore both defects) can use the $\alpha\beta$ line to switch flavors, but cannot contract through it. Therefore the loop switches flavored tori by sliding along the a tube that connects them, see also Fig.~\ref{fig:2d}c.

Furthermore, the string loop winding twice around a single dislocation, shown as a string example in Fig.~\ref{fig:2d0}, is contractible (so giving the identity operator). This is a non-trivial fact which follows from the definition of heavily edited stabilizers near the dislocation site. Going to the continuum description, this loop can be contracted on the ``connection tube'' between the tori (this geometric interpretation was first introduced in Ref.\onlinecite{Barkeshli:2013p7606} as ``wormhole''), because it does not wind around the tube (and is therefore not a new type of non-contractible string loop). Summarily, the $k=1$ system can be topologically accurately described by a torus with two holes (the original holes of the two flavored tori), a manifold whose fundamental group also has four generators and therefore the same GSD as the ideal system.

Adding another pair of dislocations creates another tube connection between the two flavored tori, essentially adding another hole to the manifold, Fig.~\ref{fig:2d}b. This adds two anticommuting non-contractible string pairs, giving a GSD enhancement by factor $2^2$. One pair has $\Sigma_{2d}$ encircling the new dislocation pair and $\Xi_{2d}$ piercing both $\alpha\beta$ lines (therefore changing flavor twice); the other pair has $\Sigma_{2d}$ and $\Xi_{2d}$ exchanged. Obviously the two strings of such pairs intersect once due to their flavor structures, see Fig.~\ref{fig:2d}a.

For $k\geqslant 1$ dislocation pairs (Fig.~\ref{fig:2d}d), the model lives on a manifold with $2+(k-1)$ holes, thus having genus $g=k+1$. We name this two-manifold $T^2_g$. Since $\pi_1(T^2_g)=\mathbb{Z}^{2g}$, there are $2g$ independent, say $\Sigma_{2d}$, strings. Another useful property of $T^2_g$ is that one can always choose the non-contractible paths in it such that they, in pairs, intersect exactly once. This ensures pairwise anticommutation with $\Xi_{2d}$ strings (which as a set do not topologically differ from the $\Sigma_{2d}$ strings), and leads to a
\begin{equation}
  \label{eq:9d}
    |GSD_{2d}|=2^{2g}=2^{2k+2},\;\;k\geqslant 1
  \end{equation}
  degeneracy, exactly in accord with Eqs.(\ref{eq:8d}) and (\ref{eq:11d}).
  
  %%%%%%%%%%%%%%%%%%%%%%%%%%%%%%%%%%%%%%% 
%%%%%%%%%%%%%%%%%%%%%%%%%%%%%%%%%%%%%%% 

\section{The three-dimensional doubled toric code}
\label{sec:doubl-three-dimens}

%%%%%%%%%%%%%%%%%%%%%%%%%%%%%%%%%%%%%%%
\begin{figure}
  \centering
\includegraphics[width=0.49\textwidth]{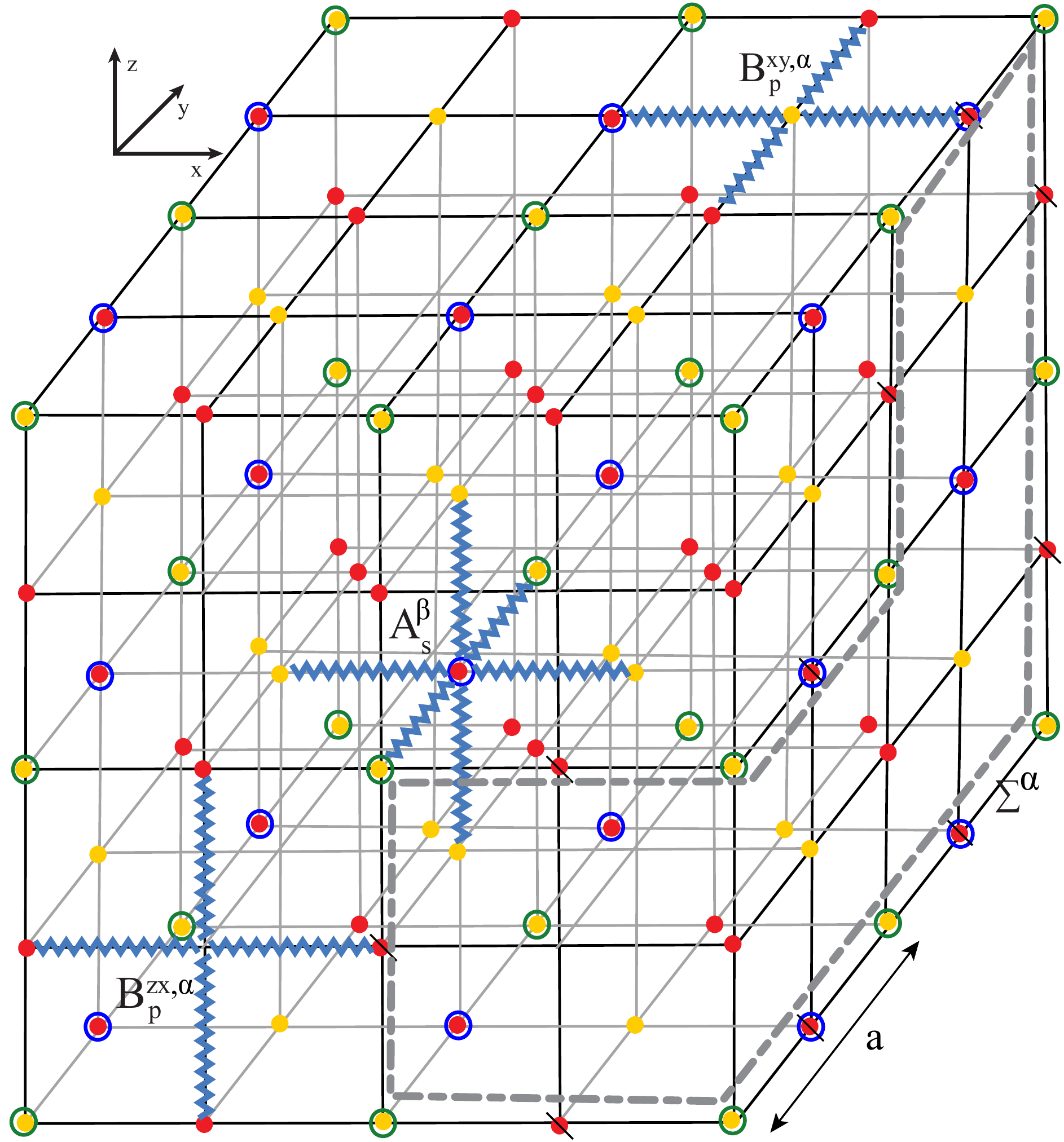}
\caption{Three-dimensional doubled toric code (compare to 2d system in Fig.~\ref{fig:2d0}). Red/yellow dots are positions of spin$-\frac{1}{2}$ of flavor $\alpha$/$\beta$, and they differ by a translation by $\yy/2$. For convenience, we draw the cubic lattice having lattice spacing $1/2$ in all directions, so that spins, stars, plaquettes and local constraints, of both flavor, are all positioned on the sites of this ``dense lattice''. Flavor $\alpha$/$\beta$ star operators (green/blue circles) are positioned on $\alpha$/$\beta$ lattice sites and act through $\sigma^x$ operators on their six neighboring spins (blue wavy lines from $A^\beta_s$).
  % Green circles also label the original cubic lattice sites of flavor $\alpha$.
  Flavor $f=\alpha,\beta$ plaquette operator $B_p^{ab,f}$ acts by $\sigma^z$ on four spins of a $f$ square lying in $\aaa\bbb$ plane, centered on $\pp$. On the shown ``dense lattice'', a center $\pp$ of $\alpha$/$\beta$ flavored square face is always positioned either: at an un-circled $\beta$/$\alpha$ spin (example $B_p^{xy,\alpha}$ shown), or at an empty position (center of $\beta$/$\alpha$ cube, having no spin, nor star), example $B_p^{zx,\alpha}$. A closed string operator (example $\Sigma^\alpha$), forms the dashed loop $\mathbb{P}$ (example lying in $\YY\ZZ$ and $\ZZ\XX$ planes of $\alpha$ lattice) and acts by $\sigma^z$ on spins in $\mathbb{P}$ (crossed-out). Multiplying $\Sigma^\alpha$ with shown $B^\alpha_p$ operators deforms the $\mathbb{P}$ loop. A center of a cube in the $f$ flavored lattice is an empty position (no spin, nor star) in the shown ``dense lattice'', and it labels a local constraint $\prod B_p^{ab,f}=\openone$ for the 6 plaquettes $\pp(\aaa\bbb)$ on the 6 sides of the given $f$ flavored cube.}
\label{fig:7a}
\end{figure}
%%%%%%%%%%%%%%%%%%%%%%%%%%%%%%%%%%%%%%%

The 3dTC has been studied in detail\cite{Hamma:2005p6133,Castelnovo:2008p6134}, but for completeness we introduce all its necessary ingredients here (see Fig.~\ref{fig:7a}) as we describe its doubled variant.

Spin$-\frac{1}{2}$ degrees of freedom belonging to one flavor (say $\alpha$) occupy the midpoints of edges in a cubic lattice, where the lattice sites $\s\equiv (i,j,k)$ are labeled by integers, and lattice constant $a\equiv 1$. Each lattice site $\s$ carries a stabilizer $A_s$ called a star operator (Fig.~\ref{fig:7a}), which acts on 6 spins neighboring $\s$, i.e. on 6 spins positioned on lattice edges which share the site $\s$:
\begin{equation}
  \label{eq:1} A_s^\alpha=\sigma^x_{\s+\xx/2}\sigma^x_{\s+\yy/2}\sigma^x_{\s+\zz/2}\sigma^x_{\s-\xx/2}\sigma^x_{\s-\yy/2}\sigma^x_{\s-\zz/2},
\end{equation}
where $\sigma^i_R$ is the $i$-th Pauli matrix acting on the spin at position $R$. Each square face of a cube is centered on $\pp(\aaa\bbb)\equiv \s+(\aaa+\bbb)/2$, for some $\s$, and for $\aaa,\bbb$ some pair from $\{\xx,\yy,\zz\}$; such a face lies in the $ab$ plane and carries a plaquette operator $B_p^{ab}$ acting on 4 neighboring spins of $\pp(\aaa\bbb)$, i.e. acting on 4 spins on the lattice edges forming the square face (Fig.~\ref{fig:7a}):
\begin{equation}
  \label{eq:2} B_p^{ab,\alpha}=\sigma^z_{\pp(\aaa\bbb)+\aaa/2}\sigma^z_{\pp(\aaa\bbb)+\bbb/2}\sigma^z_{\pp(\aaa\bbb)-\aaa/2}\sigma^z_{\pp(\aaa\bbb)-\bbb/2}.
\end{equation}
In the following, it will mostly suffice to use the shorter label $B_p^f$ for the plaquette operator, where it is understood that the position $\pp$ determines the plane $\aaa\bbb$ in which the plaquette is lying.

The $\beta$ flavor of the model is created as a copy of the lattice translated by $\yy/2$ (Fig.~\ref{fig:7a}); $\beta$ spin sites are then always at a position where there is an $\alpha$ stabilizer site, and vice versa. Notice that every second $\XX\YY$ layer of the 3dTC model is reminiscent of a 2dTC lattice.

The Hamiltonian of the 3dTC is simply given by the negative sum of all stabilizers in the lattice:
\begin{equation}
  \label{eq:45}
H=-\sum_{f\in\{\alpha,\beta\}}\left(\sum_\s A^f_s+\sum_{\pp(\aaa\bbb)} B^{\aaa\bbb,f}_p\right).
\end{equation}

Setting $H$ on a three-dimensional lattice with periodic boundary conditions, all stabilizers commute. The ground state degeneracy therefore follows from counting the number of independent stabilizers.\cite{Gottesman:1996p7600,Nielsen:2000p7597} Each center of cube in the $f$ flavored lattice marks a local ``cubic'' constraint, because the action of six $f$ plaquette operators on the sides of that cube is the identity operator:
\begin{equation}
  \label{eq:19}
\prod_{\pp\in C_i} B^f_p=\openone,
\end{equation}
with $C_i$ being one of the $i=1,\ldots,N$ cubes in the $f$ lattice, and $\pp\in C_i$ being the six centers of squares on the surface of cube $C_i$. Per flavor, in a lattice of $N$ cubes there are $3N$ spins, $3N$ plaquettes, and $N$ stars, so that the number of locally unconstrained spin$-\frac{1}{2}$ degrees of freedom is
\begin{multline}
  \label{eq:3}
  N_{spin}-(N_{stab}-N_{indep.loc.constraints})=\\
  3N-(4N-N)=0,
\end{multline}
balancing as in the 2d case of previous section. However, only $N-1$ local cubic constraints from Eq.~\eqref{eq:19} are independent, since due to ${B^f_p}^2=\openone$ the product of $N-1$ local constraints gives exactly the $N$-th constraint in the system with periodic boundary conditions. This removes one stabilizer constraint. On the other hand, the product of all stars in the lattice gives identity:
\begin{equation}
  \label{eq:35}
\prod_{\s} A^f_s=\openone,
\end{equation}
adding one global constraint (per flavor). The counting balance of Eq.~(\ref{eq:3}) therefore still seems to remain valid.

However, there are additional global constraints obeyed by the stabilizers, from which the non-trivial GSD follows. The stars are ``volume-like'', since multiplying many adjacent stars leaves only a surface of spins acted on by $\sigma^x$. The plaquettes are ``surface-like'', since multiplying many adjacent plaquettes leaves only a closed loop of lattice edges on the boundary, on which $\sigma^z$ operators act (e.g. the loop operator in Fig.~\ref{fig:7a} is equal to the product of several adjacent plaquette operators). We already counted the global star constraint obtained by filling the system volume by multiplying all stars (thereby shrinking its surface acted on by $\sigma^x$ operators to nothing). The periodic system has a shape of the three-torus, and we can multiply plaquettes so as to create three surfaces spanning the system's $\XX\YY$, $\YY\ZZ$ or $\ZZ\XX$ planes, such that they have no edges and so no $\sigma^z$ action. This gives three global constraints. Each surface can be arbitrarily deformed by putting additional plaquettes into the product. But, demanding that the surface stays boundary-less (and therefore equal to $\openone$ operator), the elementary deformation must contain the product of plaquettes forming the surface of an entire cube. The product of plaquettes on a cube surface is locally constrained (Eq.~\eqref{eq:19}), so the global constraint given by the deformed surface is, up to local constraints, equivalent to the global constraint given by the undeformed surface. The three non-contractible (and boundary-less) surfaces spanning the three-torus therefore give the only inequivalent global constraints which are independent from the local constraints. The existence of three spanning surfaces follows from $\pi_1(T^3)=\mathbb{Z}^3$. Adding up two flavors, one gets
\begin{equation}
  \label{eq:14}
  |GSD_{ideal3d}|=2^6
\end{equation}
on the three-torus, i.e. on a lattice with periodic boundary conditions.

The operators which resolve the ground state manifold are represented by non-contractible string loops (Fig.~\ref{fig:7a}) and non-contractible closed membranes\cite{Hamma:2005p6133,Castelnovo:2008p6134}. A string
\begin{equation}
  \label{eq:4}
  \Sigma^f(\mathbb{P}_f)=\prod_{i\in\mathbb{P}_f}\sigma^z_i
\end{equation}
is a product of operators on spins positioned along a path $\mathbb{P}_f$ that follows the edges of $f\in\{\alpha,\beta\}$ flavored cubic lattice, starting and ending on a lattice site $\s$, on which there is a star $A_s^f$ (Fig.~\ref{fig:7a}). The string can obviously be deformed by multiplication with plaquettes, and is a straightforward generalization of 2d strings $\Sigma_{2d}$. If the string is closed, it commutes with all the stars too, as every star acting on an edge belonging to the string will also act on another of string's edges due to the string's continuity. A membrane
\begin{equation}
  \label{eq:5}
  \Gamma^f(\mathbb{S}_f)=\prod_{i\in\mathbb{S}_f}\sigma^x_i
\end{equation}
is defined on the dual lattice, and is built from operators acting on a single spin per dual cube face (see illustration in Ref.\onlinecite{Castelnovo:2008p6134}). The membranes can be deformed when multiplied by stars, which are in the centers of dual cubes. (This deformation is geometrically the same as when the global surface constraints on plaquettes are deformed by multiplication with local cubic constraints.) Since the membranes are closed (boundary-less), they commute with all plaquettes, and are generalizations of the 2d $\Xi_{2d}$ strings hopping on the dual square lattice.

By construction, a string shares exactly one spin with a membrane at their intersection point, which leads to their anticommutation in case the string pierces the membrane odd number of times. All closed membranes must be pierced even number of times by any string loop, except in the case of non-contractible objects which exploit the three-torus topology. On the dual cubic lattice, one can construct only the three independent closed membranes $\Gamma^f(\XX\YY),\Gamma^f({\YY\ZZ}),\Gamma^f({\ZZ\XX})$. Each is pierced exactly once by $\Sigma^f(\ZZ),\Sigma^f(\XX),\Sigma^f(\YY)$ (strings spanning the three-torus), respectively, giving 3 anticommuting pairs per flavor. These strings and membranes commute with stabilizers and the Hamiltonian, so each anticommuting pair gives a two-valued quantum number labeling the degenerate ground states. Due to flavor $f$ there are in total 6 anticommuting pairs, giving the $|GSD_{ideal3d}|=2^6$.

\subsection{Dislocation loops}
\label{sec:dislocation-loops}

In this subsection we present the details of creating dislocations in the 3dTC model, which is a generalization of the 2d case from previous section. Due to our choice of displacement of flavored lattices along $\yy$, we only consider the nontrivial dislocations with $\bb=\yy/2$. The dislocation loops in $\ZZ\XX$ plane are therefore purely edge (we will call them ``edge dislocation loops''), while the $\XX\YY$ and $\YY\ZZ$ dislocation loops are of screw type along segments parallel to $\yy$ segments, and edge otherwise (these we simply call ``screw dislocation loops'').
%%%%%%%%%%%%%%%%%%%%%%%%%%%%%%%%%%%%%%%
\begin{figure}
  \centering
\includegraphics[width=0.45\textwidth]{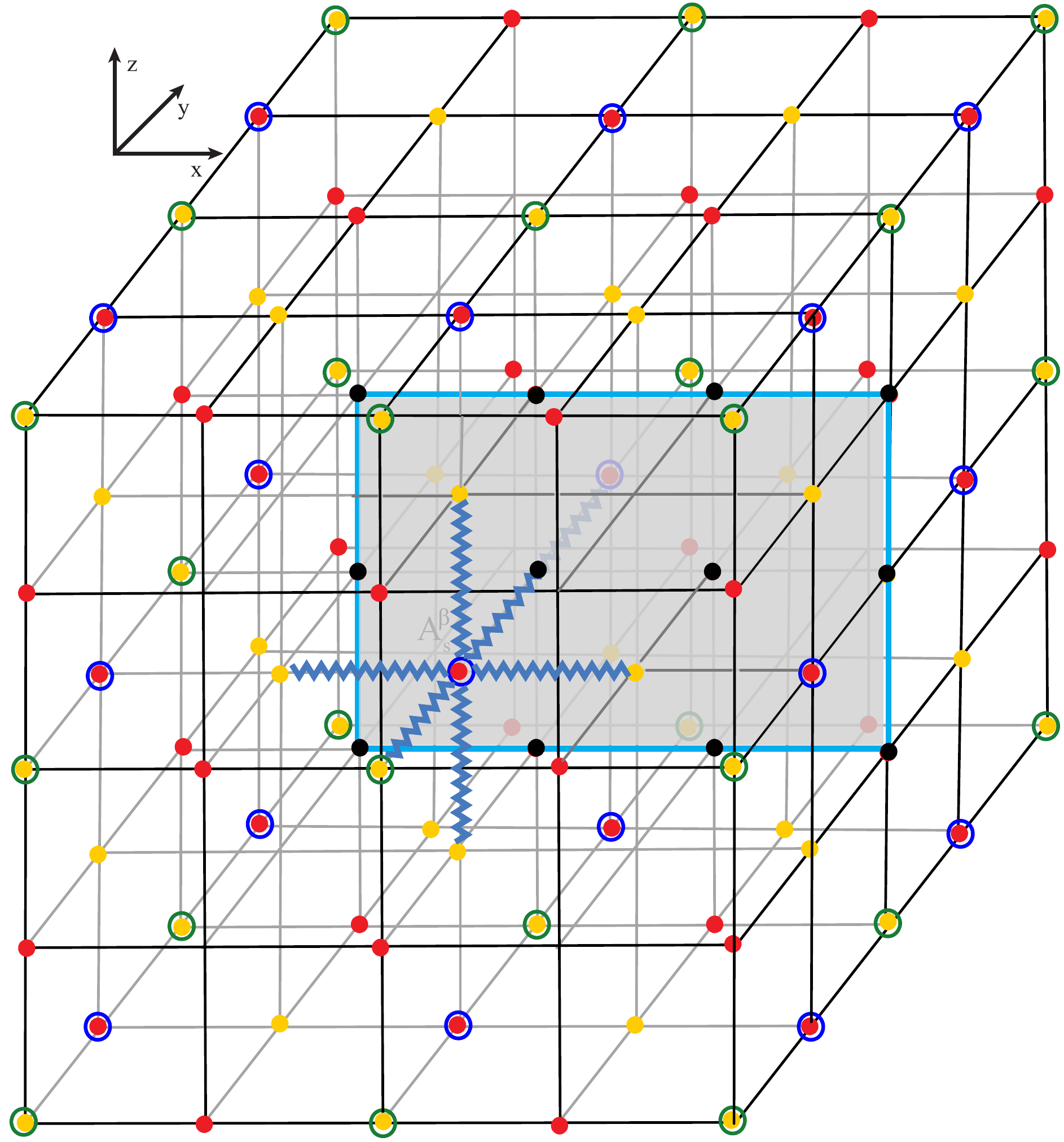}
\caption{Edge dislocation of an $\ZZ \XX$ loop (compare to Fig.~\ref{fig:2d0}). Black dots are removed sites, forming the $\alpha\beta$ surface shaded grey. The dislocation line is represented by the outmost array of black sites, marked by the blue line. The shown star operator $A_s$ is repaired by just gluing the neighboring sites together in the local neighbor network, and that is how the flavor jump occurs as a string operator which pierces the $\alpha\beta$ plane is constructed step-by-step (compare to Fig.~\ref{fig:2d0}). The exception to the seamless repair are stabilizers that contain spins on the dislocation lines, for which the explicit repair ``surgery'', such that all stabilizers commute, is needed, as shown in Fig.~\ref{fig:7b}.}
\label{fig:7g}
\end{figure}
%%%%%%%%%%%%%%%%%%%%%%%%%%%%%%%%%%%%%%%

To create an edge dislocation loop\cite{Kleinert:1989p3863}, we remove a surface of sites in a $\ZZ\XX$ plane, as in example of Fig.~\ref{fig:7g} (see Figs.~\ref{fig:7b},\ref{fig:7c} for details). Locally, the lattice, as well as the stabilizers, depend on the nearest neighbor network, which can be repaired by connecting two sites on either side, along $\yy$ direction, of a removed site. This reparation is locally indistinguishable from the original lattice, but fails at the dislocation line formed by the outermost array of removed sites, as dictated by the defect topology. The non-trivial reparation of stabilizers around the dislocation line is presented explicitly in Appendix~\ref{sec:appendixsurgery}. Once ``repaired'', all stabilizers in the lattice commute, and we use their negative sum for the Hamiltonian as in Eq.~\eqref{eq:45}.

The removal of a single layer of sites introduces a ``skip'' in the usual alternation of flavors along $\yy$, thereby inducing the flavor mixing. To be precise, the plane surface of removed sites is the Volterra surface, which is unphysical, i.e. could be arbitrarily placed in the lattice as long as it is bounded by the fixed dislocation line\cite{Kleinert:1989p3863}. This also means that the global assignment of flavor is impossible in presence of dislocations. However, we can always locally assign flavors in the lattice. Most importantly, following a string that starts out locally as flavor $\alpha$, following the local neighbor network, if it winds around a dislocation line and comes back near the start position, the ``skip'' will be felt (although impossible to say where it happened) and the string will not close on itself as it will locally be of $\beta$ flavor. (We need to perform one more wind around a dislocation line to be able to close the string.)

We can alternatively keep the original global flavor assignment, but then also keep the surface of removed sites as a fixed Volterra choice, for the simplest way for bookkeeping. Because of this, we will also call the chosen Volterra surface ``the $\alpha\beta$ surface'' as the flavor change will occur upon crossing it. This procedure only makes sense when considering closed strings and membranes which conserve the parity of number of times they cross the $\alpha\beta$ surface that spans a dislocation loop. Since our goal is analysis of GSD, the open strings and membranes carrying excitations on their endpoints/boundaries\cite{Hamma:2005p6133} will never be involved.

For the creation of screw dislocation loops, we need to perform a $\yy/2$ translation which is parallel to the chosen ``screw surface'' that spans the dislocation loop (no sites are removed), Figs.~\ref{fig:7d},\ref{fig:7e}. This translation in the local neighbor network also introduces the ``skip'' between flavors (along $\yy$), just as for the edge loops. All remarks concerning the $\alpha\beta$ surface remain as in the edge dislocation case.

The preservation of the local neighbor network preserves the commutation of stabilizers involved in the dislocation surface. However, the stabilizers involving spins on the dislocation line need to be repaired, removed or replaced in a non-trivial way. We explicitly construct the new stars and plaquettes for all types of dislocation lines and their corners. We leave the presentation of details for the Appendix, Figs.~\ref{fig:7b}-\ref{fig:7f}. The main outcome is that the recipe for lattice ``surgery'' produces the new local stars and plaquettes as well as local cubic constraints in such a way that the local counting of unconstrained degrees of freedom remains just as in Eq.~(\ref{eq:3}). This is true also for linked dislocation loops (when two $\alpha\beta$ surfaces intersect). However, the dislocations change the \textit{global} properties of the model. In our exact numerical calculations (Section~\ref{sec:numerical-results}) we corroborate the local balance of Eq.~(\ref{eq:3}) by showing that the GSD in particular does not depend on the \textit{sizes and shapes} of simple dislocation loops in the lattice. We further corroborate the prediction for the dislocation influence on global constraints (see Section~\ref{sec:cont-loops-membr}).

\section{Results}
\label{sec:results}

This section has three parts: We start by presenting our exact numerical calculation, which includes the GSD for various dislocation configurations, as well as the demonstration of independence of relevant string and membrane operators; next we present the analytical arguments that lead us to the particular choice of the strings and membranes to analyze, and we also give the prediction of the GSD; and finally we close with all results for the much simpler two-dimensional model.

\subsection{Exact numerical results}
\label{sec:numerical-results}

We analyze 3dTC on lattices with various dislocation configurations, as introduced above and in detail in the Appendix (especially captions of Figs.~\ref{fig:7b}-\ref{fig:7f}). The lattices have variable sizes, typically about 250, and up to 720 cubes, which are sufficient to study all presented scenarios.

We find the GSD directly using a quantum information theorem\cite{Gottesman:1996p7600,Nielsen:2000p7597}: Counting the number of spins in the (dislocated) lattice, and subtracting the number of independent stabilizers in that lattice gives the number of ``unconstrained'' spins; the GSD then equals the size of the spin (which is $2$ for spin$-1/2$) raised to the number of unconstrained spins.

The procedure for finding the number of independent stabilizers (stars and plaquettes separately, but including both flavors) we used is also direct. For example, consider first the stars, with $N_{star}$ being their total number. Each stabilizer is represented by the list of spins on which it acts. Since this star operator is just $\sigma^x$ or $\openone$ on any particular spin site, we just need to assign $1$ ($0$) if the $\sigma^x$ acts odd (even) number of times on the given spin. Therefore, on each spin, the operator is represented by the additive $\Zz$ algebra. There is a total of $N_s$ spins in the lattice, so we next form an $N_{star}\texttt{x}N_s$ matrix, the $i$-th row of length $N_s$ filled with 1 for the spins on which the $i$-th star acts, and filled with 0 on all other spins. Each star contributes a row to the matrix, as a candidate for an independent row under the $\Zz$ algebra. Finally, the matrix is converted to the row reduced echelon form under the algebra, and its rank found (or number of zero rows counted), which is the number of independent operators. The same procedure is used for the plaquettes.

We numerically studied up to four dislocation loops of varying sizes and shapes each. The linking did not influence the outcome, which only depends on the number of loops; among up to four loops, we could check: no linking; one linked pair; two separated linked pairs; three linked loops;  two parallel screw loops, each linked with the two parallel edge loops.

We get $|GSD_{ideal3d}|=2^6$ in the ideal lattice, as expected from Eq.~(\ref{eq:14}), and $|GSD|=2^6\cdot 2^{k-1}$ for $k$ dislocation loops in the cases $k\in\{1,2,3,4\}$.

A special case occurs when a dislocation line closes into a loop by spanning the periodic system (these lines need to occur in pairs). For one pair we get $|GSD|=2^5$, less than in the ideal lattice. The GSD does not change further by introducing spanning dislocation pairs in an orthogonal direction. However, we checked that upon introducing $k\in\{2,3,4\}$ parallel spanning pairs, the counting follows a trend similar to the 2dTC case $|GSD|=2\cdot2^{2k+2}=2|GSD_{2d}|$ (see ends of Sections~\ref{sec:gsd-stabilizers-k},~\ref{sec:cont-loops-membr} for details).

We performed the same numerical analysis on the 2dTC model, where we considered up to three dislocation pairs.

Finally, we constructed a 3dTC  lattice with a dislocation line knotted into the Trefoil knot, as shown in Fig.~\ref{fig:5}a. As our analytical arguments in the next section show, this knotted dislocation behaves as a simple dislocation loop, and indeed in the exact numerical calculation it gives the corresponding $|GSD|=2^6$.

The summary is presented in Table~\ref{tab:1}.

\begin{table}[h]
  \centering
  \begin{tabular}[c]{l|l}
    \hline\hline
    Configuration & GSD \\
    \hline
    \bf{3dTC}\\
    Ideal & $2^6$ \\
    $k\geqslant 1$ loops (independent of linking, knotting) & $2^6\cdot 2^{k-1}$ \\
    1 loop & $2^6$ \\
    1 spanning loop & $2^5$ \\
    2 orthogonal spanning loops & $2^5$ \\
    $k\geqslant 2$ parallel spanning loops & $2\cdot 2^{2k+2}$\\
    Trefoil knot dislocation & $2^6$\\
    \hline
    \bf{2dTC}\\
    Ideal & $2^4$ \\
    $k\geqslant 1$ dislocation pairs & $2^4\cdot 4^{k-1}$\\
    \hline\hline
  \end{tabular}
  \caption{GSD for $k\leqslant 4$ dislocation loops in the 3dTC, and $k\leqslant 3$ dislocation pairs in the 2dTC, found in exact numerical calculations.}
  \label{tab:1}
\end{table}

\subsection{Exact numerical analysis of stabilizers}
\label{sec:numer-analys-stab}

Using the above method, we could also check directly the independence of various strings $\Sigma^f$, membranes $\Xi^f$, and global plaquette constraints. Namely the strings can be deformed by multiplication with plaquette operators, and therefore two equivalent strings will not appear as independent additions to the set of all plaquettes. The same applies for the set of $\Xi^f$ and $A^f$. In particular, any contractible $\Sigma$ ($\Xi$) will also \textit{not} appear independent when added to the set of $B$ ($A$).

Using the analytical predictions (Section~\ref{sec:cont-loops-membr}), we could indeed check numerically, in lattices with up to $k=4$ loops, that all the predicted anticommuting non-contractible strings and membranes are indeed independent within the set of plaquettes and stars, respectively.

We delay the detailed presentation of practical construction of membranes on the lattice with dislocations to Appendix~\ref{sec:membr-oper-persp}, because the analytical proof that follows is presented through the string perspective.

\subsection{Continuum strings}
\label{sec:cont-loops-membr}

In this subsection we analyze the behavior of GSD by counting non-contractible strings, and give the prediction for GSD of unknotted dislocation loops with allowed linking.
%Strictly speaking, the non-contractible membrane part of the analysis is also needed to complete our arguments about the GSD. Although the membrane analysis provides insight, we streamline the discussion by relegating it to Appendix~\ref{sec:membr-oper-persp}.

Our present goal is to find the set of independent non-contractible string operators. Such strings necessarily pairwise anticommute with a set of independent non-contractible membranes (string piercing membrane odd number of times), thereby enumerating the GSD. To streamline the presentation, we therefore only count the strings, and delay the (in itself illuminating) analysis of membranes to Appendix~\ref{sec:membr-oper-persp}.

To this end, we have derived, by direct analysis of all situations on the 3dTC lattice (Figs.~\ref{fig:7b}-\ref{fig:7f}), a set of general rules that the strings obey as they are deformed and moved through the dislocated lattice by the action of plaquettes. We of course use the plaquettes obtained by the proper ``surgery'' in the dislocated lattice (Appendix~\ref{sec:appendixsurgery}). Having in mind the detailed discussion in Section~\ref{sec:dislocation-loops}, we fix the Volterra $\alpha\beta$ surface of each dislocation loop, and retain the global flavor assignment from the original ideal lattice.

\begin{figure}
%  \begin{widetext}  
  \centering
\includegraphics[width=0.48\textwidth]{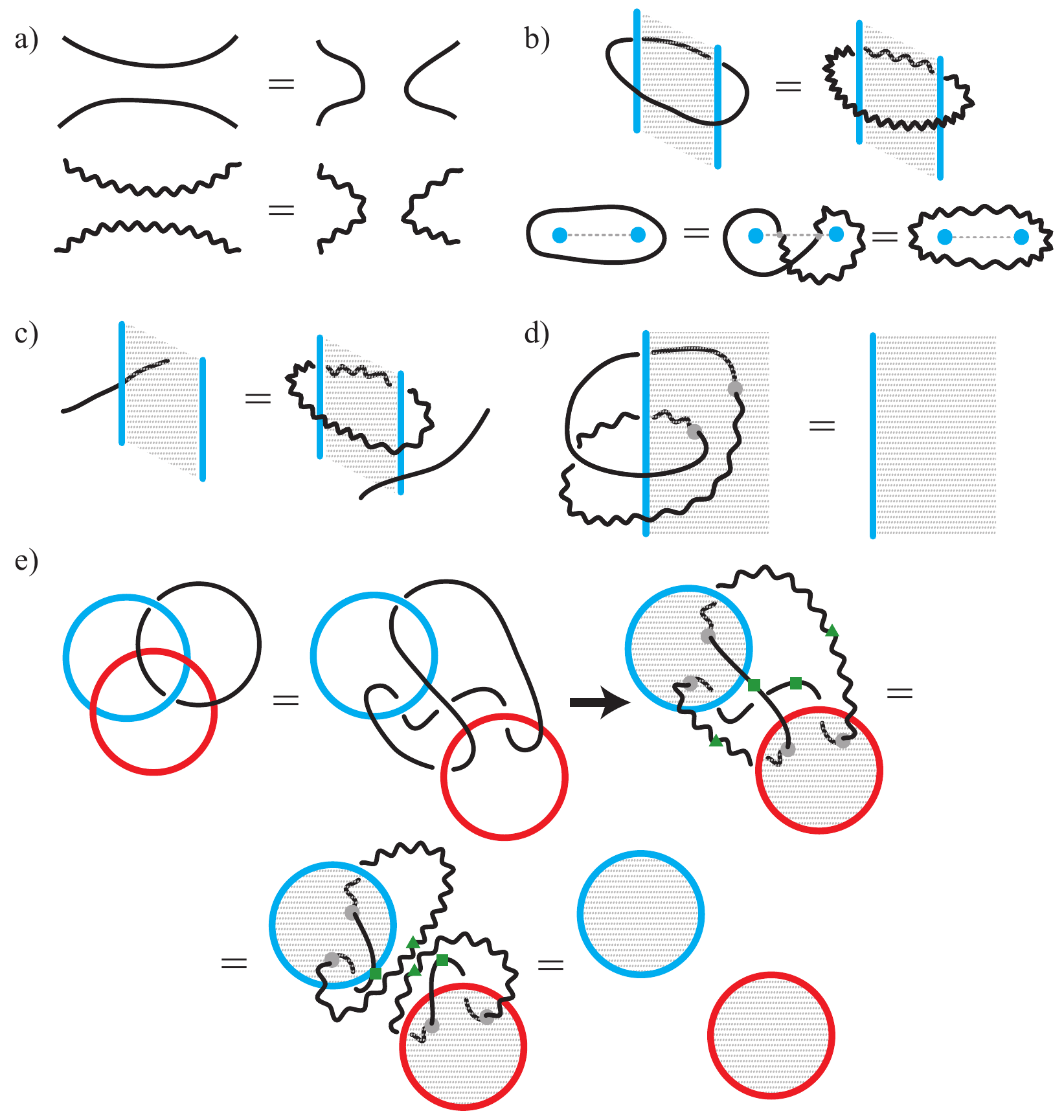}
\caption{Elementary allowed operations on string operators in presence of dislocation loops, derived from their properties on the lattice. Straight and wavy black lines represent strings of two flavors, $f=\alpha,\beta$, blue lines are dislocation lines, and the dotted surface is the Volterra $\alpha\beta$ surface (see Section~\ref{sec:dislocation-loops}). Dots mark the piercing of string through $\alpha\beta$ surface. (a) Reconnection of two string segments. (b) Local string flavor change using dislocation; a top view detailing the process is also shown. (c) Local passage of string through dislocation line. (d) String winding $w=2$ times around dislocation line is contractible; only even $w$ is possible due to flavor continuity, all contractible. (e) An example of contracting a string, using rules (a) and (d), for a string (black line, but its flavor marked only after second subfigure) which is non-trivially linked with two dislocation loops (blue and red) in a Borromean ring configuration. Green symbols mark the points of reconnection (rule (a)). The string is therefore not independent from plaquette operators, i.e. can be written as some product of those.}
\label{fig:1}
% \end{widetext}
\end{figure}

The fundamental set of lattice rules we derived for the behavior of strings is shown in Fig.~\ref{fig:1}. Intuitively, the string is an elastic reconnectable ``rubber band'' which switches flavor only upon crossing through an $\alpha\beta$ surface. \textit{The string cannot pass through a dislocation line, but can be contracted if it winds an even number ($w\in 2\mathbb{Z}$) of times around the same dislocation line.} The last property is crucial, but not intuitive. Deriving it demands the precise use of the stabilizers obtained by ``surgery'' on the lattice at the dislocation line. (A similar property holds in the 2dTC case, see Fig.~\ref{fig:2d0}.)
%%%%%%%%%%%%%%%%%%%%%%%%%%%%%%%%%%%%%%%
%%%%%%%%%%%%%%%%%%%%%%%%%%%%%%%%%%%%%%%

We analyze the consequences of these derived rules on two simple but insightful examples, obtaining outcomes that match exact numerical lattice results:
\begin{itemize}
\item A single dislocation loop does not change the GSD. First, we note that the standard strings which span the system in 3d are still independent, and free to roam through space avoiding the finite dislocation loop. However, every new candidate non-contractible string loop \textit{must} pierce the $\alpha\beta$ surface at least twice (or generally an even number of times): the string has to change flavor even number of times along its length to be able to close into a loop. But then it also winds an even number of times around the dislocation loop line, so it is contractible according to the special derived rule in Fig.~\ref{fig:1}d. Imagine an alternative candidate string that spans the system twice in one direction, piercing the $\alpha\beta$ surface twice along the way. We can combine it with both flavored (and purportedly independent) standard system-spanning strings in the same direction, to reconnect them into non-system-spanning loops, of which one of either flavor will be far from the dislocation and trivially contractible; the leftover candidate is just the type we considered (winding locally around the dislocation). Other combined possibilities can also clearly be reduced to these cases.
  \item 
    Two separate dislocation loops add \textit{one} non-contractible string (and therefore one anticommuting operator pair when an appropriate membrane is considered, leading to GSD enhancement by factor of 2).
    The new candidate \textit{non-contractible} string can now exist because there are \textit{two separate} $\alpha\beta$ surfaces. The candidate pierces both of them once, therefore winding only once around each dislocation loop line. This string is not contractible, as we cannot change the parity of times it pierces any $\alpha\beta$ surface, neither by deformation, nor by attempted (but forbidden) passage of the string across a dislocation line.
  \end{itemize}

In the two dislocation loop case, it is amusing to consider a loop that does not wind around either dislocation line, yet seems ``stuck'' because it is linked with them. We consider such a Borromean ring configuration of two dislocation loops and one string loop, Fig.~\ref{fig:1}e. Because the string pierces through each $\alpha\beta$ surface an even number of times, the string is contractible. We also check this numerically on the lattice by showing that such a string is not independent from the set of plaquettes on this lattice.

\subsection{Proof of GSD and non-contractible strings for $k$ unknotted dislocation loops}
\label{sec:gsd-stabilizers-k}

One can now generalize to $k$ separate dislocation loops. Here we consider simple dislocation loops, having oriented, non-self-intersecting $\alpha\beta$ surfaces. It is simple to see that the addition of a new separate dislocation to the system adds one independent string, increasing the GSD by factor $2$. Focus on a candidate for a new independent string; it must pierce the new $\alpha\beta$ surface an odd number of times, since we will now show that otherwise it could be deformed until completely separated from the new dislocation loop, becoming therefore equivalent to some string in the old set of strings. The odd number $2n+1$ ($n\geqslant 0$) of piercings of the new $\alpha\beta$ surface is consequently equivalent to a single piercing. To show this, consider a candidate string with an even number $2n$, $n\geqslant 1$, of piercing points through the new dislocation's $\alpha\beta$ surface. Each of $2n$ piercing points has two string segments of opposite flavor emanating to the two sides of $\alpha\beta$ surface. We can deform the string in the vicinity of the new dislocation loop to reconnect the even number $2n$ of emanating segments of either color until we obtain a collection of $n$ disconnected loops, each containing exactly two piercing points. These loops with two piercing points are contractible either trivially, or by the nontrivial rule if they wind $w=2$ times around the dislocation line (see also Fig.~\ref{fig:5}). All the $2n$ piercing points are thus removed, and the string remains completely separated from the new dislocation loop.
% (The string candidates must also pierce the old $\alpha\beta$ surfaces an odd number of times, due to flavor continuity.)

So the set of candidates for new non-contractible strings boils down to strings which pierce the new dislocation loop once. These are however all equivalent. Imagine there are two which are independent. They both pierce the new loop once, and we can deform them to coincide in the region around the piercing point. By reconnecting them at the points where they deviate from each other, they just cancel on the piece containing the piercing point (where they coincided), while the remaining piece is a string that is completely separate from the new dislocation loop. It therefore must be dependent (a combination of) the old strings. The two new candidates differ only by an old string, and are therefore not independent of each other in the enlarged set of all non-contractible strings.

The linking of dislocation line loops surprisingly leaves the situation unchanged. Consider two linked dislocation loops: their $\alpha\beta$ surfaces intersect along some finite line (see Fig.~\ref{fig:2}b for illustration). However, a string piercing each of the intersecting $\alpha\beta$ surfaces once cannot be contracted. Namely, the line of intersection of the $\alpha\beta$ surfaces does not allow the passing of the piercing points from one surface to the other. Therefore, the linked loops behave as separated loops, at least in terms of the topology of strings, membranes, and the GSD.

Adding the extra factor 2 to the GSD for each dislocation loop after the second one, gives the numerically corroborated
\begin{equation}
  \label{eq:7}
  |GSD|=2^6\cdot 2^{k-1}\quad\text{for $k\geqslant 1$ (possibly linked) loops}.
\end{equation}

Finally, a dislocation line pair which spans the system along some direction (say $\ZZ$) warrants a note. The GSD is lowered --- there are no new independent string candidates, but instead one non-contractible string spanning the system along $\ZZ$ is lost. Namely, a string spanning $\ZZ$ direction can switch flavors by passing in its entirety through the $\alpha\beta$ surface which also spans the $\ZZ$ direction. Therefore the two flavors for the $\ZZ$ spanning string are not independent anymore. The system-spanning $XY$ membranes of two flavors also merge into one when such dislocations are present. For an array of such dislocations, one can make a mapping to the 2dTC, as also discussed from the membrane viewpoint in Appendix~\ref{sec:membr-oper-persp}.

\subsection{Dislocation knotting}
\label{sec:dislocation-knotting}
  %%%%%%%%%%%%%%%%%%%%%%%%%%%%%%%%%%%%%%%
\begin{figure*}
%  \begin{widetext}  
  \centering
\includegraphics[width=0.9\textwidth]{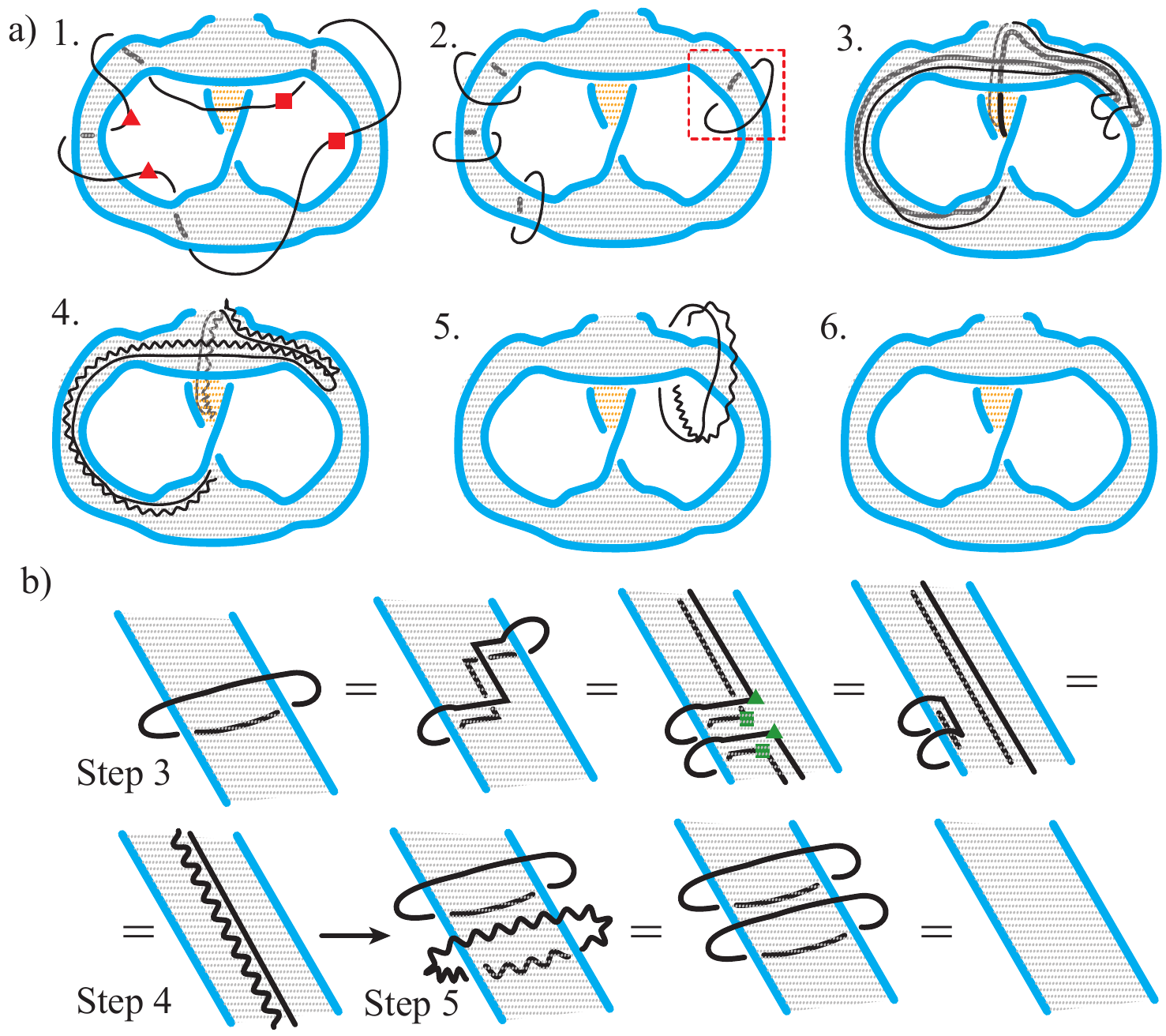}
\caption{Contracting a string that is tangled with a dislocation which forms a Trefoil knot, where the string does not pierce the $\alpha\beta$ surface. This is an example of contracting a $w=0$ string loop, needed for the proof in Section~\ref{sec:dislocation-knotting}. a) 1. A $w=0$ string, to be reconnected at the red symbols. 2. It reconnects into four elementary string loops, cycles which pass through neighboring holes in the $\alpha\beta$ surface (and do not pierce it). 3. Focusing on one (marked in previous step) of those cycles, one uses the single dislocation line to reconnect it into two non-piercing string loops. 4. A string loop changes flavor when pushed through the surface. 5. The doubled loops can again be reconnected into elementary string loops (in this example no need, as it is already such). 6. Every elementary loop is doubled, and these can be contracted locally using the local $\alpha\beta$ surface. b) Zoom-in on typical operations needed for steps 3,4 and 5.}
\label{fig:6}
 % \end{widetext}
\end{figure*}
%%%%%%%%%%%%%%%%%%%%%%%%%%%%%%%%%%%%%%% 

This subsection analyzes the case of knotted dislocation loops. Since we dealt with linking previously, it is enough to study a single knotted dislocation line loop.

Before we show why dislocation knotting in general does not influence the GSD value, we need to focus on the nature of the Volterra $\alpha\beta$ surface, which becomes non-trivial for a dislocation line twisted into a knot.

\begin{figure}
%  \begin{widetext}  
  \centering
\includegraphics[width=0.49\textwidth]{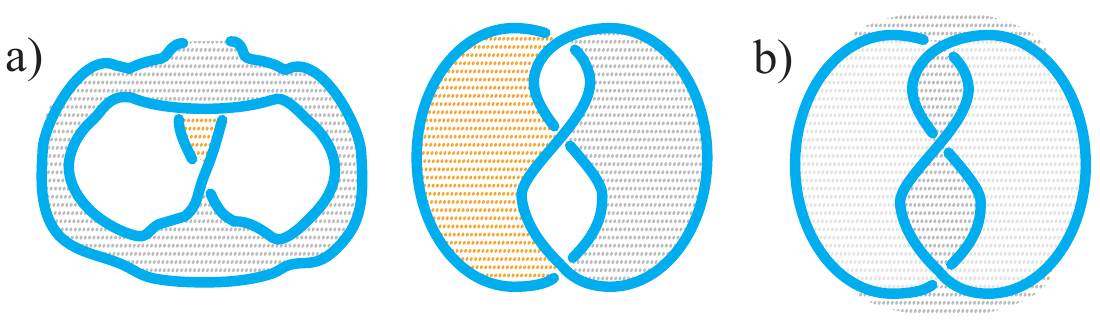}
\caption{The Volterra (``$\alpha\beta$'') surface of a dislocation line twisted into a Trefoil knot. a) A non-self-intersecting and orientable $\alpha\beta$ surface always exists, e.g. the Seifert surface (left and center, two colors for two sides of the surface). b) A non-orientable surface spanning a knot can also sometimes be found, and here it is the triple twisted M\"obius strip. See Fig.~\ref{fig:5}a for an implementation of the Trefoil knot through a self-intersecting $\alpha\beta$ surface.}
\label{fig:4}
 % \end{figure*}
\end{figure}

There are several ways of setting a general Volterra $\alpha\beta$ surface which spans a knotted loop. By the physical nature of the Burgers vector, the $\alpha\beta$ surface must be orientable, so we disregard the nonorientable option (for the Trefoil knot which is not carrying a Burgers vector charge, the triple twisted M\"obius strip could be an option\cite{Adams:1994p6411}, see Fig.~\ref{fig:4}b). However, the surface can still be self-intersecting, and we discuss such cases at the end of this subsection.

We first consider the non-self-intersecting option for the $\alpha\beta$ surface. Remarkably, such an orientable, connected, non-self-intersecting surface spanning an arbitrary knot can always be constructed\cite{Seifert:2012p6179,Adams:1994p6411}, and is called the Seifert surface. The explicit construction is based on a 2d projection of the knot, which is cut into discs, that are finally connected by twisted strips (see Fig.~\ref{fig:4}a for the Trefoil knot example). The resulting surface is non-unique and even has a non-unique genus (bounded from below), but it still satisfies the important criteria from above. Its (single) edge is topologically equivalent to the knot one started from. One can in principle construct such an $\alpha\beta$ surface even on the lattice, because it is possible to gradually twist the orientation (say in the $XY$ plane) of a strip of selected lattice sites, as one follows the strip along its length (say $Z$ axis).

Certainly, we can use the Seifert $\alpha\beta$ surface to prove the absence of non-contractible string candidates. The proof is straightforward, as all the elements have already been introduced in Section~\ref{sec:gsd-stabilizers-k}. The key properties are that:
\begin{description}
\item[P1] There is a single, orientable and connected $\alpha\beta$ surface.
  \item[P2] There is a single dislocation line (the edge of the $\alpha\beta$ surface).
  \end{description}
  The main steps of this proof are presented in Fig.~\ref{fig:6} for the example of a dislocaiton making a Trefoil knot.

We start from a candidate string that pierces the surface, necessarily even number of times, say $m=2n$ times, due to string continuity and flavor changes. Using property {\bf P1}, one can deform the string to bring all $2n$ piercing points next to each other on the $\alpha\beta$ surface, and close to some chosen point $R$ on the dislocation line, by using only smooth string deformations. (The deformation might make the string even more tangled-up, but this does not matter.) One then \textit{locally} (near the point $R$ on dislocation line) reconnects the bundle of string segments emanating from the piercing points ($2n$ points, giving $2n$ segments per flavor). The reconnection, in the sense of Fig.~\ref{fig:1}a, breaks up the string loop into a set of disconnected string loops. Only three kinds of strings loops can appear: 1) Small string loops that wind around the dislocation line near the point $R$; 2) Small string loops that pierce the $\alpha\beta$ surface near point $R$, but do not wind around the dislocation line; and 3) Potentially long and tangled-up string loops that have no contact with the $\alpha\beta$ surface. The strings of type (1) and (2) are immediately contractible, the former using the special rule from Fig.~\ref{fig:1}d, and the latter trivially. The strings of type (3) we call ``$w=0$ string loops'', since they do not pierce the $\alpha\beta$ surface, and therefore do not wind around a dislocation line in the sense of Fig.~\ref{fig:1}d.
  %%%%%%%%%%%%%%%%%%%%%%%%%%%%%%%%%%%%%%%
%%%%%%%%%%%%%%%%%%%%%%%%%%%%%%%%%%%%%%%

Our goal is to show that an arbitrary $w=0$ string loop is also contractible, which we do using a procedure demonstrated in Fig.~\ref{fig:6}. Focusing on one such string, it can either be completely detached from the knot, and therefore trivially contractible, or it can be entangled with the dislocation line knot by winding through the holes in the Seifert surface. The string is of single flavor because it does not pierce the $\alpha\beta$ surface. This makes it simple to reconnect the string loop until we obtain a set of ``elementary'' string loops, namely the string loops that are elementary cycles which enter and exit the Seifert surface through a pair of neighboring holes in the surface. This is always possible to do according to the properties of Seifert surface in {\bf P1}. An example of an elementary loop would be a string winding around one of the three twisted strips in the canonical Seifert surface of the Trefoil knot, and such a string example is marked in Fig.~\ref{fig:6}a2. Fig.~\ref{fig:6}a1,2 show the entire operation.

Let us therefore focus on a single elementary string loop, and show that it is contractible. The elementary loop is tightened close to the Seifert $\alpha\beta$ surface, and by its definition, it can be seen to wind around two short segments of dislocation line which span a short piece of Seifert surface between them. These two dislocation line segments are centered on points $A$ and $B$ of the dislocation line, respectively. Since we will use these objects in the vicinity of points $A$ and $B$, the two short segments of the dislocation line can be considered straight and parallel, and the piece of Seifert surface they span locally flat, as shown in Fig.~\ref{fig:6}b1,\ref{fig:6}a2.

We next manipulate two halves of the elementary string loop as shown in Fig.~\ref{fig:6}a3,\ref{fig:6}b1-4: one half stays hovering around the surface at point $A$; the other half of elementary string loop we drag all the way along the dislocation line starting from point $B$ until we reach the point $A$, which is always possible due to property {\bf P2}. What is created by the dragging is two parallel string segments connecting the half at $A$ to the dragged half (now also at $A$). The locally parallel string segments are positioned tightly below and above the surface (which is orientable!). This is illustrated in Fig.~\ref{fig:6}a3,b2,b3. These two segments we call $d$ and $u$, respectively, and by construction there must be a piece of Seifert surface between them all the way along their length. At point $A$ we can now reconnect the string loop, so that the ``stay-at-$A$ half'' and the ``drag-from-$B$ half'' cancel, while each $u$ and $d$ connect into a complete loop, see Fig.~\ref{fig:6}b3,4. The $u$ and $d$ are geometrically almost identical: they are offset by a short distance having the Seifert surface between them. Since the surface is orientable, we can push $d$ through it, and according to standard rules end up with two copies of the $u$ string loop, carrying opposite flavors. These twin strings are again just $w=0$ string loops and we can decompose them into elementary cycles, just as we did with the original $w=0$ string loop. The elementary string loops that we obtain now consist of two copies each, having different flavors. The earlier situation in Fig.~\ref{fig:6}b1 now becomes the one in Fig.~\ref{fig:6}b6. The elementary loops can make use of the flavor flip property (Fig.~\ref{fig:1}b), and therefore we end up with two copies of the string having the same flavor, which just cancel each other.

This completes the argument that any string loop constructed in presence of a knotted dislocation line loop is contractible, and therefore can be written as some product of plaquette operators.

\begin{figure}
%  \begin{widetext}  
  \centering
\includegraphics[width=0.48\textwidth]{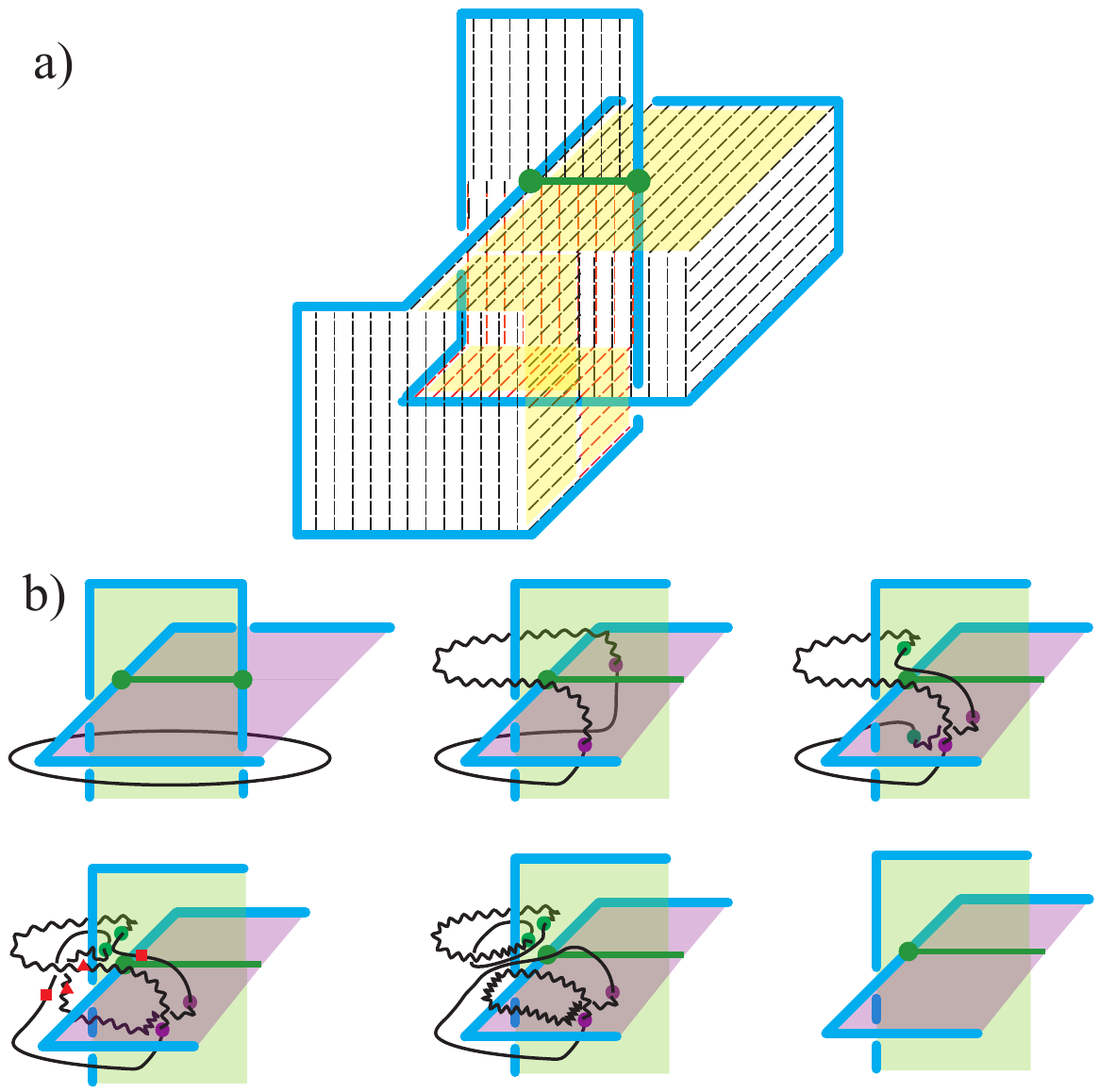}
\caption{Self-intersecting $\alpha\beta$ surface and string contraction. a) Self-intersection (along green line) of the $\alpha\beta$ surface (dashed area) in the Trefoil knot configuration we used on the lattice in Section~\ref{sec:numerical-results}. Some parts of the surface are shaded for better visibility. b) Contracting a string tangled with an $\alpha\beta$ surface that has a self-intersection (along green line). This case arises in the knot implementation from (a). The two shown surface pieces (shaded green and violet) could also be disconnected, since the string contraction is done locally. Piercing points of the string are marked with dots of corresponding color. The procedure hinges on moving piercing points close to each other and close to dislocation lines, but without passing the piercing point from one surface to the other, which is a forbidden deformation. The string is then reconnected until it gives disconnected pieces that wind $w=2$ times around a single dislocation line.}
\label{fig:5}
 % \end{widetext}
\end{figure}

So far in this subsection we focused on non-self-intersecting $\alpha\beta$ surfaces. However, for the efficient implementation in the lattice code (Section~\ref{sec:numerical-results}), as sketched in Fig.~\ref{fig:5}a, we made the opposite choice. The outcome is that the Trefoil knot constructed on the lattice using a self-intersecting Volterra surface does not change the GSD: $|GSD_{Trefoil}|=|GSD_{ideal3d}|=|GSD_{k=1disl.loops}|$.

It seems therefore that the self-intersecting case behaves just as we expect from the non-self-intersecting case. Recall from discussion in Section~\ref{sec:cont-loops-membr} and Fig.~\ref{fig:2} that from the lattice model we derived the rule saying that a string is \textit{not allowed} to move freely across an intersection line of two pieces of $\alpha\beta$ surface. This situation seems to invalidate the first step of our proof, where we demanded free movement of the piercing point of string on the $\alpha\beta$ surface. In particular, this raises the question about the seemingly non-trivial candidate for a non-contractible string, which we construct such that: 1) It is entangled with the dislocation line; 2) Does not pierce the $\alpha\beta$ surface; and 3) Seems ``stuck'' (non-contractible) because the intersection line of two pieces of $\alpha\beta$ surface is in its the way. An example relevant for the Trefoil knot on the lattice we used is shown in Fig.~\ref{fig:5}. However, such a string loop can be easily locally contracted irrespective of the $\alpha\beta$ self-intersection line, as demonstrated in Fig.~\ref{fig:5}b using the above general recipes.

%%%%%%%%%%%%%%%%%%%%%%%%%%%%%%%%%%%%%%%
%%%%%%%%%%%%%%%%%%%%%%%%%%%%%%%%%%%%%%% 

\section{Emergent non-Abelian gauge theory in dislocation melted doubled toric code in 2d}
\label{sec:emergent-non-abelian}

%%%%%%%%%%%%%%%%%%%%%%%%%%%%%%%%%%%%%%%
\begin{figure}
%\begin{widetext}
  \centering
\includegraphics[width=0.49\textwidth]{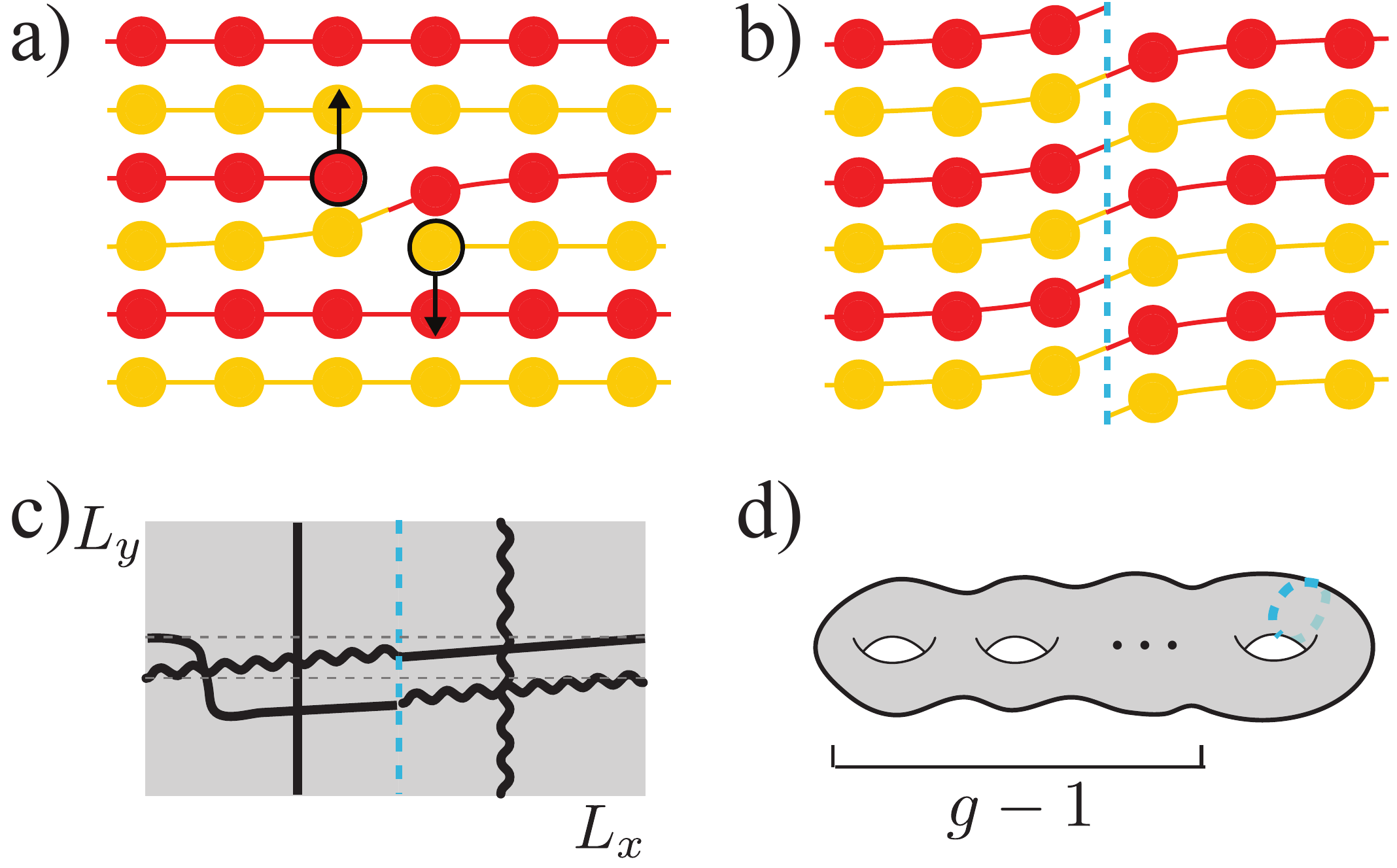}
\caption{Dynamic dislocations introduce $Z_2$ twist in the doubled toric code. In unmelted lattice (red and yellow dots schematically depict the two vertically displaced flavors of spins in the doubled model) dislocations can (a) nucleate and (b) glide until annihilation, to create a ``twist loop'' (dashed blue line) across which all quasiparticles (strings) switch flavor. (c) In the continuum description of periodic system, string loops (i.e. quasiparticle tuneling operators) of both flavor (solid and wavy line) are shown. The ``twist'' replaces two copies of a torus by a single torus copy with $L_x\rightarrow 2L_x$. The two strings along $y$ are actually equivalent due to the twist. (d) When the lattice melts, distinct topological ground state sectors are labeled by the twists they contain. In a genus $g$ surface all sectors containing any twists topologically map to the shown case. Since dislocations delocalize in the melted phase, physical ground states labeled by strings tied to any of $g-1$ holes have to be symmetrized in flavor. The physical states labeled by strings tied to the $g$-th hole are automatically symmetrized in flavor since the twist replaces the two flavor copies by a single one, as in (c).}
\label{fig:melt}
%\end{widetext}
\end{figure}
%%%%%%%%%%%%%%%%%%%%%%%%%%%%%%%%%%%%%%% 

In this section we consider the situation in which dislocations of the doubled $\Zn$ toric code become dynamical, and show that the resulting phase of matter is governed by a non-Abelian gauge theory. (The $\Zn$ toric code lattice model and its GSD, without melting, are described in Appendix~\ref{sec:appendixZn}.) The physical mechanism behind the transition is melting of the lattice through proliferation of ``double dislocations'' (lattice dislocations of length $|\bb|=a$). Condensation of point-like electric and/or magnetic charges can lead to states with different topological order\cite{Bais:2007p7602}, however in the present context the condensed double dislocation (point-like in 2d and loop-like in 3d) itself is not carrying any charge or flux in the theory.

Our first step is to describe how the dislocations $|\bb|=a/2$ obtain dynamics by lattice melting, conjecturing that the resulting state is described by a $G_n\equiv (\Zn\times \Zn)\rtimes \Zz$ gauge theory. We give a physical argument for the melted state GSD on a surface with arbitrary genus $g$, considering first the case of torus. We next obtain the exact same GSD values for states described by gauge group $G_n$ on genus $g$ surfaces. This implies that the quasiparticle contents also match, giving strong evidence that the melted state is really described by $G_n$ topological order.
% The GSD of the $G_n$ topologically ordered states is obtained here using the group structure of $G_n$, while in Appendix~\ref{sec:App_CS} it is confirmed by solving the $G_n$ Chern-Simons theory.
As a special case, this implies that the melted phase of the doubled $\Zz$ toric code is described by the non-Abelian group $D_4$.

\subsection{Dynamic dislocations and physical argument for torus GSD}
\label{sec:dynam-disl-torus}

Before considering the melted phase, we need to describe the ``twist'' effect introduced by dislocations. Consider first a periodic 2d lattice, which is topologically equivalent to a torus. Before any melting, a single dislocation---anti-dislocation pair can be created, moved around the periodic system, and annihilated, as shown in Fig.~\ref{fig:melt}a,b, specifically using the glide motion along $\bb\sim\yy$. This operation leaves a ``$\Zz$ twist loop'' (blue dashed line in Fig.~\ref{fig:melt}b,c) spanning the $\yy$ direction of the system, along which all $\xx$ rows are shifted by $a/2$ along $\yy$, and across which a flavor change is imposed. This implies that strings winding around $x$ direction are forced to change flavor twice before being able to close, while strings winding along $y$ change flavor when crossing the twist (Fig.~\ref{fig:melt}c). The two flavors have merged, and the system is described by a single copy of the toric code on a torus of size $2L_x\times Ly$.

Next we consider the effect of melting the lattice through double-dislocations. Before melting, the flavor exchange symmetry is realized through $a\yy/2$ lattice translations. The double-dislocation ($|\bb|=a$) melted phase restores full translational symmetry, so that also flavors become locally indistinguishable. Since we are not interested in liquid crystal phases, a lattice with dislocations is analogous to a $U(1)$ superfluid with vortices. Proliferation of double-vortices leads to an insulator described by $Z_2$ gauge theory, where the original single vortex becomes the deconfined gauge charge (and vortices in the double-vortex condensate are deconfined fluxes). By analogy, the original $|\bb|=a/2$ dislocations are deconfined gauge excitations in the melted phase.
% In the double-dislocation ($|\bb|=a$) melted phase, in absence of $|\bb|=a/2$ dislocations, the two flavors separated by $a/2$ are still well-defined at every point in space. However, if there are dislocations ($|\bb|=a/2$), melting makes them dynamical, as there are no restrictions on their movement; in fact, they become delocalized throughout the system.
In such a state, we can say that the flavor-changing Volterra lines connecting $|\bb|=a/2$ dislocation---anti-dislocation pairs also permeate the system, again concluding that it is unphysical to locally distinguish the two flavors. Physical states are only the ones symmetric in flavor, i.e. the flavor $\Zz$ symmetry is gauged.

The physical states can still be distinguished according to the $\Zz$ twist loops they contain. Namely, a state with a twist loop created by a non-contractible tunneling of a dislocation---anti-dislocation pair is topologically distinct from a state without that twist. In the ground state sector without twists the states just have to be symmetrized in flavor; on the other hand, in sectors having twist loops the strings (quasiparticle tunneling loops) are altered, as demonstrated in the example of Fig.~\ref{fig:melt}c.

Let us now give a physical derivation of the ground state degeneracy of the melted doubled $\Zz$ model on a torus. Before melting there were $4^2$ ground states, labeled by tunneling loops (non-contractible strings) of the quasiparticles: The $4^2$ quasiparticles were just the unconstrained pairs $(qp_\alpha,qp_\beta)$ of $\Zz$ quasiparticles $qp\in\{1,e,m,em\}$ of the two flavors $\alpha,\beta$. In the melted phase there is a single ground state sector containing no twists. In this sector the quasiparticles must be symmetrized in flavor, which gives us a total of $10$ symmetrized states: $4$ states $(qp_\alpha,qp_\alpha)$ and $6$ distinct states $(qp_\alpha,qp_{\beta\neq\alpha})+(qp_{\beta\neq\alpha},qp_\alpha)$. Next we consider a ground state sector containing one twist loop, say around the $y$ direction. According to the argument depicted by Fig.~\ref{fig:melt}c, the twist effectively merges the two flavors, leaving us $4$ ground states labeled by $4$ $\Zz$ quasiparticles $\{1,e,m,em\}$. Finally, there are exactly $3$ different twisted sectors, because the twist loop can wind around $x$, $y$, or both $x$ and $y$ (because there are only two flavors, higher winding numbers are irrelevant). All three twist sectors have the same degeneracy, because they are physically equivalent: The torus can be geometrically deformed, without changing its topology, to map different twist loops onto each other (these transformations are Dehn twists, see Appendix~\ref{sec:App_CS}). The torus GSD is therefore $10+3\cdot 4=22$.

This result is easily generalized to the doubled $\Zn$ theory on a torus. Namely, there are $n^2$ quasiparticles in a single copy of the model. The symmetrization in the untwisted sector leads to $n^2+\frac{n^2(n^2-1)}{2}$ ground states. There are still $3$ twisted sectors on the torus, each contributing $n^2$ degenerate states. The total for the torus is therefore:
\begin{equation}
  \label{eq:51}
  |GSD_{melt2d,torus}|=n^2\frac{n^2+7}{2},
\end{equation}
confirming that for doubled $\Zz$ theory $|GSD_{melt2d,torus}(n=2)|=22$.

\subsection{Physical argument for higher genus GSD}
\label{sec:higher-genus-gsd}

Generalizing arguments of the previous subsection to an arbitrary surface of genus $g$ is straightforward. Before melting, each of the $g$ holes contributes non-contractible strings, leading to a total of $(n^2)^g$ degenerate ground states of a single copy of $\Zn$ toric code.

Each copy in the doubled model has $n^{2g}$ degenerate states, which have to be symmetrized in the untwisted sector of the melted phase. This contributes $\frac{n^{2g}(n^{2g}+1)}{2}$ degenerate states.

Next, we consider the sector having a single twist loop, along the $y$ direction of the $g$-th hole (see Fig.~\ref{fig:melt}d). Focusing first on the states described by strings winding around and through the $g$-th hole only, we can apply the reasoning from the previous subsection. Namely, the twist merges the two copies of $n^2$ such states into only $n^2$ states which are automatically symmetric in flavor. The $g$-th hole therefore contributes a factor $n^2$ to the degeneracy of this sector. The strings tied to the other $g-1$ holes contribute $n^{2(g-1)}$ states per flavor, which have to be symmetrized. This gives a contribution of $\frac{n^{2(g-1)}(n^{2(g-1)}+1)}{2}$ states from these holes. Combining all the holes, the total degeneracy in the sector with a single twist loop is $\frac{n^{2g}(n^{2(g-1)}+1)}{2}$.

Again, because there are only two flavors, the twist loops that matter wind at most once. Each hole has two winding directions, giving the total of $2^{2g}$ possibilities for the winding of the twist loop. Exactly one of these possibilities is when all the winding numbers are zero, i.e. the untwisted sector. There are therefore $2^{2g}-1$ twisted sectors, all of them physically equivalent, as argued previously.

Adding the untwisted and twisted sectors, the total GSD of the melted doubled $\Zn$ toric code on a 2d manifold with genus $g$ is:
\begin{equation}
  \label{eq:52}
|GSD_{melt2d}|=\frac{n^{2g}(n^{2g}+1)}{2}+(2^{2g}-1) \frac{n^{2g}(n^{2(g-1)}+1)}{2},
\end{equation}
which reduces to the torus result in Eq.~\eqref{eq:51} when $g=1$.

\subsection{The $G_n\equiv(\Zn\times\Zn)\rtimes\Zz$ gauge theory: group structure, quasiparticle number and GSD}
\label{sec:zntim-theory:-group}

We now turn to a gauge theory continuum description of the melted lattice system. To start with, the doubled toric code is represented by the Abelian gauge group $\Zn\times \Zn$, a direct product of two flavor copies. As we argued in the previous subsections, the melting makes ($|\bb|=a/2$) dislocations dynamic, which means that also flavor-switching Volterra lines of all lengths occur in the system. This makes the two flavors physically locally indistinguishable.
% The partial restoration of translational symmetry therefore effectively gauges the $\Zz$ symmetry of exchanging flavors. In other words, ``tunneling loops'' represent a ``$\Zz$ twist'' across which the two gauge theory copies are exchanged.
The flavor-mixing operation therefore becomes a local symmetry operation, i.e. a gauge transformation.

According to these physical arguments, the gauge (structure) group of the melted doubled $\Zn$ toric code system is formed by
\begin{align}
  \label{eq:13}
  g(j_\alpha,j_\beta)&=\matl{e^{i\frac{2\pi}{n}j_\alpha}}{0}{0}{e^{i\frac{2\pi}{n}j_\beta}},\quad j_\alpha,j_\beta=0\ldots n-1\notag\\
  g_2&=\matl{0}{1}{1}{0},
\end{align}
where $j_\alpha$ and $j_\beta$ label gauge transformations in the $\alpha$ and $\beta$ flavored copy of $\Zn$ gauge group, respectively; on the other hand, the melting introduces the off-diagonal element $g_2$ which obviously exchanges the gauge fields of different flavors. The element $g_2$ generates a $\Zz$ subgroup responsible for exchanging flavors, introducing a ``twist'' into the doubled $\Zn$ gauge group we started from before melting. This twist in the gauge group is the mathematical reason for appearance of ``twist loops'' we considered earlier (see Appendix~\ref{sec:App_CS}). The gauge group of the melted theory can be written as $G_n\equiv (\Zn\times \Zn)\rtimes \Zz$, where the final $\Zz$ acts by exchanging the first two factors and is generated by $g_2$. Specializing to the $n=2$ case, the algebra of above matrices explicitly gives the non-Abelian dihedral group $D_4$ of order $8$.

The structure and physical properties of multi-component $U(1)$ Chern---Simons gauge theory containing a $\Zz$ twist have been studied in Ref.~\onlinecite{Barkeshli:2010p6844} through the example of $U(1)\times U(1)\rtimes \Zz$. In the present case, because the $\Zn \times \Zn$ gauge theory can also be described by a multi-component $U(1)$ Chern---Simons theory, one can apply the methods of Ref.~\onlinecite{Barkeshli:2010p6844} to study the properties of the $\Zz$ twisted theory. In Appendix~\ref{sec:App_CS} we study properties of the melted phase, such as GSD, using this Chern---Simons approach.

However, because the twisted theory in the present problem is described by a discrete gauge group $G_n$, one can directly study the properties of the $G_n$ gauge theory. More importantly, the multi-component U(1) Chern---Simons theory can only describe topological orders in 2+1d. But the discrete gauge theory approach also applies to higher dimensions, and we use this in Sec.\ref{sec:3dmelting}. In the rest of this subsection we obtain the GSD in 2+1d directly from considering the group structure of $G_n$, and we confirm that result by solving the Chern---Simons theory in Appendix~\ref{sec:App_CS}.

The GSD of a topologically ordered state on arbitrary genus $g$ surface, $S_g$, can be determined directly from its quasiparticle content (for instance, see Ref.\onlinecite{Barkeshli:2009p7596}):
\begin{equation}
  \label{eq:53}
  S_g=\left(\sum\limits_\gamma d_\gamma^2\right)^{g-1}\sum\limits_\gamma d_\gamma^{-2(g-1)},
\end{equation}
where $\gamma$ labels the quasiparticles species, while $d_\gamma$ is the quantum dimension of the $\gamma$ quasiparticle. On the other hand, when the topological order is described by a discrete gauge group $G$, such quasiparticle information follows directly from the group structure of $G$. Namely, a quasiparticle is a dyon labeled by the pair $\gamma\equiv (C,\mu)$, where $C$ is a conjugacy class in $G$, physically representing the flux, and $\mu$, representing the charge, labels an irreducible representation (IRREP) of the normalizer $N_C$ of a representative element in $C$.\cite{AlexanderBais:1992p6901the} The quantum dimension of a dyon equals the product of class size (i.e. number of elements in $C$) and the dimension of the IRREP $\mu$ of $N_C$.

Table~\ref{tab:2} summarizes the relevant information about the group $G_n$, which is easy to obtain using induction of representations from an Abelian subgroup. One can immediately establish from the table that there are in total: (a) $2n^2$ quasiparticle species $\gamma$ having quantum dimension $d_\gamma=1$; (b) $\frac{n^2(n^2-1)}{2}$ species having dimension $d_\gamma=2$; and (c) $2n^2$ species of dimension $d_\gamma=n$. Plugging into Eq.~\eqref{eq:53}, after trivial algebra it follows that the resulting expression precisely matches the GSD result, Eq.~\eqref{eq:52}, obtained from simpler physical reasoning:
\begin{equation}
  \label{eq:54}
  S_g(G_n)=|GSD_{melt2d}|.
\end{equation}
Since the genus $g$ is arbitrary, it also follows that the quasiparticle content of the melted doubled $\Zn$ theory matches the quasiparticle content of the $G_n$ topologically ordered theory. This is convincing proof of emergence of a non-Abelian $G_n$ topological order from a doubled Abelian $\Zn$ order upon giving dynamics to $\Zz$ ``flavor-mixing defects'' (dislocations) as a general mechanism. In particular, the simplest melted doubled $\Zz$ toric code is described by the non-Abelian $D_4$ topological order.

%    \multirow{3}{*}{$n$} & \multirow{3}{*}{$\{x^ay^a\}$, $a=0\ldots n-1$} & \multirow{3}{*}{ 1} & %\multirow{3}{*}{$G_n$} & $n$ $\times$ $n$ & \multirow{3}{*}{$\frac{n^2(n+3)}{2}$}\\
 %   & & & & $n$ $\times$ $\frac{n(n-1)}{2}$ &\\
  %  & & & & $n$ $\times$ $n$ &\\
\begin{table*}[t]
  \centering
  \begin{tabular}[c]{c|l|c|c|c|c}
        \hline
    No. of $C_g$ & $C_g$ labeled by g & Class size, $|C_g|$& Centralizer $N_g$ & No. of IRREP in $N_g$ $\times$ IRREP dim& No. of qp's \\
    \hline\hline
    $n$ & $\{x^ay^a\}$, $a=0\ldots n-1$ & 1 & $G_n$ & $2n\times 1d+\frac{n(n-1)}{2}$ $\times$ $2d$ & $\frac{n^2(n+3)}{2}$ \\
    \hline
    $\frac{n(n-1)}{2}$ & $\{x^ay^{b\neq a}\}$, $a,b=0\ldots n-1$ & 2 & $\Zn\times\Zn$ & $n^2$ $\times$ $1d$ & $\frac{n^3(n-1)}{2}$\\
    \hline
    $n$ & $\{u x^ay^b\}$, $a+b=c$; $c=0\ldots n-1$ & $n$ & $\Zn\rtimes\Zz$ & $2n$ $\times$ $1d$ & $2n^2$\\
    \hline\hline
    \multicolumn{4}{l}{Total number of quasiparticles in $G_n$} & & $\frac{n^2(n^2+7)}{2}$\\
    \hline\hline
  \end{tabular}
  \caption{Properties of (non-Abelian) gauge group $G_n=(\Zn\times\Zn)\rtimes\Zz$ ($n>1$), which describes the double-dislocation lattice melted phase of the doubled $\Zn$ toric code in 2d. An element $g\in G_n$ is written as $g=u^sx^ay^b$, where $x,y$ generate the $\alpha,\beta$ flavored $\Zn$, respectively, so that $x^n=y^n=e$ and $[x,y]=0$; the element $u$ generates the flavor-mixing $\Zz$, so $ux=yu$, $u^2=e$; therefore, $s=0,1$, and $a,b=0\ldots n-1$. $C_g$ is the conjugacy class labeled by a representative $g\in G_n$. The first column shows the total number of conjugacy classes of given type (3 types in total), and each class is labeled using parameters in second column. Centralizer $N_g$ is the subgroup of elements that commute with $g$. Every conjugacy class labels a magnetic flux. Quasiparticles are dyons, carrying a flux and a charge, where the charge is an irreducible representation (IRREP) of $N_g$, with $C_g$ labeling the flux. Hence, column six shows the total number of dyons carrying flux of the type labeled by the table row; it equals the product of the first column and the first factors in fifth column. The quantum dimension of a dyon equals the product of $C_g$ class size (third column) and the dimension of the IRREP of $N_g$ (second factors in fifth column).}
  \label{tab:2}
\end{table*}

\section{Emergent non-Abelian gauge theory in dislocation melted doubled toric code in 3d}
\label{sec:3dmelting}

Even though the systematic study of three dimensional manifolds is an open problem, we can obtain some results by focusing on the simple case of a lattice with periodic boundary conditions, i.e. a model defined on the three-torus $T^3$.

Dislocation melting in 3d is a much subtler subject than in the 2d case, as we briefly touch upon in the Discussion section. However, based on the simple physical insight that melting gauges the flavor symmetry in 2d, as described in the previous section, we conjecture that the melted phase of the doubled $\Zn$ toric code in 3d has topological order described by the $G_n\equiv(\Zn\times\Zn)\rtimes\Zz$ gauge group.

We test this conjecture only in the simplest case of $n=2$, and as already mentioned, only on the three-torus. In this case, the melted phase of the Abelian doubled $\Zz$ toric code is assumed to realize the non-Abelian $G_2=D_4$ topological order. Using physical arguments as in Section~\ref{sec:dynam-disl-torus} we will derive the GSD of the melted phase to be
\begin{equation}
  \label{eq:55}
  |GSD_{melt3d}(n=2,T^3)|=92.
\end{equation}
On the other hand, in Appendix \ref{sec:gsd-d_4-topol} we will prove that a $D_4$ topologically ordered state on a three-torus $T^3$ has GSD also equal to:
\begin{equation}
  \label{eq:56}
  S_{T^3}(D_4)=92,
\end{equation}
thereby providing some evidence for the conjecture and the fact that in three spatial dimensions dislocation melting of an Abelian phase can realize a non-Abelian phase.

Let us now consider the doubled $\Zz$ toric code in three dimensions with periodic boundary conditions, and find its GSD, which is $|GSD_{melt3d}(n=2,T^3)|$. Before melting, each flavor has GSD equal to $2^3$, due to non-contractible strings.

The melting enforces symmetrization of flavors, and also introduces ``twists''. In 3d, the twist occurs across a plane that spans the system. The twist plane can be imagined as created by a dislocation loop which is nucleated and stretched until it annihilates using the periodicity of the system. A non-contractible string on the three-torus has to switch flavors as it pierces the twist plane and spans the periodic system (see the horizontal string in Fig.~\ref{fig:melt}c for a 2d analogy). Since there are only two flavors, only strings which span three-torus directions $X,Y,Z$ up to once are relevant. There are therefore $7$ twisted sectors, each having a flavor-switching string which spans $X$, $Y$, $Z$, $XY$, $YZ$, $XZ$, or $XYZ$. The twisted sectors are physically equivalent. Analogously to the 2d case, a twisted sector has the two flavors replaced by a single $\Zz$ copy, which is automatically flavor-symmetric, and therefore has a $2^3$ degeneracy. The twisted sectors therefore contribute $7\cdot 8=56$ states. The untwisted sector has to be flavor-symmetrized, therefore $2^3\times 2^3$ leads to $2^3+\frac{2^3(2^3-1)}{2}=36$ states. The total GSD is 92, as claimed in Eq.~\eqref{eq:55}.

\section{Discussion and conclusions}
\label{sec:disc-concl}

%%%%%%%%%%%%%%%%%%%%%%%%%%%%%%%%%%%%%%%
\begin{figure}
  \centering
\includegraphics[width=0.48\textwidth]{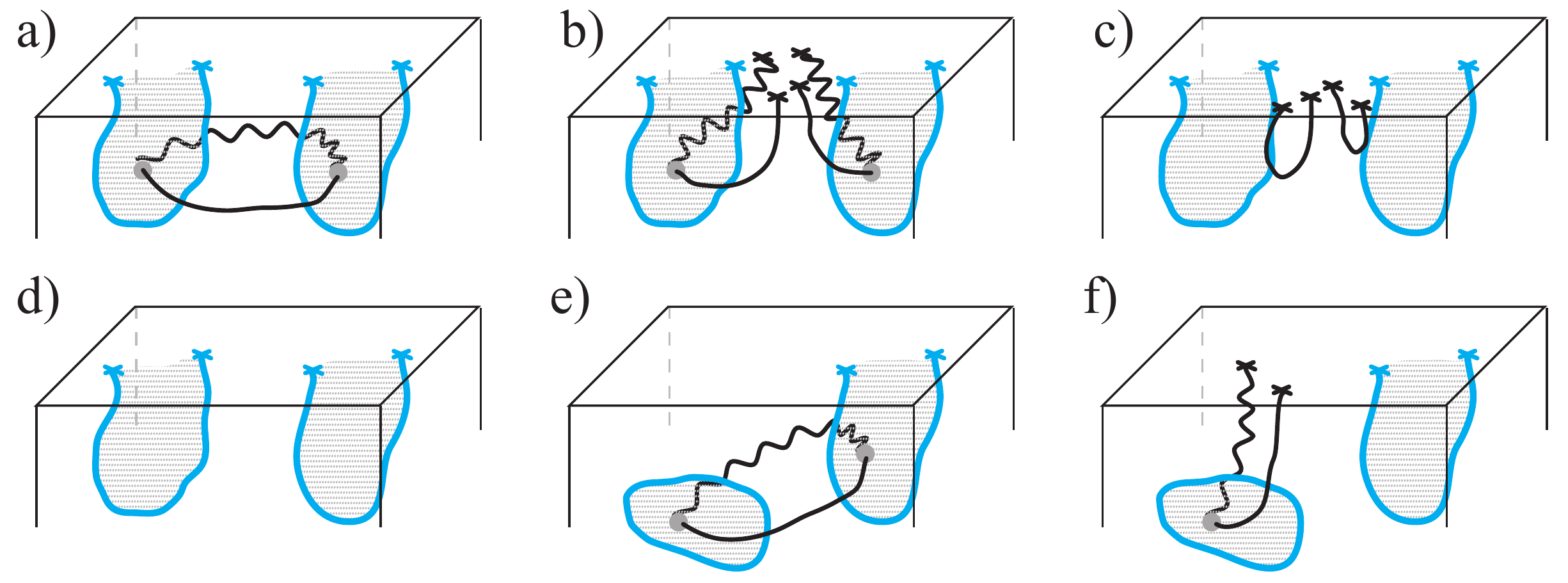}
\caption{Three-dimensional toric code in a system with open boundary conditions. Dislocation lines (blue) and flux strings (black solid/wavy for two copies) can end at the surface. (a) The string loop between two dislocations would be non-contractible if it were not for (b) the surface and the fact that (c) string endpoints on the surface can avoid dislocation line endpoints. Any number of open dislocation lines beyond the first cannot contribute to topological GSD. (e) String loop between dislocation loop (in bulk) and open dislocation line is a non-contractible string. The open string can never be contracted because it (f) intersects the dislocation loop Volterra surface an odd number of times.}
\label{fig:surface}
\end{figure}
%%%%%%%%%%%%%%%%%%%%%%%%%%%%%%%%%%%%%%% 

In this paper we explore the dependence of ground state degeneracy in simple exactly solvable topological models (in both 2d and 3d) on the complicated topology of the underlying space manifold, where that topology is provided by dislocation points (in 2d) and arbitrarily linked and knotted dislocation lines (in 3d) with Burgers vector connecting two translated copies of the model. Because we are studying topological properties, our results are valid for the entire phases to which the exactly solvable models belong.

The main new physics in this paper is about the 3d models. In 3d, if the ground state degenearcy(GSD) depends on the knotting/linking of dislocation loops, the GSD can only be a topological invariant of the links or knots of the dislocation loops. However, we find that the doubled Kitaev's toric code shows no dependence of the GSD on the knotting properties of dislocation loops. In this 3d topologically ordered phase, GSD only depends on the total number of (possibly linked) dislocation loops.

In fact, it turns out that in the limit of large number $k$ of dislocation loops (pairs in 2d), the GSD scales as $2^k$ in 3d. As a dislocation loop can always be allowed to be added into the lattice, the extra GSD introduced by one dislocation loop should be an integer.  In this context therefore the doubled Kitaev toric code is reasonably the minimal theory in 3d in which dislocations introduce extra GSD. In 2d however the scaling is $4^k$, indicating that there might be a more fundamental 2d theory, a ``square root'' of the doubled 2d Kitaev toric code, in which the GSD scales as $2^k$. Such a ``square root'' theory is indeed realized in the plaquette model version of the 2d toric code\cite{Bombin:2010p6412}, and $2^k$ scaling of the GSD was found.

The Kitaev toric code can be generalized to $\Zn$ toric code\cite{Bullock:2007,Schulz:2012}, whose low energy effective theory is described by the $\Zn$ gauge theory, and when $n=2$ the Kitaev toric code is restored. Our results can be generalized without too much difficulty to doubled $\Zn$ toric code in both 2d and 3d. Here we present the main results and sketch the strategy of proof. The detailed analysis and proof can be found in Appendix~\ref{sec:appendixZn}.

Obviously, in 2d in a doubled $\Zn$ toric code with periodic boundary conditions with $k$ pairs of dislocations, $|GSD|=n^4\cdot n^{2k-2}$ which is a direct result of the geometrical interpretation of dislocations in 2D (see Fig.\ref{fig:2d} or Ref.\onlinecite{Barkeshli:2012p7361}). A ``square root'' of such a theory has been recently studied in Ref.\onlinecite{You:2012p7608}, where GSD scales as $n^k$ when $k$ is large.

In 3d, we find that the GSD in a doubled $\Zn$ toric code with periodic boundary conditions in the presence of $k$ dislocation loops is $|GSD|=n^6\cdot n^{k-1}$, independent of linking and/or knotting of the dislocation loops. The proof of this result is a slightly non-trivial generalization of our proof for the doubled 3d Kitaev toric code. Following our strategy of using string/membrane operators to resolve the ground state degeneracy, one still can construct the string operators which can be interpreted as ``electric flux lines`` in a 3d $\Zn$ gauge theory. But in a $\Zn$ gauge theory with $n>2$, the flux line has a $\Zn$ valued strength. If we focus on the flux line with the fundamental unit of strength (strength=1), then the flux line has a well-defined orientation. Some essential steps in our proof for the doubled Kitaev toric code now need to be reconsidered carefully. For instance the step 6 illustrated in Fig.\ref{fig:6} requires annhilation of two string operators. But in a $\Zn$ theory two string operators both with strength=1 cannot annihilate if $n>2$. In Appendix~\ref{sec:appendixZn} we carefully reconsider the rules of surgeries involving the string operators and the proof still goes through (basically because dislocations allow a string of strength $s$ to be converted to strength $n-s$, which is its inverse).

Can the ground state degeneracy depend on linking or knotting of the dislocation loops in a 3D theory then? Our study of the $\Zn$ Abelian gauge models strongly suggests that at least for Abelian gauge theories this may be impossible. However for non-Abelian gauge models, for instance the discrete non-Abelian gauge theories with deconfined phases in 3D, our proof based on string operators may be invalidated and a knotting or linking dependent GSD may occur. We leave this interesting possibility as a subject of future investigation.

What happens in a system with open boundaries? It is well known that the 2d $\Zz$ toric code can have two types of edges: An open electric(magnetic) flux line ending on first(second) type of edge commutes with the Hamiltonian, while the open magnetic(electric) can not commute. The types of edges in the system therefore control the number of non-contractible electric and magnetic lines (strings). (These edge types are realized by different termination of the lattice.) However, it does not make sense to put dislocations at the edge of the 2d system, and it is easy to see that the dislocations add the same extra GSD in open as in periodic 2d systems.

In 3d, one can similarly terminate the lattice model at the surface of the system in two ways: One allows open strings to end at the surface while still commuting with the Hamiltonian, while membranes terminating at the surface cost energy. The opposite is true for the other surface type. Let us focus on the first type, for simplicity. Firstly, any non-contractible string piercing a Volterra surface of a dislocation loop in the bulk will stay non-contractible even when the system has a surface, since the number of piercings of that Volterra surface will remain odd (and the string therefore non-contractible) no matter how we manipulate the string using the surface of the system (see example in Fig.~\ref{fig:surface}e,f).

The situation is different for an open dislocation line that ends at the surface of the system (it has two endpoints on the surface). As Fig.~\ref{fig:surface}a demonstrates, the string loop that pierces Volterra surfaces of two such open dislocation lines becomes contractible. Namely, the string loop can break into two open strings by reaching the surface, and the string endpoints can avoid the dislocation lines' endpoints on the surface, allowing the open strings to contract (Fig.~\ref{fig:surface}a-d). This immediately shows that any (non-zero) number of open dislocation lines is equivalent to a single one, as far as extra topological GSD is concerned. However, as Fig.~\ref{fig:surface}e,f clearly demonstrates, the argument in the previous paragraph ensures that a single open dislocation line still behaves as a closed dislocation loop in the bulk, because a string that pierces through Volterra surfaces of a dislocation loop and an open dislocation is still non-contractible.

Therefore, in a 3d system with a boundary allowing commuting open strings, the extra GSD due to $k$ dislocation loops in the bulk (i.e. $2^{k-1}$) is multiplied by a factor $2$ if there is any (however many) open dislocation lines that terminate on the surface.

The extra topological GSD due to dislocations leads to additional zero-temperature entropy. This entropy is obviously extensive in the number of dislocation loops in the bulk, and therefore could be of experimental relevance. Let us emphasize that the GSD properties hold for the entire topologically ordered quantum phase, beyond the exactly solvable doubled model. As sketched above, the extra entropy should stay extensive in number of bulk dislocations even for realistic samples with surface.

Recently, topologically ordered phases which also have global symmetries, i.e. symmetry enriched topological (SET) phases have been studied intensively. The topological order is usually described by gauge symmetry and understanding its interplay with the global symmetry is the central issue in studying SET phases. In this work, non-Abelian topological order arises due to the promotion of the global $\Zz$ symmetry (the flavor symmetry) to a gauge symmetry. Thus defect condensation seemingly provides a physical way to make the connection between global and local symmetry, and therefore between different SET phases.

\begin{acknowledgments}
We thank Xiao-Gang Wen for pointing out Ref.~\onlinecite{Bombin:2010p6412}. This work was supported by the Alfred P. Sloan foundation and National Science Foundation under Grant No. DMR-1151440 (AM and YR), and NSERC, CIFAR, and Center for Quantum Materials at University of Toronto (YBK).
\end{acknowledgments}

\appendix

\section{Generalization to $\Zn$ toric code}
\label{sec:appendixZn}

Here we present the generalization of our GSD proof based on string operators of the 3dTC (a $\Zz$ theory) to a $\Zn$ version of the toric code. The $\Zn$ generalized toric code was introduced in Ref.\onlinecite{Bullock:2007}, and we use a description of it similar to the 2d case from Ref.~\onlinecite{Schulz:2012}. In this way the lattice and stabilizer definitions are similar to the 3d $\Zz$ toric code of the main text. We first introduce the building blocks and then identify the strategy for extending our GSD proof using string operators.

At each spin site $i$ of the cubic lattice (see Fig.~\ref{fig:7a}) of the $\Zn$ model we introduce $n$-state degrees of freedom $\ket{q_i}$, which are acted upon by generalizations of the Pauli matrices of the $\Zz$ model. These unitary operators $X_i,Z_i$ at site $i$ form a quantum rotor algebra:
\begin{equation}
    \label{eq:6}
\begin{aligned}
  Z_i\ket{q_i}&=\omega^{q_i}\ket{q_i},\\
  X_i\ket{q_i}&=\ket{q_i-1\text{ mod n}},\\
  X_i^\dag X_i&=  Z_i^\dag Z_i=\openone,\\
  X_iZ_i&=\omega Z_iX_i\;\Rightarrow\;
  X_iZ^\dagger_i=\omega^*Z^\dagger_iX_i,\\
    \omega&=e^{i2\pi/n},
\end{aligned}  
\end{equation}
with $\omega^*$ the complex conjugate. These relations generalize the anticommutation of Pauli matrices.
%%%%%%%%%%%%%%%%%%%%%%%%%%%%%%%%%%%%%%%
\begin{figure}
  \centering
\includegraphics[width=0.48\textwidth]{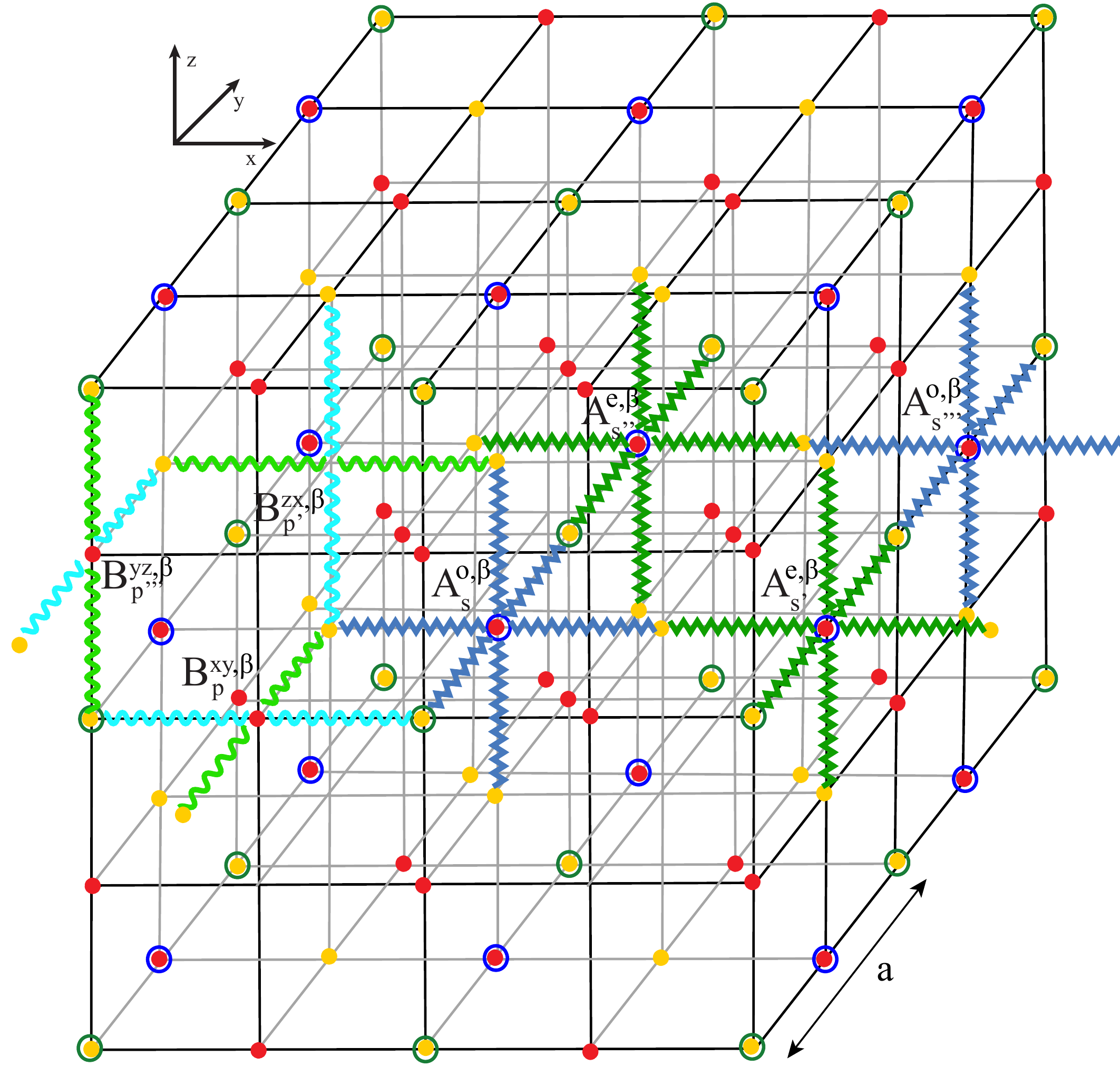}
\caption{Three-dimensional doubled $\Zn$ generalization of toric code. See caption of \ref{fig:7a} for the basic ingredients shared with $\Zz$ toric code. Three $\beta$ flavored plaquette operators are shown, acting with $Z$ (blue wavy lines) and $Z^\dag$ (green wavy lines) operators. The operators are chosen so that the local cubic constraint for the plaquettes holds. Star operators on the ``odd'' cubic sublattice act with $X$ operators (blue zag lines), and with $X^\dag$ if on the ``even'' sublattice (only four stars of $\beta$ flavor, positioned in one plane of the lattice, are shown).}
\label{fig:13}
\end{figure}
%%%%%%%%%%%%%%%%%%%%%%%%%%%%%%%%%%%%%%% 
%%%%%%%%%%%%%%%%%%%%%%%%%%%%%%%%%%%%%%%
\begin{figure*}
  \centering
\includegraphics[width=0.8\textwidth]{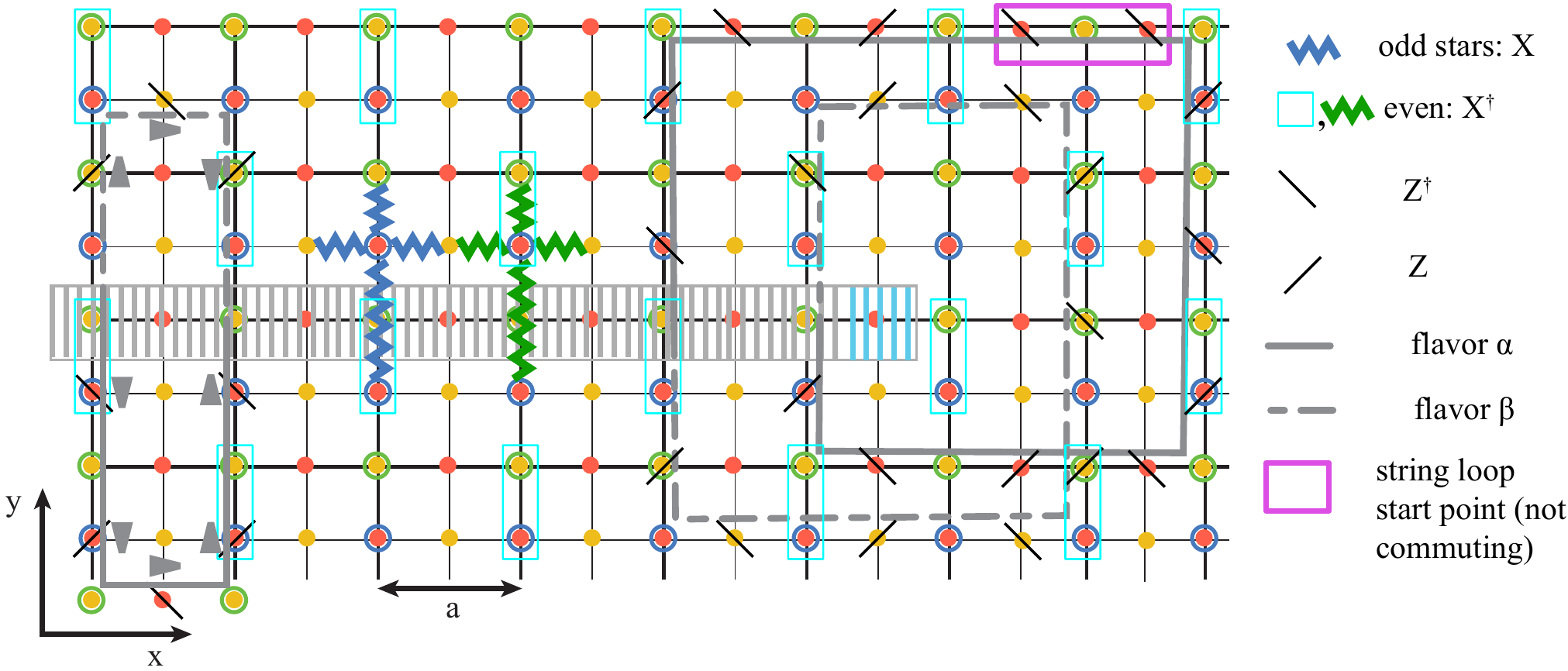}
\caption{String operators in $\Zn$ doubled toric code with dislocations. Only a single plane of 3d lattice is shown for clarity. The stabilizers are naturally repaired, because the dislocation cannot directly mix the even/odd sublattices; example shown: even and odd star. Even/odd sublattice is mixed after encircling the dislocation twice: the string constructed step-by-step clockwise on the right of the figure, starting from the star labeled by magenta is geometrically closed (flavor labeled by full/dashed line is switched twice), but does not commute with all stabilizers! Namely, the string has alternating $Z$/$Z^\dag$ operators (slanted black lines of different angle) to ensure commuting with stars, but this fails for the star at the start point. On the left of the figure is an example of a valid (commuting) string loop which changes flavor and also gets inverted, as demonstrated by its arrows (see paragraphs after Eq.~\eqref{eq:12}). This loop is contractible. There is no inversion when the flavors on the top/bottom half of the loop are chosen oppositely, as depicted in Fig.~\ref{fig:14}b(left).}
\label{fig:13a}
\end{figure*}
%%%%%%%%%%%%%%%%%%%%%%%%%%%%%%%%%%%%%%% 
%%%%%%%%%%%%%%%%%%%%%%%%%%%%%%%%%%%%%%%
\begin{figure}
  \centering
\includegraphics[width=0.48\textwidth]{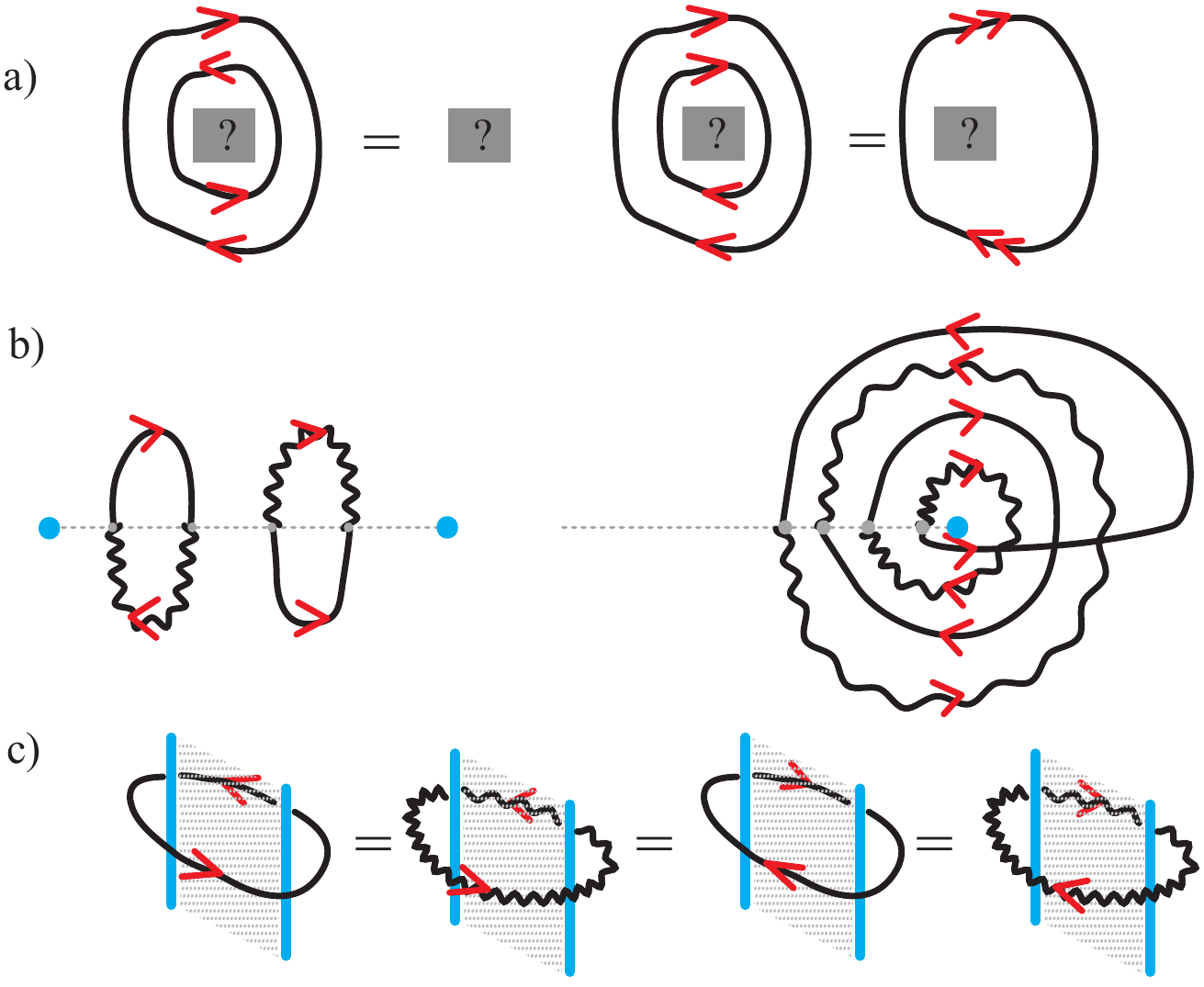}
\caption{ Continuum rules for strings in the 3d $\Zn$ doubled toric code. (a) A string carries a flux value labeled by $s\in\{0,\ldots,n-1\}$. In presence of dislocations, an arrow is a useful local representation of the flux carried by the string, because the string can get inverted ($s\rightarrow n-s$). Inverse strings cancel, while doubling a string only doubles its strength ($s\rightarrow 2s$). (b) Top view shown for clarity, see Fig.~\ref{fig:13a} for explicit lattice demonstration. (Left) Dislocation $\alpha\beta$ surface changes flavor, but also can invert the string ($s\rightarrow n-s$). (Right) Only winding $w=4k$ strings commute with stabilizers (here $k=1$ is shown), since a $|4\vec{b}|=2a$ translation is needed to avoid a jump between even/odd sublattices. These winding loops are the generalization of $w=2k$ ones from the $\Zz$ model, and are also naturally locally contractible. (c) Using rules from part (b), it is easy to show that dislocations can locally independently invert or change flavor of string loops. This suffices to ensure the possibility of step 6 of the GSD proof (see Fig.~\ref{fig:6}) within the $\Zn$ theory, upholding the entire construction.}
\label{fig:14}
\end{figure}
%%%%%%%%%%%%%%%%%%%%%%%%%%%%%%%%%%%%%%%

The stabilizers of the model of flavor $\alpha$ are presented in Fig.~\ref{fig:13}, and are defined similar to Eqs.~(\ref{eq:1},\ref{eq:2}):
\begin{align}
  A_{s\in\{o\}}^{\alpha}&\equiv A_{s}^{o,\alpha}=X_{\s+\xx/2}X_{\s+\yy/2}X_{\s+\zz/2}X_{\s-\xx/2}X_{\s-\yy/2}X_{\s-\zz/2},\\
A_{s\in\{e\}}^{\alpha}&\equiv A_{s}^{e,\alpha}=  {A_s^{o,\alpha}}^\dag,\\  B_p^{ab,\alpha}&=Z^\dagger_{\pp(\aaa\bbb)+\aaa/2}Z_{\pp(\aaa\bbb)+\bbb/2}Z^\dagger_{\pp(\aaa\bbb)-\aaa/2}Z_{\pp(\aaa\bbb)-\bbb/2},
  \label{eq:8}
\end{align}
with the key difference being the necessary bipartite decomposition of the cubic lattice such that even (``$e$'') and odd (``$o$'') cubic sites have star operators $A_s^{e,\alpha}$ and $A_s^{o,\alpha}$, respectively, positioned on them. This property also occurs in the 2d $\Zn$ model\cite{Schulz:2012}. Since the even and odd star stabilizers act with inverse operators on the spins they share ($X_i$ are unitary), the global constraint
\begin{equation}
  \label{eq:10}
    \prod_{s\in\{e,o\}} A^f_s=\openone
\end{equation}
remains as in Eq.~\eqref{eq:13d}. A plaquette stabilizer consists of operators $Z_i$ which alternate between daggered and normal as one goes around the face of a cube which defines the plaquette, so that opposite sides of the plaquette both carry either $Z$ or $Z^\dag$. This ensures that any plaquette shares two spin sites with a neighboring star such that both $Z_i$ and $Z_j^\dag$ appear, while the star contains all $X$ or all $X^\dag$ operators, leading to the commutation of the plaquette and star according to Eq.~\eqref{eq:6}. The choice of which sides of a plaquette carry the daggered operators is made in Eq.~\eqref{eq:8} such that local cubic constraints are satisfied as before (see text after Eq.~\eqref{eq:2}).

The $\beta$ flavor of the model is created by superimposing a copy of the lattice translated by $\yy/2$ (Fig.~\ref{fig:7a}), as before, and the Hamiltonian becomes
\begin{equation}
  \label{eq:9}
  H_n=-\sum_{f,s}\left(A^f_s+{A^f_s}^\dagger\right)-\sum_{f,p}\left(B^f_p+{B^f_p}^\dagger\right).
\end{equation}
Because of the commutation and unitarity, it is enough to count the stabilizers $A,B$ and fix their eigenvalues in the $\Zn$ theory. Since the constraints carry through as for the $\Zz$ toric code in the main text, the GSD counting remains the same, except that each unconstrained ``spin'' degree of freedom now carries $n$-fold degeneracy. This gives $|GSD^{(n)}_{3d,ideal}|=n^6$.

Our next step is to introduce string operators consistently in the $\Zn$ theory with dislocations. A string operator is non-trivially different from the $\Zz$ case, since its eigenvalue, i.e. the flux it carries, can be any of $n$ values $e^{is2\pi/n}$, $s\in\{0,\ldots,n-1\}$.
% Consider flavor $\alpha$ for concreteness; flavor $\beta$ is completely analogous.
The string must be defined using both daggered and normal operators: given a closed path (loop) $\mathbb{P}_f$ which stepwise connects neighboring star sites of flavor $f$, the $Z$ operator acting on the spin between two stars has to alternate between daggered and normal with each step:
\begin{equation}
  \label{eq:11}
  \Sigma^f(\mathbb{P}_f)=\prod_{i\in\mathbb{P}_f}
  \begin{cases}
    Z_i,& i\text{ odd},\\
    Z_i^\dag,& i\text{ even},
  \end{cases}
  \end{equation}
  compare to Eq.~\eqref{eq:4}. This way, a star operator crossed by the path $\mathbb{P}_f$ will share exactly a $Z$ and a $Z^\dag$ operator with the string, therefore commuting with it.

The crucial quality of the $\Zn$ theory is that although the string $\Sigma^f(\mathbb{P}_f)$ carries some flux $e^{is2\pi/n}$, $s\in\{0,\ldots,n-1\}$, i.e. it has strength $s$, it is easy to define its inverse:
  \begin{equation}
    \label{eq:12}
\bar{\Sigma}^f(\mathbb{P}_f)=  \Sigma^f(\mathbb{P}_f)^\dag,
  \end{equation}
  which of course carries flux $n-s$ while having the exact same geometrical shape. The string is canceled by its inverse, but not necessarily by itself, in contrast to the case of $\Zz$ theory (see Fig.~\ref{fig:14}(a)).

  Our main proof for the GSD in presence of dislocations crucially depends on canceling two copies of a string of a given flavor (see step 6 of Fig.~\ref{fig:6}). In the $\Zn$ theory, starting from a string of strength $s$, the two copies of the string would not necessarily cancel.

\textit{However, the proof is rescued by a remarkable fact: the $\alpha\beta$ surface of the dislocation allows inversion of a string, as well as its flavor change (Fig.~\ref{fig:14}c). As we will next show, this remarkable fact hinges on the dislocation disrupting the even/odd division of the cubic lattice, which itself is directly related to mapping a string to its inverse.}

It is important to notice that the considered dislocations with $|\vec{b}|=a/2$ directly couple the flavors, but do not simply couple the even/odd sublattices which are actually connected by an $a$ translation. The fact that the dislocation can not directly exchange even and odd sites also ensures that the stabilizers can be smoothly and consistently ``surgically'' repaired in a natural way across the $\alpha\beta$ surface (see next paragraph and Fig.~\ref{fig:13a}).

Let us now show the crucial properties of a string and its inverse in relation to dislocations (Figs.~\ref{fig:13a},\ref{fig:14}). For readability, figure \ref{fig:13a} shows only a single layer of the 3d lattice (which is similar to a 2d model), but the conclusions hold generally in 3d. As is obvious from Eqs.~\eqref{eq:11},\eqref{eq:12}, the inverse of a string is obtained by switching the odd and even steps in the string operator. The removal of star sites by the dislocation's $\alpha\beta$ surface can play exactly this role, as is explicitly shown in Fig.~\ref{fig:13a}. Given a spin site $i$ on the step between an odd and even star site (which are always well defined, as the spins they act on must be $X$ xor $X^\dag$ even after local surgery due to dislocation), the string and anti-string differ by having $Z_i$ or $Z_i^\dag$. Therefore, we are able to construct a simple pictorial way to locally distinguish the string and inverse string, exemplified in Fig.~\ref{fig:13a}. We can assign an arrow at each spin site along the geometrical line following the string, i.e.: following the path of the string loop in a fixed orientation (call it ``clockwise''), if on the step between odd and even star site the string acts with $Z$ operator, assign an arrow pointing along the path (so in the ``clockwise'' direction). Otherwise assign the opposite direction. One can easily see (e.g. using Fig.~\ref{fig:13a}) that the even/odd definition of stars and strings will preserve the arrow direction all along the string, at least in absence of dislocations. Also, deformations of the string by the action of plaquettes (and/or anti-plaquettes $B_p^\dag$) must preserve the arrows due to plaquettes' alternating $Z/Z^\dag$ geometry. By this construction, two loop segments sharing a geometrical path but having opposite arrow directions on them cancel each other, and therefore represent the string and anti-string on that path segment.

Intuitively, encircling the dislocation line twice adds up to a translation by $|2\vec{b}|=a$, which is a jump from even to odd and vice versa in the division of the cubic lattice (for both flavors). One expects that it is required therefore to encircle the single dislocation line four times to obtain a valid string loop. This is indeed true. Figure \ref{fig:13a} explicitly shows how encircling a dislocation line twice, and therefore piercing the $\alpha\beta$ surface twice in the same direction, produces a geometrically closed loop (by switching flavor twice), but also gives a non-commuting string! Namely, the star stabilizer (being all $X$ or all $X^\dag$) at the start/end point of the path shares two sites with the string, both on which the string is $Z$ (or $Z^\dag$). (See commutation relations in Eq.~\eqref{eq:6}.) The successful generalization of the double-winding string loop to the $\Zn$ theory gives the 4-winding version; this kind of loop is also naturally expected to be contractible using the locally surgically repaired stabilizers at the dislocation line, and is shown in Fig.~\ref{fig:14}b.
%Also, it does not influence the steps in our proof of contracting string loops in presence of dislocations.

Using only the properties of string loops explained above, one can find the remarkable (and useful) generalization of rules from Fig.~\ref{fig:1}. The crucial rule is that a string encircling locally parallel dislocation line segments with an $\alpha\beta$ surface stretched between them %(the 3d version of encircling a dislocation---anti-dislocation pair in 2d)
can, by using only local operations, \textit{change its arrow direction, as well as its flavor}. This rule, in the continuum representation, is depicted in Fig.~\ref{fig:14}c, and is the generalization of rule in Fig.~\ref{fig:1}b. Concerning other rules, a proof of the one in Fig.~\ref{fig:14}b(left) is shown explicitly on the lattice in Fig.~\ref{fig:13a}.

These rules allow one to use all the constructive steps in our proof (Fig.~\ref{fig:6}), with slight modifications. Firstly, the flux $s$ has to be assigned to the string loop. Once the phase between steps 3 and 4 is reached, one must also explicitly state the arrow direction for the loop, that accompanies its strength $s$. It is easy to see that the two copies of the string in step 5 are not only geometrical copies, but they also have the same arrow direction, as well as the same strength $s$. It is simple then to use the new string rules to achieve the inversion of one of the copies (i.e. switching it to strength $n-s$). The two copies then cancel, just as is required by the original proof.

\section{The membrane operator in presence of dislocations}
\label{sec:membr-oper-persp}

%%%%%%%%%%%%%%%%%%%%%%%%%%%%%%%%%%%%%%%
\begin{figure*}
%  \begin{widetext}  
  \centering
\includegraphics[width=0.8\textwidth]{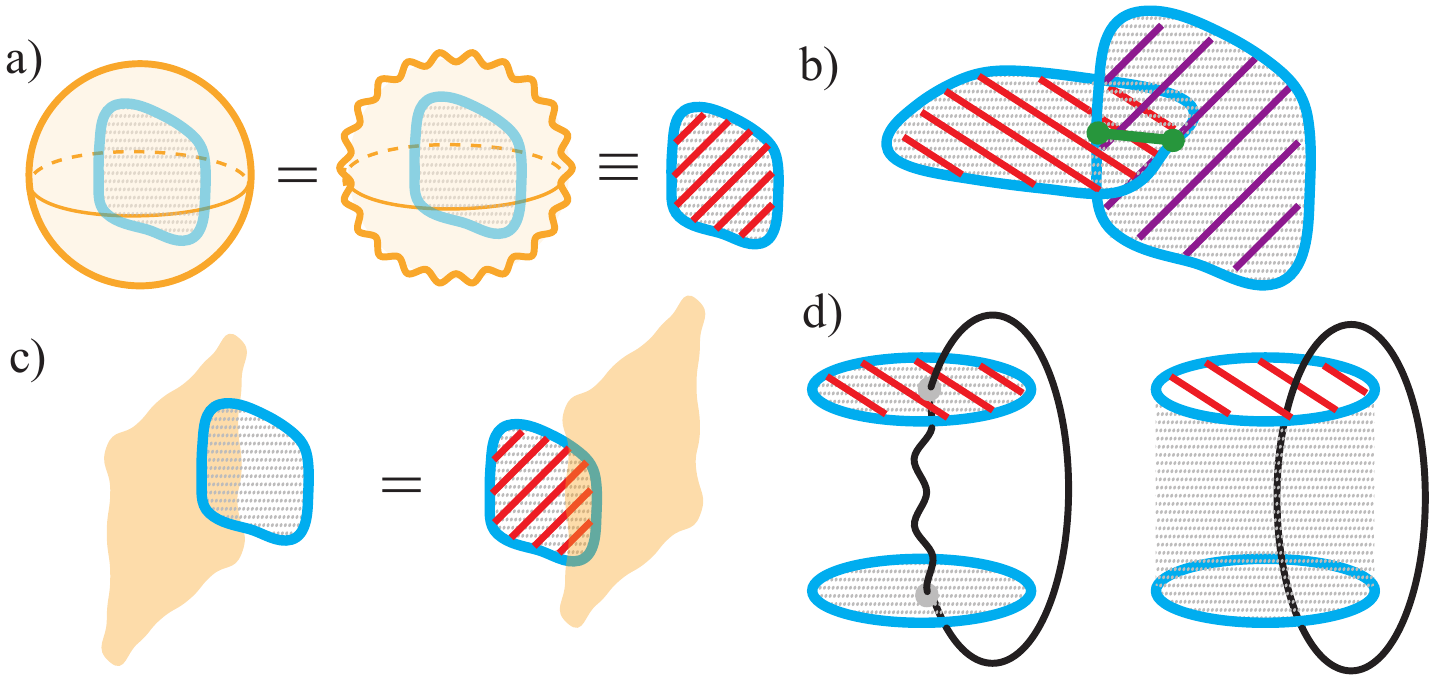}
\caption{Elementary operations on membrane operators in presence of dislocation loops, derived from their behavior on lattice. Blue lines are dislocation lines, and orange straight/wavy surface is a membrane operator of $\alpha$/$\beta$ flavor. (a) The minimal, locally non-contractible membrane (red line shading) has two halves of opposite flavor, tightly spanning a dislocation loop. It can expand into membrane of either flavor. (b) Linking dislocation loops preserves minimal membranes; local redefinition of two minimal membranes is only needed where two $\alpha\beta$ surfaces intersect (green line), see text. (c) Local passage of a piece of membrane through dislocation loop. (d) Relevant minimal membranes do not have to coincide with $\alpha\beta$ surfaces. An example of two dislocation loops is shown, with different choices for the $\alpha\beta$ surfaces. The non-contractible strings that anticommute with the chosen minimal membranes are also shown.}
\label{fig:2}
 % \end{widetext}
\end{figure*}
%%%%%%%%%%%%%%%%%%%%%%%%%%%%%%%%%%%%%%% 
Here we complete our analysis based on string operators. Strictly, a non-contractible string has to be accompanied by a non-contractible membrane it pierces an odd number of times, for our claims about the GSD in subsection~\ref{sec:gsd-stabilizers-k} to be true.

We first consider the practical construction of membranes on the lattice, and introduce the useful concept of ``minimal membrane''. Then we present the analysis in the continuum.

A particularly important membrane is one that contains the entire chosen dislocation loop. As we check directly on the lattice (see Appendix~\ref{sec:appendixsurgery}) using star operators, and discuss further later, such a membrane cannot be contracted locally, i.e. through the dislocation loop it contains. (In absence of other dislocations, the membrane can be expanded until it is contracted to nothing due to the periodicity of the lattice --- this is a ``global'' operation, as performing it depends on whether there are other dislocations in the lattice.) Using the properties of the enclosed $\alpha\beta$ surface, we show that the membrane in fact can be locally transformed to the opposite flavor. It is numerically favorable to have a simple prescription for such a membrane (whether actually it is contractible or not in given circumstances) such that it is locally of the smallest possible volume. Remarkably, such ``minimal membranes'' are easily constructed even in the case of linked dislocation loops (when two $\alpha\beta$ surfaces intersect, Fig.~\ref{fig:7f}).
% Such membranes behave independently, so we include that case in the prescription, which is as follows:
The prescription for constructing minimal membranes in presence of various dislocation loops follows:

\begin{description}
\item[(Edge loop)] Apply all star operators that are going to be removed in creation of the $\alpha\beta$ surface; after site removal, the leftover $\sigma^x$ operators form the minimal membrane; these are actually acting on nearest neighbor (opposite flavor) spins on the two sides of the $\alpha\beta$ surface. Summarily,
    \begin{equation}
    \label{eq:15}
      \Gamma_{edgemin}=\prod_{s\in V}A_s,
  \end{equation}
where $V$ is the set of star operators that are removed by the Volterra ($\alpha\beta$) surface.
\item[(Screw loop)] Apply all star operators that act on spins in screw plane, then apply the new, repaired version of same stars after screw operation; all these $\sigma^x$ operators form the minimal membrane; these are actually acting on all spins in the screw plane (both flavors), except along the dislocation edge-B (Fig.~\ref{fig:7e}) where the new stars are more heavily changed by local surgery (Fig.~\ref{fig:7e}). The procedure can be summarized by the formula:
    \begin{equation}
    \label{eq:16}
      \Gamma_{screwmin}=\prod_{s\in V}A_s\tilde{A}_s,
    \end{equation}
    where the set $V$ contains all star operators that act on spins in the Volterra surface, and $A_s$ are stars in the original lattice, while $\tilde{A}_s$ are the same stars after ``surgery''.

\item[(Linked edge and screw loops)] Apply screw loop surgery of lattice; construct the screw minimal membrane (will be edited in last step); construct edge loop minimal membrane (note that in this step the stars on the intersection of the two $\alpha\beta$ surfaces have already been edited by screw loop); finally, apply edge loop surgery on lattice (which removes some operators from the screw membrane), Fig.~\ref{fig:7f}.
\end{description}

The last case constructs two minimal membranes, for each of the linked dislocation loops. These two membranes are actually independent, i.e. they cannot cancel each other. In other words, however we modify the product of these membranes by multiplying it with stars in the final dislocated lattice, there will always remain $\sigma_x$ operators acting on some spins, and thus the two membranes cannot cancel each other.

To analyze the membranes as we did the strings, we have directly inspected all situations on the 3dTC lattice (Figs.~\ref{fig:7b}-\ref{fig:7f}) to derive a set of general rules that the membranes obey as they are deformed and moved through the lattice by the action of stars which have undergone proper ``surgery'' due to the dislocated lattice (Appendix~\ref{sec:appendixsurgery}). Having in mind the detailed discussion in Section~\ref{sec:dislocation-loops}, we fix the Volterra $\alpha\beta$ surface of each dislocation loop, and retain the global flavor assignment from the original ideal lattice.

The basic set of rules for the behavior of membranes is presented in Fig.~\ref{fig:2}, which also shows their relation to non-contractible strings. Intuitively, a membrane is an elastic closed surface which can pass through dislocation loops. However, if it is created such that it encompasses an entire dislocation loop (and its $\alpha\beta$ surface), it cannot locally be contracted, but instead can just locally (using the $\alpha\beta$ surface) change flavor. Such a membrane we will call ``minimal'' (for the given dislocation loop).
% Such a membrane can be maximally ``deflated'' until it is tightly pressed on the $\alpha\beta$ surface, and then occupies both flavors, one on each side.

Following our analysis for strings above, we again consider the two simple examples:
\begin{itemize}
\item A single dislocation loop does not change the GSD: The minimal membrane is trivial, as it can be expanded and, using system periodicity, contracted far away from the dislocation loop, as no other dislocation loop is in the way.
  \item 
    Two separate dislocation loops add \textit{one} non-trivial membrane. Here the minimal membrane of one loop can be expanded in the periodic system until it engulfs the other loop tightly. So, the two separate minimal membranes of the two dislocation loops are actually equivalent. Further, due to the flavor inversion rule the two flavors of the minimal membrane are also equivalent. We are thus left with only one independent non-contractible membrane candidate. The standard system-spanning strings must pierce this closed membrane an even number of times (possibly $0$ times). The new non-contractible string (which pierces both $\alpha\beta$ surfaces once) is topologically forced to anticommute with the minimal membrane, because the membrane can be considered as of single flavor (either one), while the string has a flavor change on the $\alpha\beta$ surface which is \textit{inside} the membrane. Notice that using system periodicity to change the meaning of word ``inside'' to ``outside'' gets us nowhere, as this operation topologically only maps one dislocation loop onto the other.
  \end{itemize}
  
Moving on to $k$ dislocation loops, one needs to consider an addition of one new dislocation loop. The appropriate candidate for the new membrane must be the minimal membrane of the new loop. Notice how the situation here is different than for strings. Namely, it is immediately obvious that by expanding in the periodic system, the new membrane just encompasses all old loops, and is therefore just the combination (product) of all the old minimal membranes. However, this does not mean it is dependent! Consider the old system, before adding the new dislocation loop. A product of all minimal membranes there is \textit{globally} contractible, by expansion away from the dislocation loops, and using the periodicity of the 3d system. With the new loop however, expanding the product of all old minimal membranes does not contract it, because the new loop is in the way. We just establish that the new minimal membrane is equivalent to the product of the old ones, but this is not a trivial membrane anymore. Therefore the product of all the old minimal membranes is now a non-contractible operator that anticommutes with the new string, due to the string's single piercing of the new dislocation loop.

From the membrane perspective it is even more remarkable how dislocation line linking does not alter the GSD. The minimal membranes of two dislocation loops stay well defined and distinguishable operators, even as the two dislocation loops are pushed into each other, and their $\alpha\beta$ surfaces intersect (see Fig.~\ref{fig:2}b, and the explicit lattice construction further in this section, as well as Fig.~\ref{fig:7f}). The two linked minimal membranes stay separate in the following sense: If we start from minimal membrane of the first $\alpha\beta$ surface, expand it using allowed deformations by stars, then once it grows through the periodic system, it will shrink back precisely into the form of the minimal membrane for the second $\alpha\beta$ surface, which belongs to the other linked dislocation loop (assuming no other dislocation loops in the system).

Finally, let us clarify the 3dTC with parallel dislocation pairs spanning the system in one periodic dimension (say $\ZZ$). In this case the membranes orthogonal to the $\XX\YY$ plane take the role of the 2dTC $\sigma^x$ strings, due to translational invariance along $\zz$. The counting in the 2dTC therefore matches the 3dTC when we consider any $\XX\YY$ cut through the 3d system. Additionally, the 3dTC still has a single $\ZZ$ spanning string (two flavors are equivalent) and a single $XY$ membrane (the two flavors are equivalent), so that there is one extra anticommuting pair in the 3dTC, leading to the
\begin{equation}
  \label{eq:50}
|GSD_{3d,spanning}|=2|GSD_{2d}|.
\end{equation}
See Table~\ref{tab:1} for exact numerical results that corroborate this, and Section~\ref{sec:gsd-stabilizers-k} for additional details about the strings.

\section{Computing the ground state degeneracy of $G_n$ gauge theory using multi-component $U(1)$ Chern---Simons theory with a twist}
\label{sec:App_CS}

The $G_n\equiv(\Zn\times \Zn)\rtimes \Zz$ gauge theory can be described by the multi-component $U(1)$ Chern---Simons theory with the following K-matrix:
\begin{align}
  \label{eq:17}
  K&=\left(
    \begin{matrix}
      K_\alpha & 0 \\
      0 & K_\beta \\
   \end{matrix}
 \right),\\
 K_\alpha&=K_\beta=\matl{0}{n}{n}{0},
\end{align}
with two blocks for two flavors (copies). The Chern---Simons theory on the space manifold $M$ has the Lagrangian
\begin{align}
  \label{eq:18} \mathsf{L}&=\frac{n}{4\pi}\int_M\sum_{f=\alpha,\beta}\left(a^f\textrm{d}\tilde{a}^f+\tilde{a}^f\textrm{d}a^f\right)=\\
  &\equiv\frac{1}{4\pi}\int_M K_{IJ}A^I\textrm{d}A^J,
  \end{align}
  with $A=A_\mu\textrm{d}x^\mu$, $A_\mu=(a_\mu^\alpha,\tilde{a}_\mu^\alpha,a_\mu^\beta,\tilde{a}_\mu^\beta)^T$ a 4-vector consisting of two flavored parts with components
  \begin{equation}
    \label{eq:30}
      A^f_\mu=(a^f_\mu,\tilde{a}^f_\mu)^T,
  \end{equation}
  while $x^\mu=t,x,y$. The Lagrangian needs to be supplied with proper boundary conditions; crucially, the gauge theory contains flavor-exchanging $\Zz$ twists. On a torus the twists are represented by
  \begin{equation}
    \label{eq:20}
    A_\mu(\rr+L_i\hat{e}_i)=(\sigma_1^{\epsilon_i}\otimes\openone_2) A_\mu(\rr) (\sigma_1^{\epsilon_i}\otimes\openone_2),
  \end{equation}
  where $\epsilon_x,\epsilon_y=0,1$ determine whether there is a $\Zz$ twist along the given direction, and $L_i$ is the system size in direction $i=x,y$. The theory on the torus therefore separates into $4$ sectors (one untwisted, three with a twist). On a genus $g$ surface, the presence of twists along the $2g$ elementary cycles (non-contractible loops) selects one out of the $2^{2g}$ topological twist sectors.

In the next two subsections we evaluate the ground state degeneracy on the torus ($|GSD|=S^{(n)}_{g=1}$) and a genus $g$ surface ($|GSD|=S^{(n)}_g$, as well as comment on the nature of the theories.

Let us note in advance that the $S^{(n)}_g$ result for $G_n$ derived below, Eq.~\eqref{eq:49} (or Eq.~\eqref{eq:54} and Eq.~\eqref{eq:52} of main text), formally coincides with the result for twisted (single flavor) $\mathbb{Z}_{n^2}$ theory described by $\mathbb{Z}_{n^2}\rtimes\Zz$.\cite{Barkeshli:2012p7607} Before the twist the latter theory has a 2x2 $K$-matrix, $K=\mat{0}{n^2}{n^2}{0}$ (contrast to Eq.~\eqref{eq:17}). This implies that the number of species and quantum dimensions of quasiparticles also coincide between $G_n$ and $\mathbb{Z}_{n^2}\rtimes\Zz$, remarkably since the theories are fundamentally different, and the difference is manifest in the derivation of $S^{(n)}_g$ itself (compare Ref.~\onlinecite{Barkeshli:2010p6844} to section~\ref{sec:degeneracy-genus-g}). However, due to the physical argument of the gauge structure (see Sec.\ref{sec:zntim-theory:-group}), we expect that the $G_n$ is the correct description for the physical state of the melted doubled toric code.

\subsection{Degeneracy on a torus}
\label{sec:degeneracy-torus}

We first consider the ground state degeneracy of the $(\Zn\times \Zn)\rtimes\Zz$ theory of melted toric code on a torus ($L_x=L_y=L$ for simplicity). Setting the gauge $A_0=0$ implies that the gauge fields of flat connections are irrotational and characterized by their holonomies along the non-contractible loops on the torus. With an additional rescaling,
\begin{align}
  \label{eq:21}
  A^f_x(t,\rr)&=\frac{2\pi}{L}(\bAx^f(t),\tilde{\bAx}^f(t))^T,\\
    A^f_y(t,\rr)&=\frac{2\pi}{L}(\bAy^f(t),\tilde{\bAy}^f(t))^T,
\end{align}
these uniform fields therefore parametrize the gauge-inequivalent configurations. We additionally need to take into account that large gauge transformations enforce the identifications
\begin{align}
  \label{eq:22}
  (\bAx^f,\tilde{\bAx}^f)&\Longleftrightarrow (\bAx^f+k_f,\tilde{\bAx}^f+k_f),\\
  (\bAy^f,\tilde{\bAy}^f)&\Longleftrightarrow (\bAy^f+l_f,\tilde{\bAy}^f+l_f), \quad k_f,l_f\in\mathbb{Z}.
\end{align}
The Lagrangian from Eq.~(\ref{eq:18}) becomes
\begin{equation}
  \label{eq:23}
\mathsf{L}=2\pi n\sum_f\left(\bAx^f \dot{\tilde{\bAy}}^f + \tilde{\bAx}^f \dot{\bAy}^f\right),
\end{equation}
leading to a vanishing Hamiltonian and therefore a wavefunction that is a linear combination of simple plane waves:
\begin{equation}
  \label{eq:24}
  \psi(\bAx^f,\tilde{\bAx}^f)=\sum_{\substack{p_\alpha,q_\alpha\\ p_\beta,q_\beta}}C_{p_\alpha,q_\alpha}^{p_\beta,q_\beta}\prod_f \delta[2\pi n\bAx^f-2\pi p_f]\delta[2\pi n\tilde{\bAx}^f-2\pi q_f],
\end{equation}
with $p_f,q_f$ being integers, $p_f,q_f=0\ldots n-1$, and $\delta[x]$ the Kronecker delta function. We can label states by
\begin{equation}
  \label{eq:37}
\ket{\psi}=\ket{p_\alpha,q_\alpha;p_\beta,q_\beta},
\end{equation}
and gauge equivalence from Eq.~(\ref{eq:22}) limits the number of independent coefficients which enumerate the independent states. Namely,
\begin{align}
  \label{eq:25}
  C_{p_\alpha,q_\alpha}^{p_\beta,q_\beta}&=C_{p_\alpha-n,q_\alpha}^{p_\beta,q_\beta}=  C_{p_\alpha,q_\alpha-n}^{p_\beta,q_\beta}=\notag\\
  &=C_{p_\alpha,q_\alpha}^{p_\beta-n,q_\beta}=  C_{p_\alpha,q_\alpha}^{p_\beta,q_\beta-n},
\end{align}
giving $n^2$ independent states per flavor, since $p_f,q_f=0\ldots n-1$.

The independent states in the untwisted sector are physical only if they are $\Zz$ symmetric. This constraint leads to $n^2$ states of the form $\ket{p,q;p,q}$ and $\frac{n^2(n^2-1)}{2}$ states of the form $\ket{p_\alpha,q_\alpha;p_\beta,q_\beta}+\ket{p_\beta,q_\beta;p_\alpha,q_\alpha}$, for the total of
\begin{equation}
  \label{eq:26}
  S^{(n)}_{g=1,untwisted}=\frac{n^2(n^2+1)}{2}.
\end{equation}

There are three twisted sectors, having a $\Zz$ twist along $x$, $y$, or both directions of the torus. There exist symmetry transformations mapping the sectors onto each other, so that all three have the same degeneracy~\cite{Barkeshli:2010p6844}. Focusing on a $y$-twist example, in accordance with Eq.~(\ref{eq:20}) we demand that
\begin{align}
  \label{eq:27}
  A^\alpha_i(x,y+L)&=A^\beta_i(x,y)\\
  A^\beta_i(x,y+L)&=A^\alpha_i(x,y),\\
  A^f_i(x+L,y)&=A^f_i(x,y).
\end{align}
We can now switch to a field $B^f_\mu$ defined on the system with double length along $y$. Physically, we glue the two tori of opposite flavors together after cutting them open completely along one direction ($y$), which is along the dislocation ``tunneling loop'' winding along $y$. (This loop is the remnant $\alpha\beta$ line of the annihilated dislocation---anti-dislocation pair.) The resulting system is topologically still a torus. The new field is constructed as:
\begin{equation}
  \label{eq:28}
  B_\mu(x,y)\equiv(b_\mu,\tilde{b}_\mu)^T=\left\{
    \begin{aligned}
      &A^\alpha_\mu(x,y),\quad &0\leq y<L\\
            &A^\beta_\mu(x,y-L),\quad &L\leq y<2L
          \end{aligned}
        \right.
\end{equation}
and satisfies
\begin{equation}
  \label{eq:29}
  B_\mu(x,y)=  B_\mu(x+L,y)=  B_\mu(x,y+2L).
\end{equation}
Similarly to the untwisted case, we can set $B_0=0$ and parametrize
\begin{equation}
  \label{eq:31}
  B_i(t,x,y)=  \frac{2\pi}{L_i} (X_i(t),\tilde{X}_i(t))^T,
\end{equation}
where $L_y=2L_x\equiv 2L$. The Lagrangian Eq.~(\ref{eq:18}) becomes
\begin{align}
  \label{eq:32}
\mathsf{L}&=\frac{n}{4\pi}\int_0^L\textrm{d}x \int_0^{2L}\textrm{d}y\left(b\textrm{d}\tilde{b}+\tilde{b}\textrm{d}b\right)=\notag\\
  &=2\pi n\left(X_1\dot{\tilde{X}}_2+\tilde{X}_1\dot{X}_2\right),
\end{align}
while the large gauge transformations consistent with Eq.~(\ref{eq:28}) introduce the periodicity
\begin{equation}
  \label{eq:33}
  X_i\Longleftrightarrow X_i+1.
\end{equation}
The trivial equations of motion and periodicity lead to the wavefunction
\begin{equation}
  \label{eq:34}
  \psi(X_i)=\sum_{p,q}c_{p,q}\delta[2\pi n X_1-2\pi p]\delta[-2\pi n X_2-2\pi q],
\end{equation}
where $ c_{p,q}=c_{p-n,q}=c_{p,q+n}$, and therefore the degeneracy of the $y$-twisted sector is $n^2$. Note that the twisted sector is equivalent to a single flavored ($\Zz$ invariant) untwisted sector (e.g. Eqs.~(\ref{eq:32}) and (\ref{eq:23})).

Since there are 3 twisted sectors, it follows $S^{(n)}_{g=1,twisted}=3n^2$, and the total ground state degeneracy on the torus is
\begin{equation}
  \label{eq:36}
S^{(n)}_{g=1}=\frac{n^2(n^2+7)}{2}.
\end{equation}

This result precisely matches the one obtained from a simpler physical argument for the melted state, Eq.~\eqref{eq:51}.

\subsection{Degeneracy on genus $g$ surface}
\label{sec:degeneracy-genus-g}

The theory on a genus $g$ surface can be analyzed in the same way as the theory on torus. The defining holonomies of gauge fields are now along $2g$ elementary cycles, i.e. non-contractible loops forming a homology basis on the surface. The canonical basis can be made by choosing loop pairs $\{\cy_i,\cyy_i\}$, with $i$-th pair tied to $i-$-th ``hole'' in the surface, $i\in\{1,\ldots,g\}$, so that only intersections of loops are between $\cy_i$ and $\cyy_i$ (see also Ref.~\onlinecite{Barkeshli:2010p6844}).

We start with the untwisted sector. As on torus, Eq.~(\ref{eq:37}), the holonomies now lead to labeling of states by
\begin{equation}
  \label{eq:38}
  \ket{\psi}=\otimes_i\ket{p^i_\alpha,q^i_\alpha;p^i_\beta,q^i_\beta},\quad i=1\ldots g,
\end{equation}
where as in Eq.~(\ref{eq:25}) the equations of motion lead to identifications:
\begin{align}
  \label{eq:39}
  (p^i_\alpha,q^i_\alpha;p^i_\beta,q^i_\beta)&\Leftrightarrow (p^i_\alpha-n,q^i_\alpha;p^i_\beta,q^i_\beta) \Leftrightarrow (p^i_\alpha,q^i_\alpha-n;p^i_\beta,q^i_\beta)\notag\\
&\Leftrightarrow(p^i_\alpha,q^i_\alpha;p^i_\beta-n,q^i_\beta) \Leftrightarrow(p^i_\alpha,q^i_\alpha;p^i_\beta,q^i_\beta-n).
\end{align}
The $\Zz$ invariance to $\alpha\leftrightarrow\beta$ allows only the combinations $\otimes_i\ket{p^i_\alpha,q^i_\alpha;p^i_\alpha,q^i_\alpha}$ and $\otimes_i(\ket{p^i_\alpha,q^i_\alpha;p^i_\beta,q^i_\beta}+\ket{p^i_\beta,q^i_\beta;p^i_\alpha,q^i_\alpha})$. There are $(n^2)^g$ of the former and $[(n^2\cdot n^2)^g-(n^2)^g]/2$ of latter, and since there is only one untwisted sector, a total of
\begin{equation}
  \label{eq:40}
  S^{(n)}_{g,untwisted}=n^{2g}\frac{n^{2g}+1}{2}.
\end{equation}

Next consider the $2^{2g}-1$ twisted sectors, labeled by the combination of cycles across which there is a twist. As discussed in detail in Ref.~\onlinecite{Barkeshli:2010p6844}, using Dehn twist transformations of the surface, $\Zz$ twists across any combination of cycles can be reduced to a twist across a single cycle. It is enough therefore to consider a $\Zz$ twist across a single elementary cycle, e.g. the $\cy_g$ loop. Picking $g$-th cycle $\cy$ for the twist means that on an $\alpha$-flavored copy of the surface, $\sug^\alpha$, the $g$-th hole is cut open and glued to its identical copy $\sug^\beta$ along the cut. The resulting surface $\su$ has genus $2g-1$. A mirror symmetry of $\su$ through the $g$-th hole maps the two flavored copies into each other, $\sigma_\su:\sug^\alpha\leftrightarrow\sug^\beta$. We can again introduce a single gauge field:
\begin{equation}
  \label{eq:41}
    B_\mu(\rr)\equiv(b_\mu,\tilde{b}_\mu)^T=\left\{
    \begin{aligned}
      &A^\alpha_\mu(\rr),\quad &\rr\in\sug^\alpha\\
            &A^\beta_\mu(\sigma_\su\rr),\quad &\rr\in\sug^\beta
          \end{aligned}
        \right.,
\end{equation}
leading to the action
\begin{equation}
  \label{eq:42}
\mathsf{L}=\frac{n}{4\pi}\int_\su\left(b\textrm{d}\tilde{b}+\tilde{b}\textrm{d}b\right).
\end{equation}
Since the flat connection $B$ is determined by its holonomies, it can be parametrized as
\begin{align}
  \label{eq:43}
    b&=2\pi\sum_{i=1}^{2g-1}(\bbx^i\omx_i+\bby^i\omy_i)\\
    \tilde{b}&=2\pi\sum_{i=1}^{2g-1}(\tilde{\bbx^i}\omx_i+\tilde{\bby^i}\omy_i)
  \end{align}
  using the dual (1-form) basis of the homology basis
  \begin{align}
    \label{eq:44}
    \int_{\cyc^v_i}\om^{(w)}_j=\delta_{vw}\delta_{ij}.
  \end{align}
  Gauge transformations again lead to $\bbx^i,\bby^i$ being defined only up to integers, while the action Eq.~(\ref{eq:42}) becomes
  \begin{equation}
    \label{eq:46}
\mathsf{L}=2\pi n\sum_{i=1}^{2g-1}\left(\bbx^i\dot{\tilde{\bby^i}}+\tilde{\bbx^i}\dot{\bby^i}\right),
\end{equation}
essentially $2g-1$ copies of the twisted torus sector theory, Eq.~(\ref{eq:32}). The states are therefore labeled by
\begin{equation}
  \label{eq:47}
\ket{\psi}=\otimes_i\ket{p^i,q^i},\quad i=1\ldots 2g-1;\quad p^i,q^i=1\ldots n.
\end{equation}
The action of $\Zz$ here exchanges holes positioned on opposite halves of the surface, i.e. $i\leftrightarrow\sigma_\su i$, where the $i=g$ variables are mapped into themselves. ``Diagonal'' states have $p^i,q^i$, $i=1\ldots g-1$ quantum numbers repeated for $i=g+1\ldots 2g-1$ on the other half of the surface, while $p^g,q^g$ are chosen independently; this gives $(n^2)^{g-1}\cdot n^2=n^{2g}$ invariant states. The ``off-diagonal'' states are symmetrized, and have $(p^i,q^i)\neq(p^{\sigma_\su i},p^{\sigma_\su i})$ for $i\neq g$, while $p^g,q^g$ are independent; this gives $n^2\cdot(n^2)^{g-1}((n^2)^{g-1}-1)/2$. In total,
\begin{equation}
  \label{eq:48}
  S^{(n)}_{g,twisted}=n^{2g}\frac{n^{2(g-1)}+1}{2}.
\end{equation}
Taking into account the number of sectors, Eqs.~(\ref{eq:40}) and (\ref{eq:48}) give
\begin{equation}
  \label{eq:49}
  S^{(n)}_g=n^{2g}\frac{n^{2g}+1}{2}+(2^{2g}-1)n^{2g}\frac{n^{2(g-1)}+1}{2},
\end{equation}
which for $g=1$ reduces to the torus results, Eqs.~(\ref{eq:26}),~(\ref{eq:36}).

The final expression for GSD in Eq.~\eqref{eq:49} precisely matches the one obtained from a simpler physical derivation of the melted state, Eq.~\eqref{eq:52}, and it also matches (as it must) the GSD result obtained directly from the structure of the gauge group $G_n$, Eq.~\eqref{eq:53}.

\section{GSD of $D_4$ topologically ordered state on the three-torus $T^3$}
\label{sec:gsd-d_4-topol}

%%%%%%%%%%%%%%%%%%%%%%%%%%%%%%%%%%%%%%%
\begin{figure}
  \centering
\includegraphics[width=0.45\textwidth]{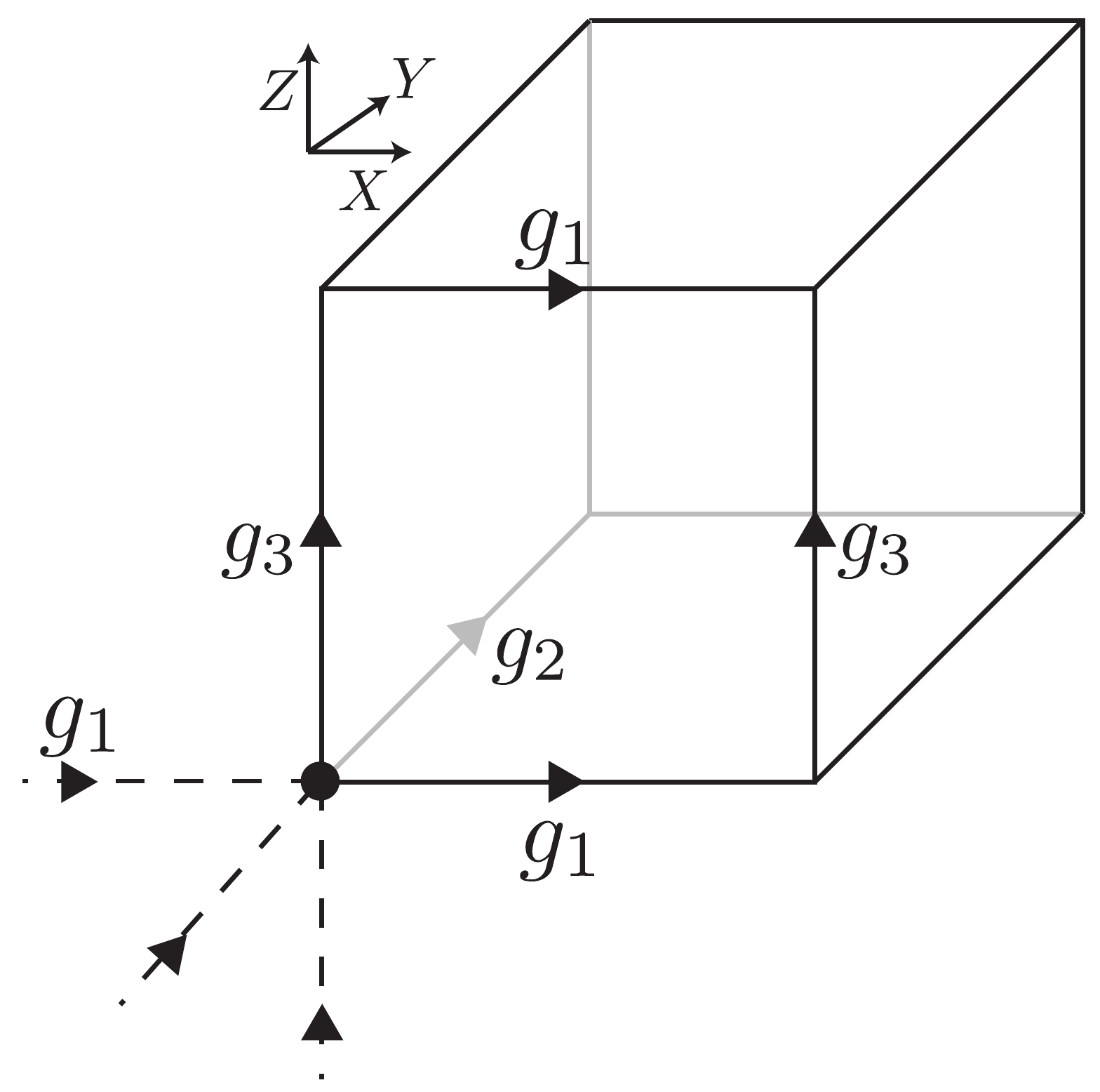}
\caption{A minimal representation of three-torus, having one site (dot) of a cubic lattice with periodic boundary conditions in all three directions. There are three independent group elements $g_1,g_2,g_3$ on directed edges (arrows) of this unit-cell, the other edges being determined by periodicity. Model with topological order with gauge group $G$ in this system contains four commuting operators: three force the flux on three cubic faces to zero, while the fourth acts similarly to a lattice gauge transformation on the six directed edges connected to the lattice site (dashed lines are periodic images of edges).}
\label{fig:D4}
\end{figure}
%%%%%%%%%%%%%%%%%%%%%%%%%%%%%%%%%%%%%%%

To build an exactly solvable model with topological order described by discrete (possibly non-Abelian) gauge group $G$ in three dimensions with periodic boundary conditions (i.e. on a three-torus $T^3$), it is enough to consider a single unit-cell of a cubic lattice, Fig.~\ref{fig:D4}. We follow the model construction from Refs.\onlinecite{Mesaros:2013p7698,Hu:2012p7528}. There is only one lattice site and three directed edges in the unit-cell cube, the rest being automatically determined by the periodicity. Each edge is assigned a direction (marked by arrow in Fig.~\ref{fig:D4}) and a group element, so the degrees of freedom are $g_i\in G$, $i=1,2,3$. The orthonormal basis of the Hilbert space is $\ket{g_1,g_2,g_3}$. When an edge with group element $g$ is considered in direction opposite to its arrow, the assigned group element is considered as the inverse $g^{-1}$.

The model is defined using operator $A$ (tied to the single lattice site) and operators $B_{ij}$ tied to the three faces of the unit-cell cube. The $B_{ij}$ operator projects the total flux on the $ij$ face of the cube to zero, e.g. $B_{XZ}$ enforces the identity $g_1\cdot g_3\cdot g_1^{-1}\cdot g_3^{-1}=\openone$, where direction of edges is taken into account as one goes around the $XZ$ square plaquette. (Formally, the zero-flux definition leads to the expression $B_{XZ}\ket{g_1,g_2,g_3}=\sum\limits_{\substack{h_1,h_2\in G\\ [h_1,h_2]=0}}\delta(h_1,g_1) \delta(h_2,g_3)\ket{g_1,g_2,g_3}$.)

Next we consider the $A_g$ operator, which acts as a gauge transformation on the six edges connected to the cubic lattice site: Element $g_i$ on edge $i$ is transformed into $g\cdot g_i$ [$g_i\cdot g^{-1}$ ] if the edge is directed away [towards] the site. Note that three of the six edges are periodic images (dashed lines in Fig.~\ref{fig:D4}). The site operator $A=1/|G|\sum_{g\in G}A_g$ therefore acts as:
\begin{equation}
  \label{eq:58}
  A \ket{g_1,g_2,g_3}=\frac{1}{|G|}\sum_{g\in G}\ket{g g_1 g^{-1},g g_2 g^{-1},g g_3 g^{-1}}.
\end{equation}

It is easy to check that $A$ and $B_{ij}$ operators are all projectors and commute with each other. The exactly solvable Hamiltonian is therefore given by
\begin{equation}
  \label{eq:57}
  H=-A-B_{XY}-B_{YZ}-B_{XZ},
\end{equation}
and describes a phase with $G$ topological order.\cite{Hu:2012p7528,Mesaros:2013p7698}
The ground state manifold is given by solving the set of equations:
\begin{align}
  \label{eq:59}
  A\ket{\psi}&=\ket{\psi}\\\notag
  B_{ij}\ket{\psi}&=\ket{\psi}.
\end{align}
An arbitrary wavefunction is $\ket{\psi}=\sum\limits_{h_1,h_2,h_3\in G}c(h_1,h_2,h_3)\ket{h_1,h_2,h_3}$, and we consider the set of equations for the coefficients $c(h_1,h_2,h_3)$ which result from applying Eqs.~\eqref{eq:59}. (Note that the three $B_{ij}$ equations just enforce that $c(h_1,h_2,h_3)=0$ unless $h_1,h_2,h_3$ commute with each other.)

By solving the equations we find that out of $|D_4|^3=512$ coefficients $c(h_1,h_2,h_3)$ the ground state manifold has exactly $92$ independent ones, thus the GSD of $D_4$ topologically ordered state on the three-torus is $92$.

\section{Local ``surgery'' for operators in lattice with dislocations}
\label{sec:appendixsurgery}
  
\begin{figure*}
  \centering
\includegraphics[width=1\textwidth]{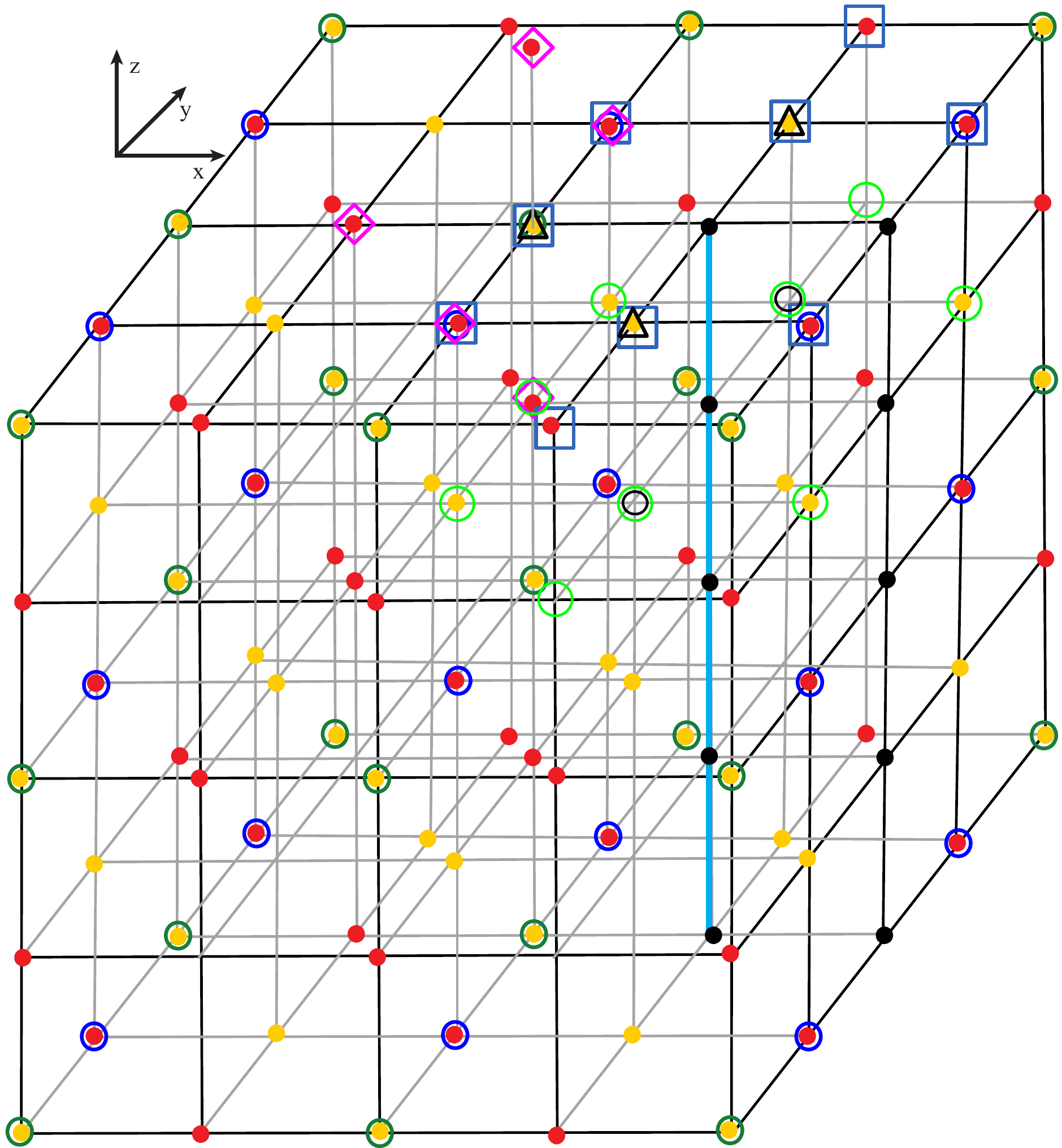}
\caption{Edge dislocation of an $\ZZ \XX$ loop, example of edge-1 segment along $\zz$. Black dots are removed sites, forming the $\alpha\beta$ surface. The dislocation line is represented by the outmost array of black sites, and in this edge-1 example does not contain star operators. All stabilizers that contain removed spins are repaired by just gluing their neighboring sites together in the local neighbor network, and that is how the flavor jump occurs as a string operator which pierces the $\alpha\beta$ plane is constructed step-by-step. The exception to the seamless repair are stabilizers that contain spins on the dislocation lines, for which the explicit repair ``surgery'', such that all stabilizers commute, is shown. The two plaquettes (two black triangles) lying in a star containing plane orthogonal to the dislocation line are merged into a new plaquette, with $\sigma^z$ acting on 9 blue square spins. Notice that this stabilizer locally mixes the flavors. The star operator (third black triangle) just loses the removed spin, acting on 5 purple diamonds. The surgery is repeated along edge, shifting by a lattice constant (two planes). A new local cubic constraint for plaquettes, containing the merged plaquette operator, can be constructed, using $\sigma^z$ on the light green spins. Two black circles mark the removed local constraints. The corner site of a square-shaped dislocation loop made of edge-1 is on an empty site, and therefore needs no special surgery. Locally in total, one plaquette and one constraint are removed, consistent with independence of GSD on the length of dislocation lines.}
\label{fig:7b}
\end{figure*}

\begin{figure*}
  \centering
\includegraphics[width=1\textwidth]{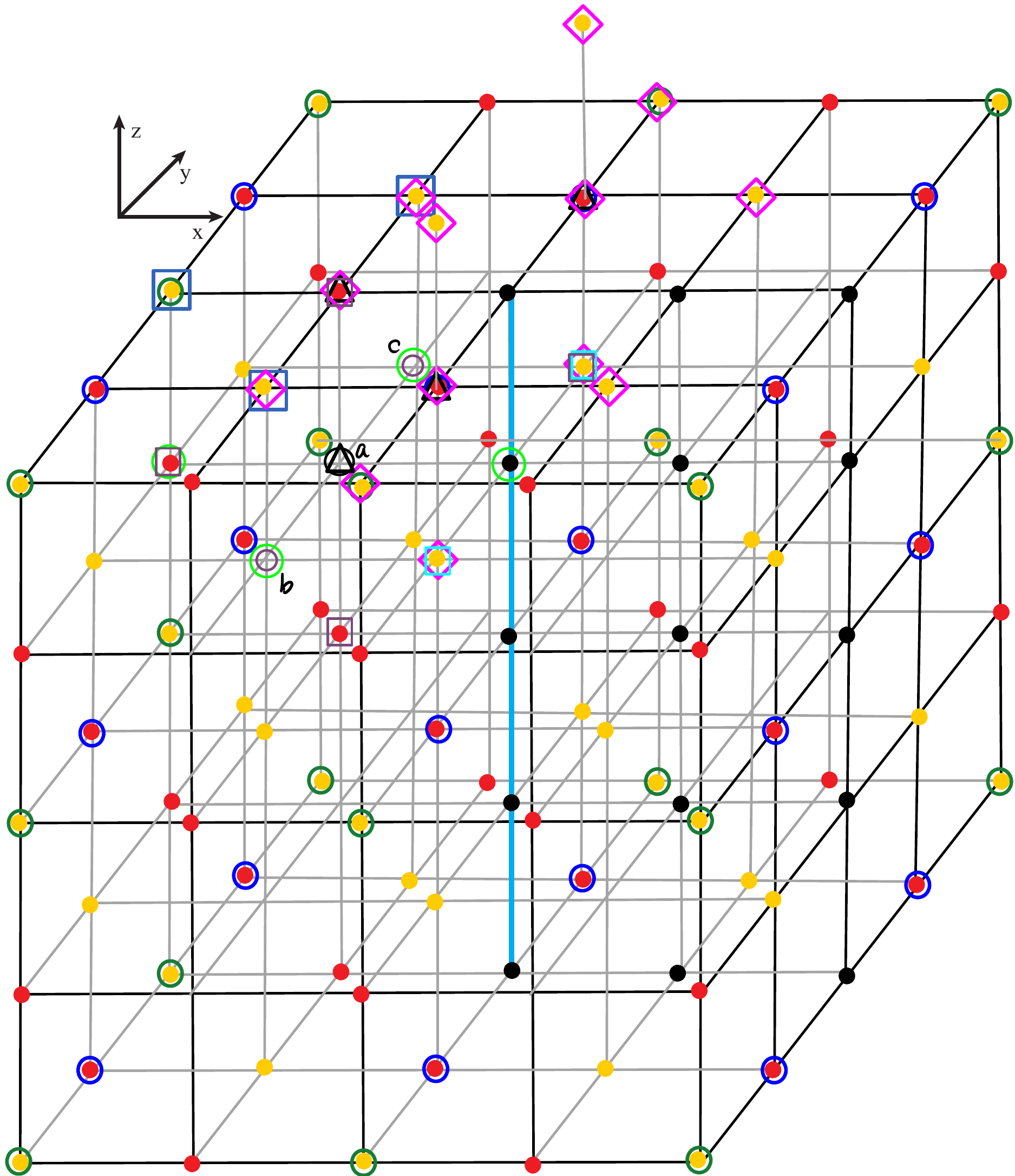}
\caption{Edge dislocation of an $\ZZ \XX$ loop, example of edge-2 segment along $\zz$. In this case the dislocation line contains star operators. See caption of Fig.~\ref{fig:7b} for details. The two star operators (black triangles) are merged into a new one acting on 13 purple diamonds. Two plaquettes are replaced by three new ones $B_1,B_2,B_3$, marked by 3 blue, 2 cyan, and 4 purple squares, respectively. Three cubic constraints (green circles) are replaced by new ones marked $a,b,c$, each at a center of a ``modified cube'' formed by neighboring plaquettes. The $a,b,c$ constraints include: two $B_1$ type plaquettes as ``cube'' top and bottom (in total 6 plaquettes); a $B_3$ as a ``cube'' back side (in total 6); and both $B_2$ and $B_3$ as ``cube'' sides (in total 7), respectively. }
\label{fig:7c}
\end{figure*}

\begin{figure*}
  \centering
\includegraphics[width=1\textwidth]{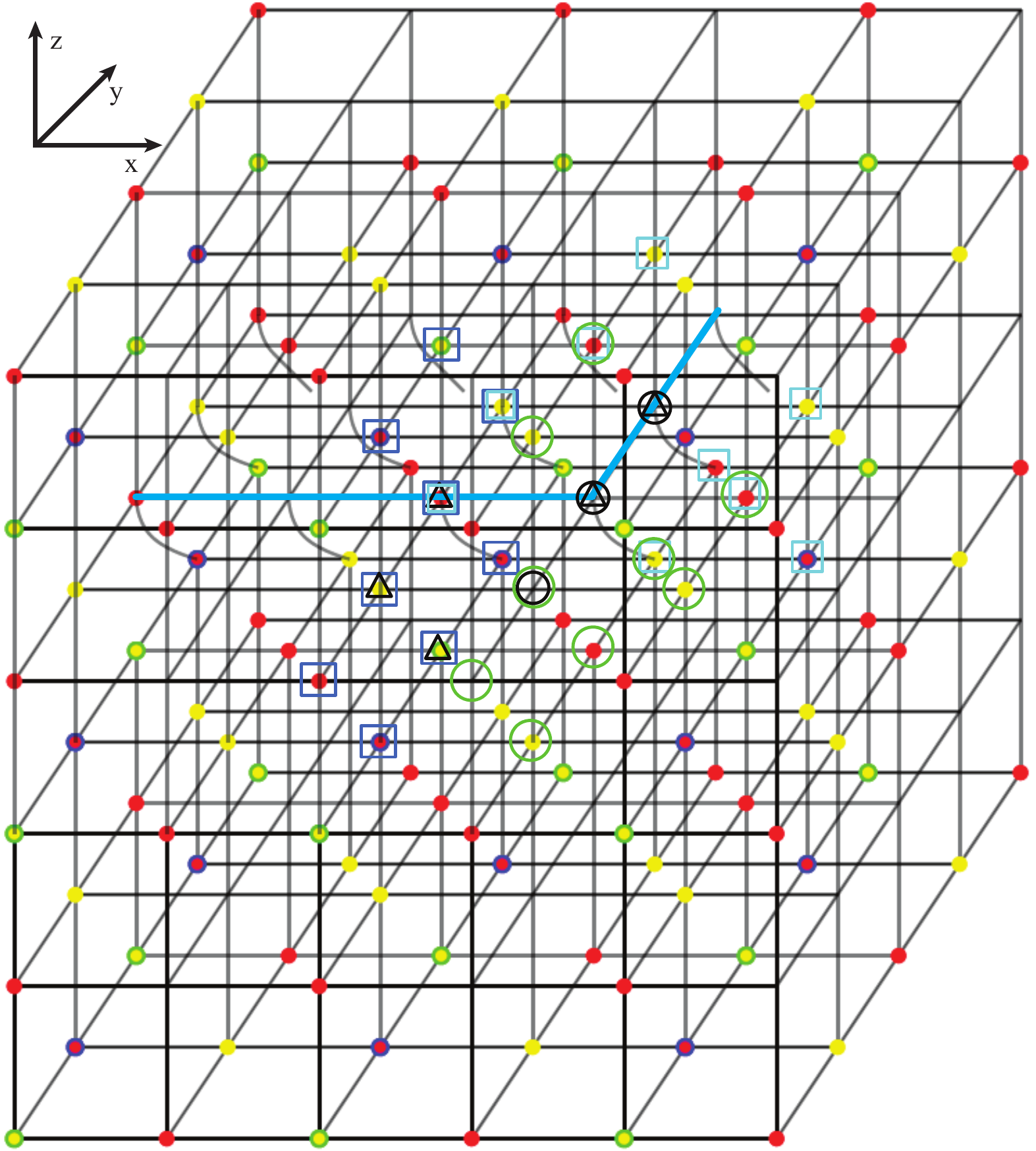}
\caption{Screw and edge-A dislocation segments of an $\XX\YY$ loop ($\YY\ZZ$ is equivalent). There are no removed sites in this dislocation loop, just a shearing of bonds site along $\yy$ within the loop. All stabilizers are redefined with such a new neighbor network on the lattice. 1) The screw dislocation line segment passes through the empty sites of the lattice, while edge-A passes through line without star operator sites. An $\ZZ\XX$ plane plaquette positioned on a screw dislocation line segment (example: rightmost black triangle for flavor $\beta$) is repaired by replacement by a 9 site $\sigma^z$ operator $B_{s1}$ (cyan squares). The cubic constraints on this segment (rightmost black circle for flavor $\alpha$) now use the repaired plaquette operators on front and back cube sides. 2) For each spin site $R$ on the edge-A segment, two plaquettes (black triangles) are merged into one new $B_{s2}$ (blue squares, mixing flavors), while the star positioned below $R$ is edited by removing its action on spin at $R$. Two local cubic constraints on either side of edge-A sites are merged into a single constraint containing two $B_{s2}$ plaquettes. Green circles mark plaquettes which are involved, together with $B_{s1}$ and $B_{s2}$, in the cubic constraint (black circle) at the loop corner. The corner should be included in the screw segment. In total locally, the number of stabilizers and constraints is unchanged, consistent with independence of GSD on the size of dislocation loops.}
\label{fig:7d}
\end{figure*}

\begin{figure*}
  \centering
\includegraphics[width=1\textwidth]{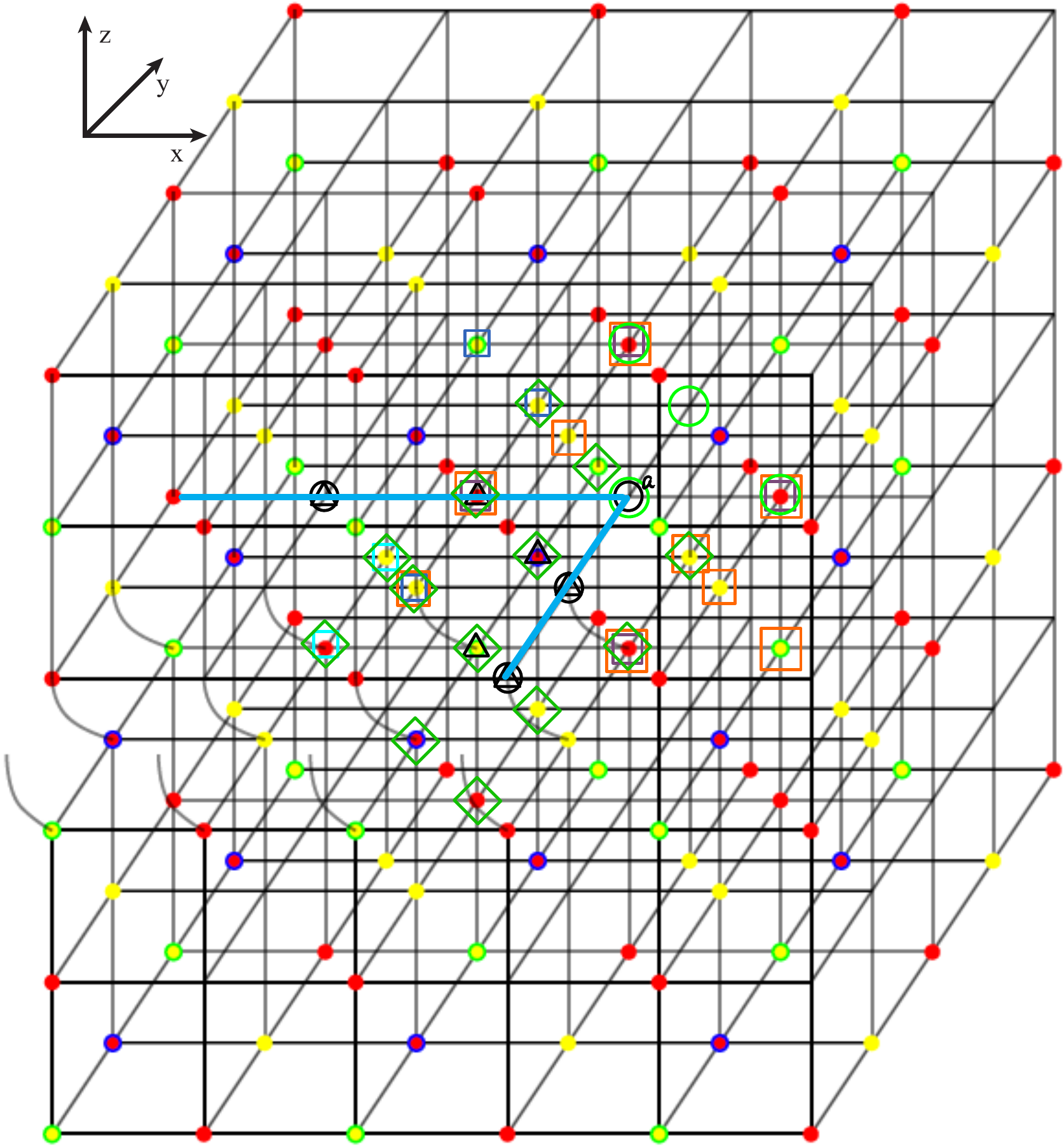}
\caption{Edge-B dislocation segment of an $\XX\YY$ loop ($\YY\ZZ$ is equivalent), lying along $\xx$ (inequivalent to edge-A due to $\bb=\yy$). See caption of Fig.~\ref{fig:7d}. The edge-B does not pass through star operator sites. The plaquettes labeled by spins on the dislocation line sites (example: black triangle for flavor $\beta$) have to lose one spin operator (blue squares). The plaquettes on empty sites of edge-B (examples are black circles) are replaced by two (purple and cyan squares). Pairs of star operators below edge-B (black triangles) are merged (green diamonds). The corner (marked with $a$) should be included into the edge-B segment. In total locally, the number of stabilizers (stars and plaquettes) and local cubic constraints is unchanged, consistent with independence of GSD on the size of dislocation loops.The cubic constraint belonging to corner ($a$) includes plaquettes labeled by green circles.}
\label{fig:7e}
\end{figure*}

\begin{figure*}
  \centering
\includegraphics[width=1\textwidth]{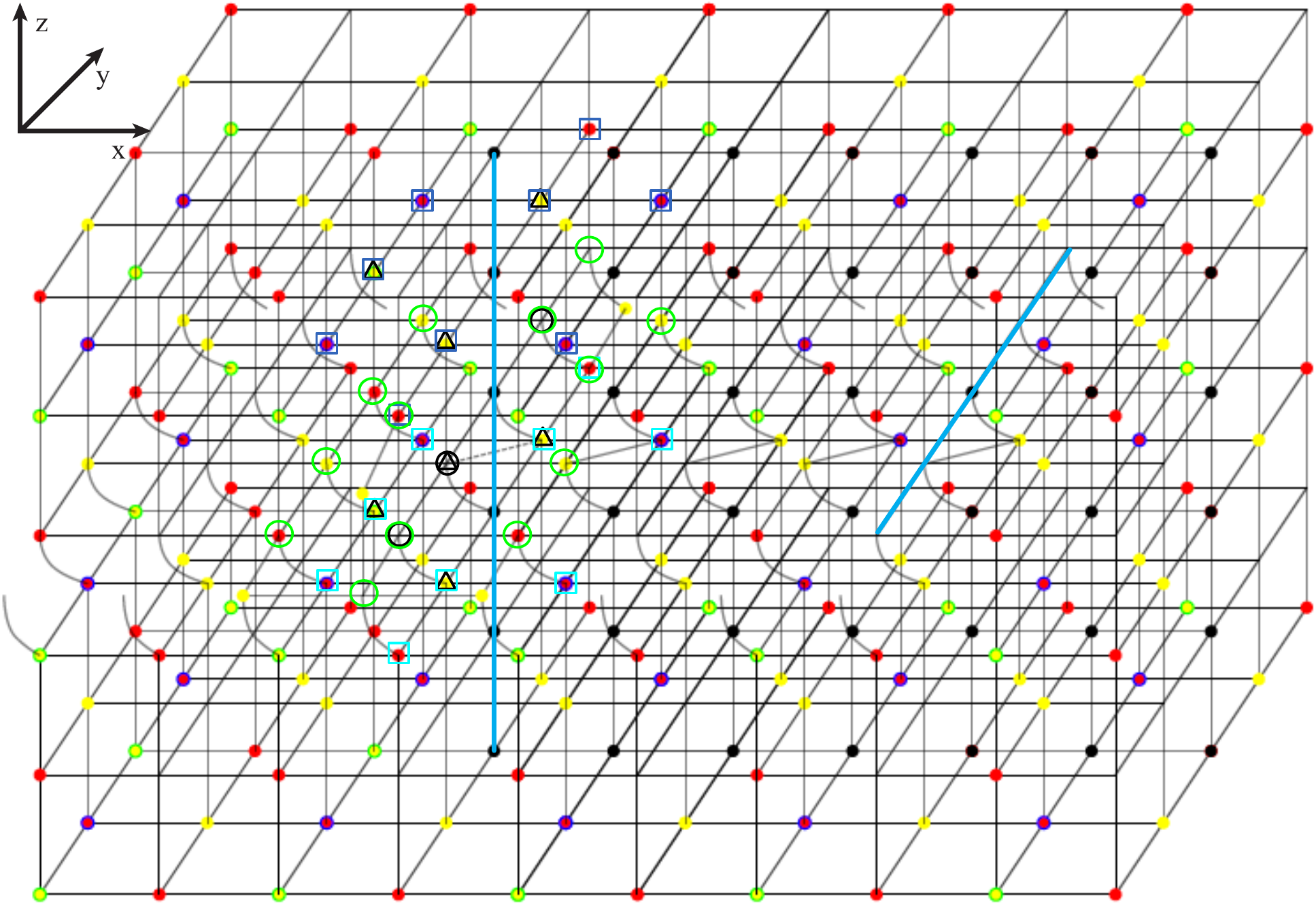}
  \caption{Dislocation loop linking, example of $\ZZ\XX$ and $\XX\YY$ loops. After the screw loop is created by editing the local neighbor network, the sites for the $\ZZ\XX$ loop should be removed, and network edited again per rules of Fig.~\ref{fig:7b}. A single plaquette in the screw plane (black triangle in circle) must be completely removed. The cubic constraint at that same site now contains $B_1$ plaquettes (Fig.~\ref{fig:7b}) and the ones marked by green circles.}
\label{fig:7f}
\end{figure*}

Here we present the explicit local ``surgery'' operations for redefining local star and plaquette operators, as well as other details for the 3d toric code analyzed in the main text. All the information is given in Figures~\ref{fig:7b} to \ref{fig:7f}.
%%%%%%%%%%%%%%%%%%%%%%%%%%%%%%%%%%%%%%%
%%%%%%%%%%%%%%%%%%%%%%%%%%%%%%%%%%%%%%%

%%%%%%%%%%%%%%%%%%%%%%%%%%%%%%%%%%%%%%%
%%%%%%%%%%%%%%%%%%%%%%%%%%%%%%%%%%%%%%%

%%%%%%%%%%%%%%%%%%%%%%%%%%%%%%%%%%%%%%%
%%%%%%%%%%%%%%%%%%%%%%%%%%%%%%%%%%%%%%%

%%%%%%%%%%%%%%%%%%%%%%%%%%%%%%%%%%%%%%%
%%%%%%%%%%%%%%%%%%%%%%%%%%%%%%%%%%%%%%%

%%%%%%%%%%%%%%%%%%%%%%%%%%%%%%%%%%%%%%%
%%%%%%%%%%%%%%%%%%%%%%%%%%%%%%%%%%%%%%%

\bibliography{ToricDislFinal1}

\end{document}